\documentclass[12pt]{article}

\pdfoutput=1

\usepackage[usenames,dvipsnames,svgnames,table]{xcolor} 
\usepackage[obeyspaces,hyphens,spaces]{url}
\usepackage{jcapmod}
\usepackage{verbatim}
\usepackage{graphics}
\usepackage{epsfig}
\usepackage{booktabs}
\usepackage{booktabs}
\usepackage[english]{babel}
\usepackage{amsmath, amssymb, amsbsy, amstext, amsthm, simplewick}
\usepackage{hyperref}
\usepackage{graphicx}
\usepackage{amsfonts}
\usepackage{upgreek}
\usepackage{framed}
\usepackage{tensor}
\usepackage{pifont}
\usepackage{latexsym, mathrsfs}
\usepackage{array}
\usepackage{hyperref}
\usepackage{xspace}
\usepackage{longtable}
\usepackage{multirow}
\usepackage{cases}
\usepackage{empheq}

\usepackage{bm}
\usepackage{bbm}
\usepackage{float}
\usepackage{afterpage}
\usepackage{setspace}
\usepackage{caption}
\usepackage[load-configurations=astronomy, range-units=brackets, range-phrase=-, per-mode=reciprocal, mode=math]{siunitx}
\usepackage{threeparttable}
\usepackage{subfigure}
\usepackage{tcolorbox}
\allowdisplaybreaks 

\usepackage{braket}

\newcommand{\be}{\begin{equation}}
\newcommand{\ee}{\end{equation}}
\newcommand{\bea}{\begin{eqnarray}}
\newcommand{\eea}{\end{eqnarray}}
\newcommand{\bml}{\begin{subequations}}
\newcommand{\eml}{\end{subequations}}
\newcommand{\bfig}{\begin{figure}}
\newcommand{\efig}{\end{figure}}

\newcommand{\eps}{\epsilon}

\newcommand{\slashed}{\hspace{-1.1ex}/}

\newcommand{\bmat}{\begin{pmatrix}}
\newcommand{\emat}{\end{pmatrix}}

\usepackage{ragged2e}

\usepackage{hhline}
\usepackage{array}
\newcolumntype{P}[1]{>{\centering\arraybackslash}p{#1}}
\usepackage{booktabs}
\usepackage{tikz}
\usetikzlibrary{calc,shadings,patterns,tikzmark,fadings}
\usepackage{cleveref} 
\Crefname{equation}{Eq.}{Eqs.}
\Crefname{section}{Sec.}{Secs.}
\Crefname{figure}{Fig.}{Figs.}
\Crefname{table}{Table}{Tables}

\definecolor{Blue}{rgb}{0.25, 0.41, 0.88}
\definecolor{Red}{rgb}{0.92,0.,0.}
\definecolor{darkorange}{rgb}{1.0,0.549,0.}
\definecolor{cobalt}{RGB}{44, 98, 120}
\definecolor{Mathematica1}{rgb}{0.368417, 0.506779, 0.709798}
\definecolor{Mathematica2}{rgb}{0.880722, 0.611041, 0.142051}
\definecolor{Mathematica3}{rgb}{0.560181, 0.691569, 0.194885}
\definecolor{Mathematica4}{rgb}{0.922526, 0.385626, 0.209179}
\definecolor{Mathematica5}{rgb}{0.528488, 0.470624, 0.701351}
\definecolor{Mathematica6}{rgb}{0.772079, 0.431554, 0.102387}
\definecolor{Mathematica7}{rgb}{0.363898, 0.618501, 0.782349}
\definecolor{Mathematica8}{rgb}{1, 0.75, 0}
\definecolor{Mathematica9}{rgb}{0.647624, 0.37816, 0.614037}
\definecolor{plotBlue}{RGB}{94, 130, 181}
\definecolor{plotRed}{RGB}{233, 85, 54}
\definecolor{plotGreen}{RGB}{142, 176, 50}
\definecolor{plotPurple}{RGB}{135, 120, 178}

\newcolumntype{C}[1]{>{\centering\let\newline\\\arraybackslash\hspace{0pt}}m{#1}}

\def\l{{\ell}}

\def\G{\Gamma}

\def\r{{\bf r}}

\def\x{{\bf x}}

\makeatletter
\newlength{\apb@width}
\newcommand{\autoparbox}[2][c]{\settowidth{\apb@width}{#2}\parbox[#1]{\apb@width}{#2}}

\makeatother

\makeatletter
\newsavebox\myboxA
\newsavebox\myboxB
\newlength\mylenA

\newcommand*\xoverline[2][0.75]{
	\sbox{\myboxA}{$\m@th#2$}%
	\setbox\myboxB\null
	\ht\myboxB=\ht\myboxA%
	\dp\myboxB=\dp\myboxA%
	\wd\myboxB=#1\wd\myboxA
	\sbox\myboxB{$\m@th\overline{\copy\myboxB}$}
	\setlength\mylenA{\the\wd\myboxA}
	\addtolength\mylenA{-\the\wd\myboxB}%
	ifdim\wd\myboxB<\wd\myboxA%
	\rlap{\hskip 0.5\mylenA\usebox\myboxB}{\usebox\myboxA}%
	\else
	\hskip -0.5\mylenA\rlap{\usebox\myboxA}{\hskip 0.5\mylenA\usebox\myboxB}%
	\fi}
\makeatother

\numberwithin{equation}{section}
\numberwithin{figure}{section}
\numberwithin{table}{section}

\def\beq{\begin{equation}}
\def\eeq{\end{equation}}

\def\bea{\begin{eqnarray}}
\def\eea{\end{eqnarray}}

\def\beq{\begin{equation}}
\def\eeq{\end{equation}}
\def\bea{\begin{eqnarray}}
\def\eea{\end{eqnarray}}

\numberwithin{equation}{section}
\def\beq{\begin{equation}}
\def\eeq{\end{equation}}

\def\bea{\begin{eqnarray}}
\def\eea{\end{eqnarray}}

\def\l{{\ell}}

\def\G{\Gamma}

\def\r{{\bf{r}}}

\DeclareRobustCommand{\SkipTocEntry}[4]{}

\usepackage{colortbl}
\definecolor{blue2}{cmyk}{1, 0.1, 0.1, 0.1}

\definecolor{pyBlue}{RGB}{31, 119, 180}
\definecolor{pyRed}{RGB}{214, 39, 40}
\definecolor{pyGreen}{RGB}{44, 160, 44}
\definecolor{pyBlue2}{RGB}{0, 111, 237}
\definecolor{pyRed2}{RGB}{224, 52, 36}

\newcolumntype{P}[1]{>{\centering\arraybackslash}p{#1}}
\newcolumntype{M}[1]{>{\centering\arraybackslash}m{#1}}

\usepackage{tikz,xcolor,hyperref}

\usepackage{slashed}

\definecolor{lime}{HTML}{A6CE39}
\DeclareRobustCommand{\orcidicon}{
	\begin{tikzpicture}
	\draw[lime, fill=lime] (0,0) 
	circle [radius=0.2] 
	node[white] {{\fontfamily{qag}\selectfont \tiny ID}};
	\draw[white, fill=white] (-0.0625,0.095) 
	circle [radius=0.007];
	\end{tikzpicture}
	\hspace{-2mm}
}

\foreach \x in {A, ..., Z}{\expandafter\xdef\csname orcid\x\endcsname{\noexpand\href{https://orcid.org/\csname orcidauthor\x\endcsname}
			{\noexpand\orcidicon}}
}

\begin{document}

\tcbset{colframe=black,arc=0mm,box align=center,halign=left,valign=center,top=-10pt}

\renewcommand{\thefootnote}{\fnsymbol{footnote}}

\pagenumbering{roman}
\begin{titlepage}
	\baselineskip=2.5pt \thispagestyle{empty}
	
	
\begin{center}
 ~{\Huge 
	 {\fontsize{50}{35} \textcolor{Sepia}{\bf\sffamily ${\cal CFT}$ ${\cal P}$erspective ${\cal O}$n de-${\cal S}$itter ${\cal C}$osmological ${\cal C}$orrelators }}}\\  \vspace{0.25cm}
\end{center}

		\begin{center}
	
	{\fontsize{14}{18}\selectfont Sayantan Choudhury\orcidA{}${}^{\textcolor{Sepia}{1}}$\footnote{{\sffamily \textit{ Corresponding author, E-mail}} : {\ttfamily sayantan\_ccsp@sgtuniversity.org,  sayanphysicsisi@gmail.com}}}${{}^{,}}$
	\footnote{{\sffamily \textit{ NOTE: This project is the part of the non-profit virtual international research consortium ``Quantum Aspects of Space-Time \& Matter" (QASTM)} }. }${{}^{,}}$.~

\end{center}


\begin{center}
	
{ 
	
	\textit{${}^{1}$Centre For Cosmology and Science Popularization (CCSP),
SGT University, Gurugram, Delhi-NCR, Haryana- 122505, India.}
	}
\end{center}

\vspace{1.2cm}
\hrule \vspace{0.3cm}
\begin{center}
\noindent {\bf Abstract}\\
\end{center}

We investigate the principles of quantum field theory using a stiff de Sitter space. We demonstrate that a non-unitary Lagrangian on a Euclidean AdS geometry can produce the perturbative expansion of late-time correlation functions to all orders. This discovery greatly simplifies perturbative computations while also allowing us to prove fundamental features of these correlators, which are part of a Euclidean CFT. This allows us to construct an OPE expansion, limit the operator spectrum, and deduce the analytic structure of the spectral density that captures the conformal partial wave expansion of a late-time four-point function. In general, the standard CFT concept of unitarity does not apply to dimensions and OPE coefficients. Rather, the positivity of the spectral density represents the unitarity of the de Sitter theory. This assertion is non-perturbative and does not depend on the use of Euclidean AdS Lagrangians. In a scalar theory, we compute tree-level and entire one-loop-resummed exchange diagrams to demonstrate and verify these characteristics. In the spectrum density, an exchanged particle shows up as a resonant characteristic that may be helpful in experimental searches.

\vskip10pt
\hrule
\vskip10pt

\text{Keywords:~~Conformal Field Theory, Quantum field theory in curved space,}\\ \text{AdS-CFT Correspondence.}

\end{titlepage}

\newpage
\setcounter{tocdepth}{2}

\tableofcontents

\newpage
	
	\clearpage
	\pagenumbering{arabic}
	\setcounter{page}{1}
	
	\renewcommand{\thefootnote}{\arabic{footnote}}

\clearpage

\section{Introduction}

  The most symmetric and, therefore, the most straightforward cosmic spacetime is De Sitter (dS) space. The cosmological evolution of our own universe appears to include at least two epochs with spacetime metrics that are strikingly comparable to dS, based on a wealth of observable evidence. These two periods are the current accelerated expansion and the inflationary period, which is the earliest in the universe's history that we can directly observe through experimentation. Compared to its generally symmetric cousins, anti-de Sitter (AdS) and Minkowski spaces, dS space is far less explored despite its fundamental status. The cause is a number of philosophical and technical issues it displays, some of which we will discuss below. Our aim in this research is to investigate the characteristics of correlation functions in quantum field theory on a stiff dS space. Regretfully, not much is currently known about them. Some of the questions we pose in this study, for example, what imprint an exchanged heavy particle left on the observed quantities, were analyzed in \cite{Arkani-Hamed:2015bza}. Cosmology still lacks a thorough grasp of such basic problems, while having a very clear solution when it comes to flat space scattering. It should be noted that the closed-form expression for the exchange diagram, even at the tree level, was only obtained \cite{Sleight:2020obc}, and a systematic treatment of infrared divergences of light particles was also understood \cite{Gorbenko:2019rza}. Our primary attention is on the fields' boundary correlation functions, which are assessed at dS space's future infinity. These objects are strongly related to inflationary correlation functions observed by an observer situated in the future from the reheating surface and who simultaneously observes a very high number of Hubble patches, despite the fact that they are formally not observables.

  There are three reasons behind this undertaking, which are appended below:
  \begin{itemize}
      \item First, calculating these correlation functions is a crucial step in determining the primordial non-Gaussianities in the context of inflation.

      \item Second, the pursuit of a fundamental description of gravitational theories in the spirit of AdS/CFT, which is currently illusory in the cosmological setting, may be aided by a mastery of perturbative techniques.

      \item Third, another, possibly far less well-known path to cosmology in quantum gravity is provided by quantum field theory on a stiff dS. A gravitational spacetime with a FRW-like shape and a crushing singularity is known to be holographically dual to some large $N$ quantum field theories on a stiff dS backdrop. See refs. \cite{Maldacena:2010un,Hertog:2004rz,Hertog:2005hu,Turok:2007ry,Craps:2007ch,Barbon:2011ta,Barbon:2013nta,Kumar:2015jxa} to see the instances in the context of String Theory. In this research, we establish tools of quantum field theory on a stiff dS to analyze quantum gravity in a cosmological scenario. This relationship to cosmology is less direct than what was previously discussed.
  \end{itemize}

  In this work we try to address the following issues in detail:
  \begin{itemize}
      \item We create a systematic methodology for reducing every perturbative calculation in dS to that of Euclidean AdS (EAdS), where numerous computing approaches have been established. We accomplish this by constructing a Lagrangian in EAdS that directly computes boundary correlation functions in dS.
Our findings systematize and expand on previous observations about the relationship between Feynman diagrams in dS and EAdS.

\item We analyze the analytic structure of boundary correlation functions in dS using the EAdS Lagrangian we developed, with a special emphasis on the operator product expansion (OPE). We then examine how the OPE could be interpreted in terms of quasi-normal modes in a static dS patch.

\item We investigate the impact of bulk unitarity on the border correlators and relate it to specific positivity criteria. The derivation is valid at a non-perturbative level because it is not based on perturbation theory.

  \end{itemize}
  This research does not take into account gravity or the potential for dS isometries to break down during inflation. Much of our discussion can be broadened to encompass either or both, despite numerous additional conceptual and technical complexities. As a result, we believe it is critical to first grasp the simplest conceivable instance and then defer generalizations until later.
  
The organization of this paper is as follows:
 In \underline{\textcolor{purple}{\bf Section \ref{ka1}}}, we briefly review the important basics of de Sitter (dS) space, which plays the central role of the study of this paper.
In \underline{\textcolor{purple}{\bf Section \ref{ka2}}}, we  discuss the general overview on the in-in formalism. In \underline{\textcolor{purple}{\bf Section \ref{ka3}}}, we discuss the Feynman rules and diagrammatics for Euclidean AdS (EAdS) space. 
In \underline{\textcolor{purple}{\bf Section \ref{ka4}}}, we discuss the Wick rotation in a very simple tractable language.  Further in \underline{\textcolor{purple}{\bf Section \ref{ka5}}},  we discuss the buliding block of EAdS. Next in \underline{\textcolor{purple}{\bf Section \ref{ka6}}},  we discuss about basics of conformal block.  Further in \underline{\textcolor{purple}{\bf Section \ref{ka7}}},  we discuss about the details of constructing four-point function in terms of the conformal block. Then in \underline{\textcolor{purple}{\bf Section \ref{ka8}}} and \underline{\textcolor{purple}{\bf Section \ref{ka9}}},  we study the state operator correspondence and operator product expansion (OPE) in great detail. Further \underline{\textcolor{purple}{\bf Section \ref{ka10}}}, we study the analyticity of the late-time correlators in dS space. Then in \underline{\textcolor{purple}{\bf Section \ref{ka11}}}, we study the connection between OPE and Quasi Normal Modes (QNMs). Then in \underline{\textcolor{purple}{\bf Section \ref{ka12}}} and \underline{\textcolor{purple}{\bf Section \ref{ka13}}}, we study some implications of analyticity, positivity and unitarity. Further \underline{\textcolor{purple}{\bf Section \ref{ka14}}} and \underline{\textcolor{purple}{\bf Section \ref{ka15}}}, we study the positivity constraints on dS two-point and four-point functions. Next \underline{\textcolor{purple}{\bf Section \ref{ka16}}}, we study the possibility of having exchange diagram in EAdS. Further in \underline{\textcolor{purple}{\bf Section \ref{ka17}}}, we compute the expression for the four point function at the tree level. Next \underline{\textcolor{purple}{\bf Section \ref{ka18}}} and \underline{\textcolor{purple}{\bf Section \ref{ka19}}}, we discuss the resummation technique and resonance in four-point function. Finally in \underline{\textcolor{purple}{\bf Section \ref{ka20}}},  we conclude with the future prospects.

\section{Some important basics}\label{ka1}
	In this section, we first cover the basics regarding the dS spacetime, such as the coordinate systems, isometries, and unitary representations. Further discussion introduces wave functions and correlation functions of fields in dS. In particular, we make a clear distinction between two different concepts of late-time CFTs: one that represents the wave function and another that describes the correlation function. Although the contents covered in these two subsections are well-known, we chose to include them in order to correct the norms and notations and to prevent any confusion in the sections that follow.
Lastly, we propose a potential interpretation of the correlator CFT as two wave-function CFTs coupled by nearly marginal double-trace operators, each corresponding to a distinct boundary condition.
\subsection{Brief review on de Sitter space}
\subsubsection{Coordinate representation}

In order to correct the notations and conventions for subsequent sections, we briefly review the fundamentals of dS here. For more details see refs. \cite{Spradlin:2001pw,Anninos:2012qw}. Defining the $(d+1)$-dimensional dS space as a hypersurface inside $\mathbb{R}^{d+1,1}$ that satisfies the following equation is the most straightforward method:
\be
\eta_{AB}X^{A}X^{B}=l^2,
\ee
with the following structure of the metric: 
\be
\eta_{AB}={\rm diag}(-1,\underbrace{1,\ldots, 1}_{d+1}).
\ee
The Hubble radius, $l$, determines the size of dS in this case. For convenience, we shall set it to $l=1$ for the remainder of this study. Three coordinate slicings of dS are commonly utilized in the literature. They are appended below:
\begin{enumerate}
    \item \underline{\bf Global coordinates:}\\ This corresponds to parametrizing $X^{A}$ as:
\be
X^{A}=\left(\sinh\tau, \cosh \tau\, n_{1},\ldots  ,\cosh \tau\, n_{(d+1)}\right)~,\ee
subject to a constraint condition:
\be \sum_{j=1}^{d+1}(n_j)^2=1.\ee
The corresponding metric structure in the above-mentioned global coordinate is given by the following expression:
\bea ds^2=\bigg(-d\tau^2+\cosh^2 \tau d\Omega_{d}^2\bigg).\eea
Here the tomporal coordiante spans from, $-\infty<\tau<\infty$. Also it is to be noted that, $d\Omega_d^2$ represents the part of the total metric which geometrically specifies a $d$ dimensional sphere. The term comes from the fact that this coordinate system encompasses the whole hypersurface. This encompasses the whole Lorentzian AdS and is an analog of the global coordinates in AdS.

 \item \underline{\bf Poincar\'e coordinates:}\\ This is represented by:
 \be
X^{A}=\left(\frac{1-\eta^2+|x|^2}{2\eta},\frac{x^{\mu}}{\eta}, \frac{1+\eta^2-|x|^2}{2\eta}\right).
\ee
Here the coordinate $\eta$ spans from $-\infty<\eta<0$. In this specific type of coordinates the dS metric takes the following simplified structure:
\be ds^2=\frac{1}{\eta^2}\bigg(-d\eta^2+dx.dx\bigg).\ee
A half of the global dS is covered by these coordinates. The Poincar\'e patch of AdS is analogous to this.

\item \underline{\bf Static coordinates:}\\
In this coordinate system the geometrical structure of the metric is given by the following expression:
\bea
ds^2=-\left(1-r^2\right)dt^2+\frac{dr^2}{\left(1-r^2\right)}+r^2 d\Omega_{d-1}^2~~~~~{\rm where}~~~~~r=\sin\theta.
\eea
Here the angular parameter $\theta$ is lying within the window $0<\theta<\pi/2$ and $d\Omega_{d-1}^2$ represents the part of the metric for the $(d-1)$-dimensional sphere. In terms of the embedding the following parametrization is used in the present context of discussion:
\be
X^{A}=\left(\cos\theta\sinh  t,\cos\theta\cosh t,\, \sin\theta\,  \tilde{n}_1,\ldots , \,\sin\theta\,  \tilde{n}_{d}\right)~,
\ee
subject to the following constraint condition:
\be \sum_{j=1}^{d}(\tilde{n}_j)^2=1.\ee
\end{enumerate}

\subsubsection{Embedding formalism}
The study of correlation functions on the late-time surface—a co-dimension $1$ surface described by $\eta\to 0^{-}$ in the Poincar\'e coordinates—is the primary topic of this article, as stated in the introduction. A ``boundary version" of the embedding coordinates, provided by a projective null cone, can be used to parametrize a point on the surface:
\be
\eta_{AB}P^{A}P^{B}=0.
\ee
For the Poincar\'e coordinates, the relationship is provided by:
\be
P^{A}=\left(\frac{1+|y|^2}{2},y^{\mu}, \frac{1-|y|^2}{2}\right)
\ee
We can create invariants under the dS isometry using $X$ and $P$. The two crucial quantities are the bulk-boundary distance $X\cdot P$ and the two-point invariant ${\cal F}(X_1,X_2)$:
\bea
X\cdot P&=&\bigg(\frac{\eta^2-|x-y|^2}{2\eta}\bigg),\\
{\cal F}(X_1,X_2)&\equiv& X_1\cdot X_2=\bigg(\frac{\eta_1^2+\eta_2^2-|x_{12}|^2}{2\eta_1\eta_2}\bigg).
\eea
Observe that the following relation exists between the two-point invariant and the chordal distance $\zeta (X_1,X_2)$ between the two points:
\be
\zeta(X_1,X_2)=2\left(1-{\cal F}(X_1,X_2)\right)=\bigg(\frac{-\eta_{12}^2+|x_{12}|^2}{\eta_1\eta_2}\bigg).
\ee

\subsubsection{Connection with EAdS}

In Euclidean AdS (EAdS) space, embedding coordinates have equivalents. Here, we provide an overview and demonstrate how they relate to the Poincar‘e coordinate:
\bea
\eta_{AB}X_{\rm AdS}^{A}X_{\rm AdS}^{B}=-1,
\eea
where
\bea X_{\rm AdS}^{A}=\left(\frac{1+z^2+x^{\mu}x_{\mu}}{2z},\frac{x^{\mu}}{z},\frac{1-z^{2}-|x|^{2}}{2z}\right).\eea
Here $(z,x^{\mu})$ are the Poincar\'e coordinate which is described by the following geomtrical structure of the metric:
\bea ds^2=\frac{1}{z^2}\bigg(dz^2+dx.dx\bigg).\eea
hence the two-point invariant in its EAdS form is given by the following expression:
\bea {\cal F}_{\rm AdS}(X_1,X_2)=X_1\cdot X_2=-\left(\frac{z_1^2+z_2^2+|x_{12}|^2}{2z_1z_2}\right).\eea

\subsubsection{The vacuum state}
In this study, we examine correlation functions for the so-called Bunch-Davies vacuum state in de Sitter at late times. Other names for it include the Hartle-Hawking state or the ``no-boundary" state, depending on the situation, especially when gravity is present. It is defined for quantum field theory in de Sitter as a state that can be produced by analytically continuing the path-integral on a hemisphere $HS^{d+1}$ to de Sitter. The static patch or global coordinates are the easiest places to observe the analytic continuation. The sphere's metric can be expressed as:
\bea ds^{2}=\bigg(d\theta^2 +\cos^2d\phi^2+\sin^2\theta d\Omega_{d-1}^2\bigg).\eea
The metric for the global dS is obtained by performing the continuation $\theta\to i\tau$, whereas the static patch is obtained by considering $\phi\to it$. Similar to the Lorentzian continuation of the thermo-field double or the eternal AdS black hole, the continuation to the static patch needs a periodic coordinate $\phi$, whereas the continuation to the global dS does not. This is one significant distinction between the two. This is the fundamental explanation for why the physics in the global dS does not clearly appear to be thermal whereas the physics in the static patch does.
\subsubsection{$SO(d+1,1)$ Isommetries}

The Euclidean conformal group $SO(d+1,1)$ isomorphic to the isometry group of dS. The embedding coordinates, where the isometries are provided by the following straightforward differential operators, provide the clearest example of this.
\bea \mathcal{L}_{AB}&=&\bigg(X_{A}\frac{\partial}{\partial X^{B}}-X_{B}\frac{\partial}{\partial X^{A}}\bigg)=\bigg(P_{A}\frac{\partial}{\partial P^{B}}-P_{B}\frac{\partial}{\partial P^{A}}\bigg).\eea
In the present context of discussion it is simple to verify that the commutation relations are isomorphic to $SO(d+1,1)$. States in a unitary QFT in dS are categorized by unitary irreducible representations of the isometries, much like in flat space. There is literature on both the math and physics sides of this rich but well-researched topic \cite{newton1950note,thomas1941unitary,Dobrev:1976vr,dobrev1977harmonic}. Specifically, the primary series, complementary series, and discrete series are the three representations that are crucial to dS. For further information on this matter, see ref. \cite{Baumann:2017jvh}.
Using the terminology of conformal field theory, it is helpful to characterize these representations by parametrizing them in terms of a conformal dimension $\Delta$, which is connected to the quadratic Casimir value as follows:
\be
\mathcal{C}_2=-\frac{1}{2}\sum_{A}\sum_{B}\mathcal{L}_{AB}\mathcal{L}^{AB}=\bigg(\Delta (\Delta-d)+J(J+d-2)\bigg)~~~~~{\rm where}~~~J={\rm  spin~ of~ the~ state}.
\ee
States with conformal dimention: \be \Delta=\bigg(\frac{d}{2}+i\nu\bigg),\ee in this description correspond to the primary series, where $\nu$ is a positive real number. A ``heavy" field, or one whose mass meets: \be m>\frac{d}{2}\ee in units of the Hubble parameter, is what is meant to be applied to a free scalar theory in dS. The relationship between the conformal dimension and the mass in dS allows one to verify this:
\be
m^2=\Delta(d-\Delta)~~~\Longrightarrow~~~\Delta_{\pm}=\bigg(\frac{d}{2}\pm i\nu\bigg)~~~~{\rm with}~~~\nu=\sqrt{m^2-\frac{d^2}{4}}.
\ee
Since they appear as two distinct power-law decays in correlation functions, we frequently retain both of the signatures of the conformal dimension solution in the computation. 

States with conformal dimension are associated with the complementary series with zero spin: \be 0<\Delta<\frac{d}{2}~~~~~~\Longrightarrow ~~~~~~\frac{d}{2}>m>0,\ee which is equivalent to a ``light" field in a free scalar theory in dS with mass. 

Lastly, states with integer or half-integer conformal dimensions correspond to a discrete series. They relate to so-called partially massless fields and are only present for a particle with spin in free theory. They may have an impact on the inflationary observables, but as we are concentrating on scalar correlation functions in this research, we will not examine them. When we summarize the findings for zero-spin states, we have:
\bea
\text{\bf Principal series:}&& \qquad \Delta=\bigg(\frac{d}{2}+i\nu\bigg)~~~~~\Longrightarrow~~~~~ m>\frac{d}{2},\\
\text{\bf Complementary series:}&& \qquad 0<\Delta<\frac{d}{2}~~~~~\Longrightarrow~~~~~ \frac{d}{2}>m>0.
\eea
In the remainder of this paper, we frequently parametrize the representation of states and particle mass using $\Delta$ and $\nu$ interchangeably. Specifically, we show that using $\nu$ even for the light particles is convenient. Here we choose the negative signature in the conformal dimension which satisfy the following condition in the complementary series representation:
\bea \frac{d}{2}>\Delta_{-}=\bigg(\frac{d}{2}-i\nu\bigg)>0.\eea

\subsection{Wave function of the Universe}

The concepts of wave functions and correlation functions of fields in dS, their interrelation, and their fundamental characteristics are presently reviewed. We also briefly discuss the static patch technique and the dS S-matrix. All of these topics are well-known among researchers in the field, but we hope that our explanation will help people who are not familiar with the subject avoid any confusion. Highlighting the nuances involved in identifying basic physical observables in dS space and, more broadly, in cosmology is another objective of this section. Determining observables is a fairly easy task because we are working with perturbative massive quantum field theories on a fixed background in this paper. However, it is ideal to do so in a way that can stay consistent once gauge fields—most notably, the metric—are also dynamical. We provide a method for linking the wave functions and correlators in the subsequent section, which might help with some of these nuances.

We can attempt to draw a lesson from asymptotically flat or AdS spacetimes, where well-defined observables in dynamical gravity theories are asymptotic, that is, specified on the border of corresponding space. The past and future asymptotic bounds of the global dS space are both space-like. Several scholars were prompted by this to examine the dS S-matrix, which is the transition amplitude from the past to the future border \cite{Witten:2001kn,Strominger:2001pn}. However, the next barrier is not precisely on the same level as the previous one. In the future, one can specifically work with a specific Euclidean continuation of the geometry that has a single boundary \cite{Hartle:1983ai}. One advantage of the no-boundary geometry is that it does not include the contracting section of dS, which is barely relevant to cosmology. According to this perspective, the universe's wave function in the no-boundary condition is a natural object to research \cite{Maldacena:2002vr}. This led a number of authors to examine two objects: the universe's wave function in the no-boundary state \cite{Maldacena:2002vr} and the transition amplitude from the past to the future boundary, sometimes known as the dS S-matrix \cite{Witten:2001kn,Strominger:2001pn,Choudhury:2026pus,Choudhury:2025pzp}. The values of every field on the edges of the related geometries are fixed by these two objects. An alternative way to conceptualize the wave function is as a transition amplitude from the fixed state. Further generic transition amplitudes will not be covered in the following; readers who are interested are directed to some recent research on the scattering technique \cite{Marolf:2012kh,Cotler:2019dcj,Albrychiewicz:2020ruh}. Numerous pleasant characteristics of the late time wave function are similar to those of the AdS partition function, which is the main object in the AdS/CFT correspondence. Specifically, it has the following form modulo specific local terms:
\bea
\Psi\left[\phi\right]&=&\exp\bigg(\int dx_1 dx_2~ \phi(x_1)\phi(x_2)~G_2(x_1,x_2)\nonumber\\
&&\qquad +\int dx_1 dx_2 dx_3~ \phi(x_1)\phi(x_2)\phi(x_3)~G_3(x_1,x_2,x_3)\nonumber\\
&&\qquad +\int dx_1 dx_2 dx_3 dx_4~\phi(x_1)\phi(x_2)\phi(x_3)\phi(x_4)~G_4(x_1,x_2,x_3,x_4)+\ldots\bigg)\,.~~~~~
\eea
In this case, $x$'s parametrize the future boundary, and $\phi(x)$ represents the late time value of a field propagating in dS. In a form of the Euclidean CFT, which we refer to as CFT$_{\Psi}$, the function $G_n\forall n\geq 2$ possesses all the characteristics of a correlation function.
The wave function $\Psi$ may now be expressed as a generating function of CFT$_{\Psi}$, which will be extremely helpful for the remaining calculations performed in this work:
\be
\Psi [\phi]=\left<\exp\bigg(\int dx~ \mathcal{O}(x)~\phi(x)\bigg)\right>_{\text{CFT}_{\Psi}}.
\ee
In this case, $\mathcal{O}(x)$ is an operator in CFT$_{\Psi}$ whose associated correlation functions provide $G_n$. In numerous recent articles \cite{Arkani-Hamed:2017fdk,Hillman:2019wgh,Benincasa:2019vqr,Meltzer:2021zin,Pajer:2020wnj,Cespedes:2020xqq,Goodhew:2021oqg,Goodhew:2020hob,Jazayeri:2021fvk,Baumann:2021fxj}, these $G_n$s have been discussed; they are sometimes called correlators. It is important to emphasize that these are not what we refer to as cosmological correlators. We will refer to $G_n$'s as the wave function coefficients in this study.

In fact, in no physical system is the wave function itself an observable. Rather, the correlation functions of the fields themselves, $\langle\phi(x_1)\phi(x_2)\ldots\rangle$, computed at late times, are the objects of main interest to observational cosmologists and phenomenologists. We refer to these items as cosmological correlators. Both Poincar\'e and global slicing can be used to study them. A system must be prepared and measured numerous times in order to identify correlation functions empirically, which is regrettably not achievable in an observable universe. Additionally, correlation functions are not directly detectable in a single copy of a system. However, cosmologists make use of the universe's approximate translational and rotational symmetry. The experiment is then essentially repeated numerous times by measuring correlation functions in a variety of locations and orientations. There are a number of natural slicings of dS, even if the notion of time is well-defined because we operate with a fixed metric. We will utilize both global and Poincar\'e coordinates in this work, and the metric will look like this:
\be
ds^2=-d\tau^2+\cosh^2\tau~ d \Omega^2\,, \qquad \tau \in (-\infty,\infty)
\ee
and 
\be
ds^2=\frac{1}{\eta^2}\bigg(-d\eta^2+d{\bf x}^2\bigg)\,, \qquad \eta \in(-\infty,0).
\ee
Importantly, when time is taken to future infinity, there is no difference between the global and Poincar\'e slices because they almost overlap at late times.

Formally speaking, in dS, as in quantum mechanics, knowing the wave function guarantees knowing all correlation functions:
\bea
\left\langle \phi(x_1)\phi(x_2)\ldots\phi(x_n)\right\rangle = \int D\phi~ \Psi\left[\phi\right] \Psi^*\left[\phi\right] \phi(x_1)\phi(x_2)\ldots\phi(x_n)\,,
\eea
At the boundary, $\int D\phi$ represents the path integral over all dynamical fields in the theory. Actually, knowing the wave function is not all that different from knowing an action or a Lagrangian of some theory; the correlation functions still require a lot of work. Furthermore, it is a highly non-local action. The extraction of the correlators from the wave function coefficients is obviously simple at the lowest orders in perturbation theory, but it is unclear how this process is progressing non-perturbatively. This is particularly problematic when the metric is dynamical, therefore solving some Euclidean quantum gravity is required to take the above path integral, which is not always feasible. too see some application in the context of higher spin theories please look at the ref. \cite{Anninos:2017eib}. In this study, we will find it more practical to use the in-in formalism to calculate correlators directly, avoiding the wave function computation, even though this is the perturbative calculation technique.

The characteristics of these correlation functions will be anticipated. The dS isometry group $SO(1,d+1)$, which is also the Euclidean conformal group in $d$ dimensions, makes it evident that the correlators of fields on the future boundary will be invariant in our instance of QFT on a rigid dS. Therefore, the boundary correlators of a QFT in dS should form the correlators of some Euclidean CFT, which we shall refer to as CFT$_C$. We will validate this anticipation in the remainder of this work. Let's emphasize for the time being that the correlators CFT$_C$ and the wave function CFT$_{\Psi}$ are significantly different. As we shall demonstrate, CFT$_C$ contains about twice as many single trace operators than CFT$_{\Psi}$, and both CFTs have an analog of a large-$N$ counting parameter at weak coupling in the bulk. Given that field boundary values are not set in the computation of correlators, two operators representing two modes of each basic field with dimensions connected by the shadow transform should be expected:
\be \bigg(\Delta_1+\Delta_2\bigg)=d,\ee
which is further expected to be corrected in the context of perturbation theory.

For a moment, let us comment on how dynamical gravity changes this image. CFT$_{\Psi}$ is endowed with the stress tensor, and the wave function calculation is preserved, at least in perturbation theory. The change on the correlator side is significantly more dramatic: from the boundary point of view, it is unclear how to define local and gauge invariant observables. The operator positions in some systems cannot be parametrized by coordinates alone, as some sort of boundary gravity still plays a role in the computation. Let's make a broad analogy since we don't currently have a specific proposal: the relationship between the wave function and correlators in dS is similar to that between the S-matrix and inclusive cross-sections in flat space. Furthermore, a partial trace over soft gravitational modes is occasionally required to make it well-defined. In fact, the cross section is what is actually visible in Minkowski as well. It would be intriguing to observe whether the methods created to address these modes can assist us in establishing appropriate observables in cosmology as well.

Let us return at last to another problem concerning asymptotic observables in dS. Indeed, not a single physical observer could perceive the full future infinity of dS space, even without gravity. The claim that correlators are helpful physical objects to study may therefore be contested. There are solutions for this issue. An asymptotic flat space observer with access to the entire reheating surface is created by supposing that dS spacetime is actually the limit of an extremely long inflation and that the cosmos reheats into a flat space after a certain amount of late time. One way to conceptualize our boundary correlators is as the limit of correlators measured on the reheating surface, which is then carried to infinity. See ref. \cite{Anninos:2011af,Anninos:2012qw} for more discussions on this issue. In reality, since all observed modes are super-horizon, this is the limit used in the majority of phenomenological inflationary assessments. Alternatively, we can state that the only basic physical observables in dS are those that are available to a single causal observer, which is limited to a static patch.

\subsection{Double trace deformation}

For now, we'll assume that we know something about CFT$_\Psi$ on our own, perhaps at the non-perturbative level. This connection can then be seen as creating a new theory by linking two CFTs, one corresponding to $\Psi$ and the other to $\Psi^*$, via an integral over the common sources $\phi$. One way to conceptualize an integral over sources for perturbative theories at leading order is as the addition of double-trace operators composed of single-trace operators from the two theories, despite the fact that it is not a well-defined procedure due to the requirement of knowing the CFT partition function for arbitrary values of the sources. Such operators will be largely unimportant in general. 

We therefore suggest a different method of connecting the two CFTs, one in which the RG flow can continue to be perturbative.
The first step is to write $\Psi^{\ast}$ in terms of its canonical conjugate momentum $\pi (x)$ rather than $\phi(x)$. From a dS perspective, this equates to imposing a different boundary condition at future infinity, which is what is referred to as an alternate boundary condition in AdS \cite{Isono:2020qew}.  However, the double-trace deformation covered in this section, which directly explains the correlator CFT, should not be mistaken with this; it is merely an analog of what was addressed in EAdS \cite{Witten:2001ua}. In a semi-classical manner, the Fourier-transformation links the new wave function $\tilde{\Psi} [\pi]$ to the original wave function. It was recently described in \cite{Klebanov:1999tb} how the wave function in dS with an additional boundary condition relates to the double-trace deformation:
\be
\Psi^{\ast}[\phi]\sim\int D\pi \, \exp\bigg(\int dx\phi(x)\pi (x)\bigg)~\tilde{\Psi}[\pi].
\ee
Writing a counterpart of the above equation for $\tilde{\Psi}[\pi]$ is the second step, which gives:
\be
\tilde{\Psi}[\pi]=\left<\exp\bigg(\int dx \tilde{\mathcal{O}}(x)\pi (x)\bigg)\right>_{\text{CFT}_{\tilde{\Psi}}}.
\ee
Here it is important to note that, $\text{CFT}_{\tilde{\Psi}}$ represents a CFT which is the generator of the correlation functions of the $\pi (x)$, which is given by:
\bea
\tilde \Psi\left[\pi\right]&=&\exp\bigg(\int dx_1 dx_2~ \pi(x_1)\pi(x_2)~\tilde G_2(x_1,x_2)\nonumber\\
&&\qquad+\int dx_1 dx_2 dx_3~ \pi(x_1)\pi(x_2)\pi(x_3)~\tilde G_3(x_1,x_2,x_3)\nonumber\\
&&\qquad+\int dx_1 dx_2 dx_3 dx_4~ \pi(x_1)\pi(x_2)\pi(x_3)\pi(x_4)~\tilde G_4(x_1,x_2,x_3,x_4)+\ldots\bigg)\,.~~~~~
\eea
 After integrating out the contributions from the source fields $\phi (x)$ and $\pi (x)$, in the semi-classical limit, it turns out that we are left with computing a Gaussian integration after performing which we get the following expression:
 \be
\langle\phi (x_1)\ldots \phi (x_n)\rangle\sim \left<\tilde{\mathcal{O}}(x_1)\ldots \tilde{\mathcal{O}}(x_n)\,\,\exp\bigg(\int dx~ \mathcal{O}(x)\tilde{\mathcal{O}}(x)\bigg)\right>_{\text{CFT}_{\Psi}\times \text{CFT}_{\tilde{\Psi}}}.
\ee
The following shadow transform, \be \bigg(\Delta_{\mathcal{O}_i}+\Delta_{\tilde{\mathcal{O}_i}}\bigg)=d,\ee would relate the single-trace operator spectrum of CFT$_{\Psi}$ and CFT$_{\tilde{\Psi}}$ in the semi-classical limit, namely free fields in dS, and the double-trace operators $\mathcal{O}_i\tilde{\mathcal{O}}_i$ become marginal \cite{Witten:2001ua}. Thus, the correlation function represented by the equation above can be thought of as a marginal double-trace deformation of two linked CFTs. See also ref. \cite{DiPietro:2021sjt}.

In contrast, the operators $\mathcal{O}_i$ and $\tilde{\mathcal{O}}_i$ in an interacting but weakly coupled theory have dimensions that do not need to sum up to $d$ i.e. \be \bigg(\Delta_{\mathcal{O}_i}+\Delta_{\tilde{\mathcal{O}_i}}\bigg)\neq d.\ee At the very least, the operators $\mathcal{O}_i \tilde{\mathcal{O}}_i$ will not be precisely marginal, but rather only marginally relevant or irrelevant. The possibility that CFT$_C$ can still be represented by conformal perturbation theory around CFT$_{\Psi}\times \text{CFT}_{\tilde{\Psi}}$ by suitably adjusting double-trace and perhaps higher-trace couplings to critical values, however, is alluring:
\be
\label{3cfts}
\text{CFT}_C\sim \bigg(\text{CFT}_{\Psi}\times \text{CFT}_{\tilde \Psi}\bigg)+\sum_i c_i \mathcal{O}_i \tilde{\mathcal{O}}_i \neq \bigg(\text{CFT}_{\Psi}\times \text{CFT}_{\tilde \Psi}\bigg)\,.
\ee
We will leave this as an intriguing future challenge as this conjecture may be easily checked in the perturbative examples of QFTs in dS that we examine in this paper. The individual CFTs i.e. CFT$_{\Psi}$ and CFT$_{\tilde \Psi}$ and their direct product are complex according to the definition provided in \cite{Gorbenko:2018ncu}, but CFT$_C$ must be real. As long as they occur in complex conjugate pairs, real non-unitary CFTs can have complex anomalous dimensions. Therefore, assuming the connection above is accurate, the double-trace deformation must restore reality.

For the purposes of this discussion, assume that a relationship akin to the factorization formula above is true. How can it be used to examine bulk dynamical gravity cosmological theories? There is a glimmer of light because the new approach mentioned in the factorization relationship above does not require a path integral over the metrics any more. Naturally, the issue with residual border gravity persisted, but it was subsumed in the computation of $\tilde\Psi$. It is true that CFTs connected to induced boundary gravity correspond to AdS theories with different boundary conditions for the metric \cite{Leigh:2003ez,Compere:2008us,Giombi:2013yva}. Such theories have been shown to be power-counting renormalizable and do not contain the typical Einstein-Hilbert term in the action \cite{Compere:2008us}. As a result, they could be UV complete and potentially non-perturbatively defined. The next query is if we can understand the equivalent coupling between the relevant field in the theory CFT$_{\tilde \Psi}$ and the stress tensor of CFT$_{\Psi}$. To tackle these problems, it is a good idea to think of either a dS-like solution in two-dimensional JT gravity, for which the alternate boundary conditions were also recently addressed \cite{Goel:2020yxl}, or a three-dimensional bulk, where the boundary gravity is tractable \cite{Giombi:2013yva,Cotler:2019nbi}. An alternative approach involves examining abelian or non-abelian gauge theories in dS, wherein some of the nuances covered here also apply. In the next section, we return to the issues that form the core of this work and that we comprehend incomparably better, leaving all these intriguing conundrums for later.

	\section{In-In formalism: A conceptual overview}
	\label{ka2} 

The in-in formalism, which is frequently employed in inflationary computations, should be briefly reviewed first. Refs. \cite{Maldacena:2002vr, Weinberg:2005vy, Seery:2007we, Adshead:2009cb, Senatore:2009cf} defines an in-in correlation function of a group of fields situated at a certain time slice at time $t$:
\bea
\label{inin1}
&&\left\langle \phi(x_1,t)\phi(x_2,t)\ldots \phi(x_n,t) \right  \rangle\nonumber\\
&&=
\left\langle \Omega(t_0)\right\vert\bar T \bigg(\exp\bigg(i\int_{t_0}^t dt' H\bigg)\bigg)\phi(x_1,t)\phi(x_2,t)\ldots \phi(x_n,t) T \bigg(\exp\bigg(-i\int_{t_0}^t dt' H\bigg)\bigg)\left\vert\Omega(t_0) \right  \rangle\,.\nonumber\\
&&
\eea
In this case, $T$ and $\bar{T}$ stand for time and anti-time ordering, and the state $\left\vert\Omega(t_0) \right \rangle$ is the generalization of the Bunch-Davies vacuum to the interacting theory, suggesting that the theory's Minkowski vacuum contains very high energy modes. It is simple to utilize the interaction picture, which is defined by dividing the whole Hamiltonian into a free and an interacting half, to calculate these correlation functions in perturbation theory:
\be H = H_0 + H_I,\ee 
where $H_0$ is the free part of the Hamiltonian which basically evolve the operator. On the other hand, $H_I$ represents the interacting part of the Hamiltonian which specifically evolve the quantum states. As a consequence the previously written equation (\ref{inin1}) can be further recast into the following form:
\bea\label{ininpert}
&&\left\langle \phi(x_1,t)\phi(x_2,t)\ldots \phi(x_n,t) \right  \rangle\nonumber\\
&&=\frac{1}{\left\langle 0 \right\vert\bar T \bigg(\exp\bigg(i\int_{-\infty(1+i \epsilon)}^t dt' H_I^{\rm int}\bigg)\bigg)\,T \bigg(\exp\bigg(-i\int_{-\infty(1-i \epsilon)}^t dt' H_I^{\rm int}\bigg)\bigg)\left\vert 0 \right  \rangle}\,\nonumber\\
&&\times \left\langle 0 \right\vert\bar T \bigg(\exp\bigg(i\int_{-\infty(1+i \epsilon)}^t dt' H_I^{\rm int}\bigg)\bigg)\, \nonumber\\
&&~~~~~~~~~~~~~\phi^{\rm int}(x_1,t)\phi^{\rm int}(x_2,t)\ldots \phi^{\rm int}(x_n,t) \,T \bigg(\exp\bigg(-i\int_{-\infty(1-i \epsilon)}^t dt' H_I^{\rm int}\bigg)\bigg)\left\vert 0 \right  \rangle.
\eea
Here the interaction picture is denoted by the superscript `$\rm int$', $|0\rangle$ is the free Bunch-Davies vacuum, also known as the Fock vacuum, and the denominator is used to cancel out disconnected vacuum bubble diagrams. The perturbative expansion is then constructed simply by expanding the anti-time-ordered or time-ordered exponentials on the left and right, respectively, and computing the resulting correlation function using Wick contractions. In the following, we shall use equation \ref{ininpert} for perturbative computations and remove the superscript `$\rm int$'. The same formula is used for insertions of composite operators rather than fields. Three different types of propagators are found when the Wick theorem is applied to \eqref{ininpert}. These are denoted by the letters $G^{LL}$, $G^{RR}$, and $G^{LR}$, depending on whether the Wick contraction occurs between the left and right Hamiltonians, within the left Hamiltonian, or both. Since they coincide for spacelike separations, any propagator can be used when a contraction occurs between two external fields. For a mass $m$ scalar field, the propagators are:
\bea
&& \label{dSprops1}{\rm Anti-time-ordered~propagator:}~~~G^{LL}_\nu=V_\nu\left({\cal F}+i\eps\right)\,, \\&& \label{dSprops2} {\rm Time-ordered~propagator:}~~~G^{RR}_\nu=V_\nu\left({\cal F}-i\eps\right)\,,\\ && \label{dSprops3} {\rm Wightman~ function:}~~~G^{LR}_\nu=G^{RL}_\nu=V_\nu\left({\cal F}-i\eps\,{\text{sgn}}(t_L-t_R)\right)\,,
\eea
where the function $V_\nu$ is defined by the following expression:
\bea
\label{defW1}
V_\nu({\cal F})=\frac{\displaystyle\Gamma\left(\frac{d}{2}+ i \nu\right)\Gamma\left(\frac{d}{2}- i \nu\right)}{\displaystyle(4 \pi)^{ \frac{d+1}{2}}\Gamma\left(\frac{d+1}{2}\right)}{}_2F_1\Bigg(\frac{d}{2}+i\nu,\frac{d}{2}-i\nu;\frac{d+1}{2};\frac{1+{\cal F}}{2}\Bigg)~.
\eea
In addition, the subscript factor $\nu$ and the two-point invariant ${\cal F}$ is defined as:
\bea &&\nu=\sqrt{m^2-\frac{d^2}{4}},\\
&& {\cal F}(X_1,X_2)\equiv X_1\cdot X_2=\bigg(\frac{\eta_1^2+\eta_2^2-|x_{12}|^2}{2\eta_1\eta_2}\bigg).\eea
Adding imaginary portions to the time variables in the manner described below yields the $i\eps$ prescription:
 \bea
 \label{ieps1}
&& \eta^L=\eta^L(1+i\eps),\\ && \label{ieps2} \eta^R=\eta^R(1-i\eps)\,.
 \eea
Naturally, the correlation function response is independent of the coordinate system employed as long as we are describing the same state in dS.

It is further important to note that the singularity of the hypergeometric function in \eqref{defW1} occurs at ${\cal F}=1$. As a result, the difference between the two spots becomes zero. Their imaginary component is different in the time-like region ${\cal F}>1$ because of varied $i\eps$ prescriptions, but for ${\cal F}<1$ the points are space-like separated and all propagators agree and are real. For ${\cal F}>1$, we have a branch-cut on the real axis for the major branch of the hypergeometric function, which leads to these features. The expansion of \eqref{defW1} to the large-${\cal F}$ yields the asymptotic behavior of the propagators:
\be
\lim_{|{\cal F}|\to\infty}V_\nu({\cal F})\approx \frac{1}{4 \pi^{ \frac{d}{2}+1}}\left(\frac{1}{\left(-2 {\cal F}\right)^{\frac{d}{2}-i \nu}}\G\left(\frac{d}{2}-i\nu\right)\G(i \nu)+\frac{1}{\left(-2 {\cal F}\right)^{\frac{d}{2}+i \nu}}\G\left(\frac{d}{2}+i\nu\right)\G(-i \nu)\right)\,.\,
\ee
Only the two leading power-law behaviors—each accompanied by a sequence of subleading integer-shifted powers—were displayed here. In the limit of enormous time, the light field will correspond to an operator of definite scaling dimension \be \Delta_-=\bigg(\frac{d}{2}-i\nu\bigg),\ee since the power $\frac{d}{2}-i\nu$ prevails for light fields.
However, both powers have the same absolute value for heavy fields. Correlators of operators with definite scaling dimension will be handy for our discussion. For the heavy fields, this can be accomplished by inserting suitable linear combinations of the fields and their time-dependent first derivatives, or conjugate momenta. This makes defining the bulk-to-boundary propagator convenient, which is described by the following expression:
\bea
\label{KdS}
{\cal Y}^{l(r)}_\nu({\cal F}) & =& \frac{1}{4 \pi^{\frac{d}{2}+1}}\frac{1}{ \left(-2 {\cal F}+(-)i\eps\right)^{\frac{d}{2}-i \nu}}\G\left(\frac{d}{2}-i\nu\right)\G(i \nu)\,.
\eea
The Wick contraction, which is parametrized by $\eta_c\approx 0$ in the Poincar\'e coordinates, will be performed using this propagator on a fixed time slice at a long time with field insertions. More specifically, we have the following, suppressing the $i\epsilon$ prescription:
\begin{equation}\label{eq:Ketac1}
{\cal Y}_\nu({\cal F}) =\frac{1}{4 \pi^{\frac{d}{2}+1}} \left(\frac{-\eta_c \eta_1}{ \eta_1^2 - x_{12}^2}\right)^{\frac{d}{2}-i \nu}\G\left(\frac{d}{2}-i\nu\right)\G(i \nu)~.
\end{equation}
In this case, the coordinates of the bulk point are $(\eta_1,x_1)$, while the spatial coordinate of the insertion on the time slice at $\eta_c$ is $x_2$. For the bulk-to-boundary propagator, this expression holds true for both light fields and heavy field insertions with dimension $\Delta_-$. However, for heavy field insertions with dimension: \be \Delta_+ = \bigg(\frac{d}{2}+i\nu\bigg),\ee we just need to take into account the complex conjugate. By using the aforementioned propagators, the Feynman rules of the in-in perturbation theory in dS reduce to joining a set of left and right vertices along with external operators.

In theory, the perturbative formalism in dS space is simple, but in practice, computations become cumbersome. Conventionally, the time integrals remain explicit while moving to momentum space with regard to spatial coordinates. The direct assessment of these integrals is rapidly unfeasible, even for basic tree-level layouts. As a result, several other methods have been created, such as those based on the Mellin-Barnes representation \cite{Sleight:2019mgd, Sleight:2019hfp, Sleight:2020obc, Sleight:2021iix} and analyticity and factorization qualities in momentum space \cite{Arkani-Hamed:2015bza, Arkani-Hamed:2018kmz, Baumann:2019oyu, Baumann:2020dch}. An additional option is to use the fact that the Euclidean representation of dS space is a sphere, calculate on the sphere, and then continue the solutions to dS analytically \cite{Marolf:2010zp, Marolf:2010nz, Marolf:2011sh}. This method makes it possible to compute specific diagrams and deduce certain significant general characteristics of correlators that we will utilize later. Nevertheless, it also seems hard to directly evaluate higher-point correlation functions on the sphere. The sphere's absence of an asymptotic border and the expectation of large simplifications in the computation of asymptotic observables are the reasons. Rather, we will employ the Euclidean Anti-de Sitter space (EAdS), a relation of de Sitter space to an alternative Euclidean manifold. Originally investigated in the context of wave function computations \cite{Maldacena:2002vr, Harlow:2011ke, Mata:2012bx, Anninos:2014lwa}, this connection was more recently applied in the Mellin-Barnes method to dS correlators \cite{Sleight:2019mgd, Sleight:2019hfp, Sleight:2020obc,Choudhury:2025pzp}. Since the relationship between AdS and dS tree-level exchange diagrams was established in this series of publications, the authors were able to get explicit formulas for arbitrary masses and spins for the first time. We assert that a linear combination of Witten diagrams in EAdS can be used to express any Feynman diagram that appears in the in-in formalism. Furthermore, a local field theory on EAdS that produces all of these diagrams has twice as many fields as the original theory. This suggests that one can use computational methods, such as harmonic analysis on hyperbolic spaces, that were created for EAdS computations. Let's take a quick look at the fundamentals of perturbation theory in EAdS for scalar field theories before delving into this in more detail. We will mostly use the notations of \cite{Carmi:2018qzm}.

\section{Revisiting Feynrules for EAdS}
\label{ka3}
In EAdS, we will examine the field theory for a scalar particle with mass $m_{\rm AdS}$. In this case, the Green's function is the propagator:
\bea
\label{GAdS1}
G_\nu^{\rm AdS}({\cal F}^{\rm AdS})& =& \frac{\G\left(\frac{d}{2}+i\nu\right)}{2\pi^{\frac{d}{2}} \G\left(1+i\nu\right)\left(-2({\cal F}^{\rm AdS}+1)\right)^{\frac{d}{2}+i\nu}}{}_2F_1\left(\frac{d}{2}+i\nu,\frac{1}{2}+i\nu,1+2i\nu,\frac{2}{{\cal F}^{\rm AdS}+1}\right)\,.\nonumber\\
&&
\eea
The definition of the two-point invariant ${\cal F}^{\rm AdS}$ in this case is:
\be\label{fads}
{\cal F}^{\rm AdS}(X_1,X_2)={\cal F}^{\rm AdS}_{12}(X_1,X_2)=-\bigg(\frac{z_1^2+z_2^2+x_{12}^2}{2 z_1z_2}\bigg)\,,
\ee
In this case, $\nu$ is associated with the particle's mass by: \be \nu=\sqrt{-m_{\rm AdS}^2-\frac{d^2}{4}}.\ee
The behavior of this propagator at infinity is $|{\cal F}^{\rm AdS}|^{-\frac{d}{2}-i\nu}$, which is consistent with the standard boundary conditions. We will assume more broad complex $\nu$'s with ${\text{Im}}\, \nu<0$ below, although $\nu$ is fully imaginary for masses over the BF bound, which is often investigated in AdS. The propagator with different boundary conditions for a field of the same mass, $G_{-\nu}^{\rm AdS}({\cal F}^{\rm AdS})$, will also be required and described by the following expression:
\bea
\label{GAdS2}
G_{-\nu}^{\rm AdS}({\cal F}^{\rm AdS})& =& \frac{\G\left(\frac{d}{2}-i\nu\right)}{2\pi^{\frac{d}{2}} \G\left(1-i\nu\right)\left(-2({\cal F}^{\rm AdS}+1)\right)^{\frac{d}{2}-i\nu}}{}_2F_1\left(\frac{d}{2}-i\nu,\frac{1}{2}-i\nu,1-2i\nu,\frac{2}{{\cal F}^{\rm AdS}+1}\right)\,.\nonumber\\
&&
\eea
The bulk-to-boundary propagators, which are defined as follows, link operators inserted at the AdS boundary to canonically normalized bulk fields:
\bea
\label{KAdS}
{\cal Y}_{\nu}^{\text{AdS}}(X_1,P_2) = \sqrt{\left(\frac{\Gamma(\frac{d}{2} + i\nu)}{2 \pi^{d/2} \Gamma(1+i\nu)}\right)}~\left(\frac{z_1}{z_1^2 + x_{12}^2}\right)^{\frac{d}{2} + i \nu}.
\eea
The bulk point $X_1$ has Poincar\'e coordinates $(z_1,x_1)$, the boundary point $P_2$ has coordinate $x_2$, and the coefficient factor $\sqrt{\left(\frac{\Gamma(\frac{d}{2} + i\nu)}{2 \pi^{d/2} \Gamma(1+i\nu)}\right)}$ is such that the boundary operators have unit-normalized two-point functions.
A finite normalization of boundary operators was achieved by multiplying the propagator by a divergent factor $z_2^{-\frac{d}{2} -i \nu}$, as opposed to \eqref{KdS}. The EAdS harmonic function is another significant item that can be expressed using the following formula:
\be
\label{OmegaAdS}
\Omega_\nu^{\rm AdS}({\cal F}^{\rm AdS})=\frac{1}{\G(i\nu)\G(- i\nu)}V_\nu({\cal F}^{\rm AdS})\,,
\ee
where the function $V_\nu$ and its argument ${\cal F}^{\rm AdS}$ are defined in equation \ref{defW1} and \ref{fads} respectively. It should be noted that since ${\cal F}^{\rm AdS}<-1$, \eqref{GAdS1} and \eqref{OmegaAdS} can be defined without a $i\eps$-prescription.
For ${\rm Im}\,\nu<0$, the propagators and the harmonic function have the following relationships:
\bea
\label{GOmegarel}
&&{\cal H}_\nu^{\rm AdS}=\frac{i\nu}{2\pi}\left(G^{\rm AdS}_\nu-G^{\rm AdS}_{-\nu}\right),\eea where the factors $G^{\rm AdS}_\nu$ and $G^{\rm AdS}_{-\nu}$ are given by:
\bea 
&&G^{\rm AdS}_\nu=\int_{-\infty}^{\infty} d\nu'\bigg(\frac{1}{\nu'^2-\nu^2}\bigg)~{\cal H}_\nu^{\rm AdS}\\
&&G^{\rm AdS}_{-\nu}=\int_{-\infty}^{\infty} d\nu'\bigg(\frac{1}{\nu'^2-\nu^2}\bigg)~{\cal H}_\nu^{\rm AdS}+\frac{2 \pi i}{\nu}~{\cal H}^{\rm AdS}_\nu\,=\bigg(G^{\rm AdS}_\nu+\frac{2 \pi i}{\nu}~{\cal H}^{\rm AdS}_\nu\,\bigg). 
\eea
Depending on the boundary conditions for the fields, the propagators $G^{\rm AdS}_\nu$ and ${\cal Y}^{\rm AdS}_\nu$, or $G^{\rm AdS}_{-\nu}$ and ${\cal Y}^{\rm AdS}_{-\nu}$, connect interaction vertices and boundary sources to generate perturbation theory in EAdS.

\section{Wick rotation demystified}
\label{ka4}
We will now demonstrate that the EAdS harmonic analysis may be applied to linearly combine EAdS Feynman diagrams by Wick rotating integrals over vertex locations corresponding to any Feynman diagram in dS. In the theory of in-in perturbations, three distinct propagators are observed. To prevent any Feynman diagram singularities from being crossed, we wish to rotate as follows:
\bea
\label{Wick1}
&&\eta^L\to \exp\bigg(\frac{i\pi}{2}\bigg)\eta^L\,,\\
\label{Wick2} &&\eta^R\to \exp\bigg(-\frac{i\pi}{2}\bigg)\eta^R\,,
\eea
followed by the determination of $\eta$'s absolute value using the EAdS radial coordinate $z$. We will begin with $G^{LR}_\nu$. This propagator has a straightforward relationship with the EAdS harmonic function \cite{Choudhury:2025pzp,Sleight:2019hfp}. Yes, under \eqref{Wick1} and \eqref{Wick2}, we obtain:
\be\label{eq:lrint}
{\cal F}(X_1^L,X_2^R)\to {\cal F}^{\rm AdS}(X_1,X_2)\,, 
\ee
and as a result further using \eqref{OmegaAdS}, we get:
\bea
\label{splitW}
&&G^{LR}_\nu({\cal F})=G^{RL}_\nu({\cal F})\to\G(i\nu)\G(- i\nu)\Omega_\nu^{\rm AdS}({\cal F}^{\rm AdS})\nonumber\\
&&~~~~~~~~~~~~~~~~~~~~~~~~~~~~~~~~~=\frac{i\nu  }{2\pi}~\G(i\nu)\G(- i\nu)~\bigg(G^{\rm AdS}_\nu ({\cal F}^{\rm AdS})-G^{\rm AdS}_{-\nu}({\cal F}^{\rm AdS})\bigg)\,.
\eea
Consequently, any left-right propagator following the Wick rotation is substituted, up to an overall coefficient, by the harmonic function or the difference of two EAdS propagators with differing boundary conditions. With the left-left and right-right propagators, the issue is a little more nuanced. In fact, we obtain the following under \eqref{Wick1} and \eqref{Wick2}:
\be\label{eq:llint}
{\cal F}(X_1^{L(R)},X_2^{L(R)})\to - {\cal F}^{\rm AdS}(X_1^{\rm AdS},X_2^{\rm AdS})\,, 
\ee
Thus, we do not simply obtain the requisite EAdS propagators. However, because the wave function of a QFT in dS can be gotten from a comparable analytic continuation, we should assume that left-left and right-right propagators, which fundamentally encapsulate the temporal evolution of the state in the interaction image, can similarly be obtained from EAdS as:
\bea
\label{splitGreen}
&&G_\nu^{LL}({\cal F})\to\frac{i \nu}{2 \pi} {\G(i\nu)\G(- i\nu)}\nonumber\\
&&~~~~~~~~~~~~~~~\times\bigg(G^{\rm AdS}_{\nu}({\cal F}^{\rm AdS})\exp\bigg( i \pi \left(\frac{d}{2}+i\nu \right)\bigg)\nonumber\\
&&~~~~~~~~~~~~~~~~~~~~~~~~~~~~~~~-G_{-\nu}^{\rm AdS}({\cal F}^{\rm AdS})\exp\bigg( i \pi \left(\frac{d}{2}-i\nu \right)\bigg)\bigg)\,,\\
&&G_\nu^{RR}({\cal F})\to\frac{i \nu}{2 \pi} {\G(i\nu)\G(- i\nu)}\nonumber\\
&&~~~~~~~~~~~~~~~\times\bigg(G^{\rm AdS}_{\nu}({\cal F}^{\rm AdS})\exp\bigg( -i \pi \left(\frac{d}{2}+i\nu\right)\bigg)\nonumber\\
&&~~~~~~~~~~~~~~~~~~~~~~~~~~~~~~~-G_{-\nu}^{\rm AdS}({\cal F}^{\rm AdS})e\exp\bigg( -i \pi \left(\frac{d}{2}-i\nu\right)\bigg)\bigg)\,.
\eea
AdS Green's functions are thus linearly combined to replace these propagators as well. Also equivalent relations can be obtained for the propagators' Mellin-Barnes representations~\footnote{We choose this normalization to highlight the fact that, instead of using boundary operators, we typically compute expectation values of bulk fields introduced at a considerable time $\eta_c$ in dS.}:
\bea
\label{Krot1}
&& {\cal Y}^{L}_\nu({\cal F})\to(-\eta_c)^{\frac{d}{2}-i\nu} \exp\bigg(i\frac{\pi}{2}\left(\frac{d}{2}-i\nu\right)\bigg)\frac{ \frac{1}{4 \pi^{\frac{d}{2}+1}}\G\left(\tfrac{d}{2}-i\nu\right)\G(i \nu)}{ \left(\frac{\Gamma(\frac{d}{2} - i\nu)}{2 \pi^{d/2} \Gamma(1-i\nu)}\right)^{\frac{1}{2}}}{\cal Y}^{\rm AdS}_{-\nu}(X,P),\\
\label{Krot2}
&&{\cal Y}^{R}_\nu({\cal F})\to(-\eta_c)^{\frac{d}{2}-i\nu} \exp\bigg(-i\frac{\pi}{2}\left(\frac{d}{2}-i\nu\right)\bigg)\frac{ \frac{1}{4 \pi^{\frac{d}{2}+1}}\G\left(\tfrac{d}{2}-i\nu\right)\G(i \nu)}{ \left(\frac{\Gamma(\frac{d}{2} - i\nu)}{2 \pi^{d/2} \Gamma(1-i\nu)}\right)^{\frac{1}{2}}}{\cal Y}^{\rm AdS}_{-\nu}(X,P)\,.
\eea
We have already demonstrated that each in-in diagram is substituted by a linear combination of EAdS Witten diagrams, which is adequate for our computational needs.

\section{Constructing the building block of EAdS}\label{ka5}
In this part, we shall present a purely perturbative argument to any order in perturbation theory, but at the level of Feynman diagrams. To keep things simple, we will use the example of a single scalar field with the Lagrangian to illustrate this process:
  \be
{\cal{L}}=-\frac{1}{2} \left(\partial\phi\right)^2- \frac{m^2}{2}\phi^2-V(\phi)~.
\ee  
Finding an EAdS Lagrangian whose related perturbative expansion exactly matches this theory's in dS is the main goal. First, let's talk about the internal lines. By adding independent fields that propagate on the left and right sides of the contour, it is common practice in in-in computations to double the field content. As demonstrated above, each propagator is swapped out for a pair of EAdS propagators that correspond to distinct boundary conditions following our Wick rotation. Therefore, we start by substituting four AdS fields for a single dS field. Specifically, we take the following parametrization of the AdS potential for the double field content:
\be
V^{\rm AdS}\left(\phi^L_+,\phi^L_-,\phi^R_+,\phi^R_-\right)=\exp\bigg(-i\frac{\pi}{2}(d-1)\bigg)V(\phi^L_++\phi^L_-)+\exp\bigg(i\frac{\pi}{2}(d-1)\bigg)V(\phi^R_++\phi^R_-)\,,
\ee
where the double field content is identified by the following expressions written in terms of the left and right field contents:
\bea
&&\phi_+\equiv\Bigg[\phi^L_+~ \exp\bigg(i\frac{\pi}{2}  \left(\frac{d}{2}+i\nu\right)\bigg)+~\phi^R_+~\exp\bigg(-i\frac{\pi}{2}  \left(\frac{d}{2}+i\nu\right)\bigg)\Bigg]\,,\\
&&\phi_-\equiv\Bigg[\phi^L_-~ \exp\bigg(i\frac{\pi}{2}  \left(\frac{d}{2}-i\nu\right)\bigg)+~\phi^R_-~\exp\bigg(-i\frac{\pi}{2}  \left(\frac{d}{2}-i\nu\right)\bigg)\Bigg]\,.
\eea
Here for the + and - field contents the corresponding propagator matrices are given by the following expressions:
\bea
\label{GM1}
G_+({\cal F}^{\rm AdS})&=&\begin{pmatrix}
    G_+^{LL}({\cal F}^{\rm AdS}) & G_+^{LR}({\cal F}^{\rm AdS}) \\
    G_+^{RL}({\cal F}^{\rm AdS}) &  G_+^{RR}({\cal F}^{\rm AdS})
 \end{pmatrix}\nonumber\\
 &=&\frac{i \nu}{2 \pi} {\G(i\nu)\G(- i\nu)}G_\nu^{\rm AdS}({\cal F}^{\rm AdS})
 \begin{pmatrix}
    \exp\bigg(i\pi \left(\frac{d}{2}+i\nu\right)\bigg) & 1 \\
    1 &  \exp\bigg(-i\pi \left(\frac{d}{2}+i\nu\right)\bigg) 
 \end{pmatrix}\,,~~~~~~
\eea\bea
\label{GM2}
G_-({\cal F}^{\rm AdS})&=&\begin{pmatrix}
    G_-^{LL}({\cal F}^{\rm AdS}) & G_-^{LR}({\cal F}^{\rm AdS}) \\
    G_-^{RL}({\cal F}^{\rm AdS}) &  G_-^{RR}({\cal F}^{\rm AdS})
 \end{pmatrix}\nonumber\\
 &=&-\frac{i \nu}{2 \pi} {\G(i\nu)\G(- i\nu)}G_{-\nu}^{\rm AdS}({\cal F}^{\rm AdS})
  \begin{pmatrix}
    \exp\bigg(i\pi \left(\frac{d}{2}-i\nu\right)\bigg) & 1 \\
    1 &  \exp\bigg(-i\pi \left(\frac{d}{2}-i\nu\right)\bigg) 
 \end{pmatrix}\,.~~~~~~
\eea
The original dS theory is precisely reproduced by this four-field theory. Next, observe that the propagators above have degenerate matrices. It indicates that in each sector, only a single linear combination truly spreads.

By designating these linear combinations as $\Phi_+$ and $\Phi_-$, we deduce that the following EAdS Lagrangian generates all Wick rotated Feynman diagrams:
\bea
\label{LAdS}
 {\cal{L}}^{\rm AdS}(\Phi_+,\Phi_-) &=&
 -i\,{\sinh\pi\nu} \left((\partial\Phi_+)^2- m^2\Phi_+^2\right) +  i\,{\sinh\pi\nu} \left((\partial\Phi_-)^2-- m^2\Phi_-^2\right)\, \\
&& -\,\exp\left(-i\pi\frac{d-1}{2}\right)V\left( \exp\left(i\frac{\pi}{2}  \left(\frac{d}{2}+i\nu\right)\right)\Phi_++\exp\left(i\frac{\pi}{2}  \left(\frac{d}{2}-i\nu\right)\right)\Phi_-\right)\nonumber\\
&&-\exp\left(i\pi\frac{d-1}{2}\right)V\left(  \exp\left(-i\frac{\pi}{2}  \left(\frac{d}{2}+i\nu\right)\right)\Phi_++\exp\left(-i\frac{\pi}{2}  \left(\frac{d}{2}-i\nu\right)\right)\Phi_-\right)\,,\nonumber
\eea
where the propagators in \eqref{GM1}, \eqref{GM2} are matched by the following boundary conditions for the fields:
\bea
&& \Phi_{+}\approx z^{\frac{d}{2} + i\nu}\,,\\
&& \Phi_{-}\approx z^{\frac{d}{2} - i\nu}\,.
\eea
We now need to talk about how our dS and AdS theories map external fields. We examine the asymptotic behavior of the dS propagators under the assumption of light external fields. We can also take heavy fields in foreign states into consideration at the leading rank. A linear combination of $\phi$ and its time derivative is necessary in this situation for the operator with definite scaling dimension. But these fields are more nuanced because their correlation functions contain contact components, and because they deviate from the principal series when interactions are taken into account. It is evident that in order to replicate correlation functions in dS, the operators dual to the field $\Phi_-$ on the edge of EAdS must be inserted. This operator, which we assume has a unit-normalized two-point function, will be called ${\cal M}_-$. Next, we must determine the normalization constant $\theta$ so that:
\be
\left\langle \theta {\cal M}_-(x_1)\theta{\cal M}_-(x_2)\ldots\right\rangle_{\rm AdS}=\left\langle\phi(x_1,\eta_c)\phi(x_2,\eta_c)\ldots\right\rangle_{\rm dS},
\ee
which gives:
\be
\theta=(-\eta_c)^{\frac{d}{2}-i\nu}\sqrt{\frac{1}{4 \pi^{\frac{d}{2}+1}}\G\left(\frac{d}{2}-i\nu\right)\G(i \nu)}\,.
\ee
It should be emphasized that the theory is not unitary in the EAdS sense because the fields have opposing sign kinetic terms, which means that one of them is always a ghost regardless of the value of $\nu$. This is not an issue as we are not attempting to understand this Lagrangian as a theory on Lorentzian AdS's Euclidean rotation. From the EAdS perspective, both fields have negative mass-squared, as indicated by the signs of the mass terms. Below the BF constraint, EAdS fields correspond to heavy dS fields, i.e.
\be m^2 > \frac{d^2}{4}.\ee
Even if light dS fields are above the bound, they fall outside of the region where other boundary conditions are permitted in AdS for \be 0\leq m^2<\left(\frac{d^2}{4}-1\right).\ee The usage of this formulation is challenging outside perturbation theory due to these complexities, but it does not cause any problems in perturbation theory at any level.  Since the Lagrangian is essentially a real functional, for light fields \eqref{LAdS} corresponds to a real, rather than complex. Naturally, this supports the existence of correlation functions in dS.

\section{Revisiting basics of conformal block}\label{ka6}
Constructing conformal partial waves is helpful in discussing the features of the boundary four-point function. Conformal partial waves are single-valued solutions to the Casimir differential equation in $x_j$:
\be\label{eq:difeqcpw}
 -\frac{1}{2}\left(\mathcal{L}_1^{AB}+\mathcal{L}_2^{AB}\right)\left(\mathcal{L}_{1,AB}+\mathcal{L}_{2,AB}\right)\cdot \widehat{\mathcal{E}}_{\Delta^{\prime},J}^{\{\Delta_j\}}(x_1,x_2,x_3,x_4)=c_{2}^{(\Delta^{\prime},J)}\widehat{\mathcal{E}}_{\Delta^{\prime},J}^{\{\Delta_j\}}(x_1,x_2,x_3,x_4).
\ee
In this case, the differential operator of $x_j$ that describes an action of a generator of~$SO(d+1,1)$ is $\mathcal{L}_{j}^{AB}$. The conformal dimension $\Delta^{\prime}$ and spin $J$ are used to determine the eigenvalue $c_{2}^{\Delta^{\prime},J}$ as follows:
\be\label{eq:eigenc}
c_{2}^{\Delta^{\prime},J}=\Bigg(\Delta^{\prime}(\Delta^{\prime}-d)+J(J+d-2)\Bigg).
\ee
Another significant quantity is the conformal block $\widehat{\mathcal{Q}}_{\Delta^{\prime},J}^{\{\Delta_j\}}$, which may be used to express the conformal partial wave by the following expression:
\be\label{eq:CPW}
\widehat{\mathcal{E}}^{\{\Delta_j\}}_{\Delta^{\prime},J}(x_i)=\Bigg(\mathcal{C}_{d-\Delta^{\prime},J}^{\Delta_3,\Delta_4}\widehat{\mathcal{Q}}_{\Delta^{\prime},J}^{\{\Delta_j\}}(x_i)+\mathcal{C}_{\Delta^{\prime},J}^{\Delta_1,\Delta_2}\widehat{\mathcal{Q}}_{d-\Delta^{\prime},J}^{\{\Delta_j\}}(x_i)\Bigg),
\ee
where the coefficients $\mathcal{C}_{\Delta^{\prime},J}^{\Delta_1,\Delta_2}$ and $\mathcal{C}_{d-\Delta^{\prime},J}^{\Delta_3,\Delta_4}$ are given by the following expressions:
\bea
&&\mathcal{C}_{\Delta^{\prime},J}^{\Delta_1,\Delta_2}=\frac{\pi^{\frac{d}{2}}\Gamma(\Delta^{\prime}-\frac{d}{2})\Gamma (\Delta^{\prime}+J-1)\Gamma (\frac{d-\Delta^{\prime}+\Delta_1-\Delta_2+J}{2})\Gamma (\frac{d-\Delta^{\prime}+\Delta_2-\Delta_1+J}{2})}{(-2)^{J}\Gamma (\Delta^{\prime}-1)\Gamma (d-\Delta^{\prime}+J)\Gamma (\frac{\Delta^{\prime}+\Delta_1-\Delta_2+J}{2})\Gamma (\frac{\Delta^{\prime}+\Delta_2-\Delta_1+J}{2})},\\
&& \mathcal{C}_{d-\Delta^{\prime},J}^{\Delta_3,\Delta_4}=\frac{\pi^{\frac{d}{2}}\Gamma(\frac{d}{2}-\Delta^{\prime})\Gamma (d-\Delta^{\prime}+J-1)\Gamma (\frac{\Delta^{\prime}+\Delta_3-\Delta_4+J}{2})\Gamma (\frac{\Delta^{\prime}+\Delta_4-\Delta_3+J}{2})}{(-2)^{J}\Gamma (d-\Delta^{\prime}-1)\Gamma (\Delta^{\prime}+J)\Gamma (\frac{d-\Delta^{\prime}+\Delta_3-\Delta_4+J}{2})\Gamma (\frac{d-\Delta^{\prime}+\Delta_4-\Delta_3+J}{2})}.
\eea
Conventional definitions of the conformal blocks and conformal partial waves in the literature are based solely on the conformal cross ratios $(z,\bar{z})$. One way to do this is to:
\be\label{eq:stripping}
\begin{aligned}
\widehat{\mathcal{E}}^{\{\Delta_j\}}_{\Delta^{\prime},J}(x_i)&=\frac{1}{(x_{12}^2)^{\frac{\Delta_1+\Delta_2}{2}}(x_{34}^2)^{\frac{\Delta_3+\Delta_4}{2}}}\left(\frac{x_{14}^2}{x_{24}^2}\right)^{\frac{\Delta_2-\Delta_1}{2}}\left(\frac{x_{14}^2}{x_{13}^2}\right)^{\frac{\Delta_3-\Delta_4}{2}}\mathcal{E}^{\{\Delta_j\}}_{\Delta^{\prime},J}(z,\bar{z}),\\
\widehat{\mathcal{Q}}_{\Delta^{\prime},J}^{\{\Delta_j\}}(x_i)&=\frac{1}{(x_{12}^2)^{\frac{\Delta_1+\Delta_2}{2}}(x_{34}^2)^{\frac{\Delta_3+\Delta_4}{2}}}\left(\frac{x_{14}^2}{x_{24}^2}\right)^{\frac{\Delta_2-\Delta_1}{2}}\left(\frac{x_{14}^2}{x_{13}^2}\right)^{\frac{\Delta_3-\Delta_4}{2}}\mathcal{Q}_{\Delta^{\prime},J}^{\{\Delta_j\}}(z,\bar{z}).
\end{aligned}
\ee

\section{Four point function as a conformal block}\label{ka7}
We refer to the following representations as the spectral decomposition because they are known to be admitted by the boundary four-point functions in EAdS:
\be\label{eq:EAdSCP}
\begin{aligned}
\langle \mathcal{O}(x_1)\mathcal{O}(x_2)\mathcal{O}(x_3)\mathcal{O}(x_4)\rangle&=\frac{1}{|x_{12}|^{2\Delta_{\mathcal{O}}}|x_{34}|^{2\Delta_{\mathcal{O}}}}\sum_J\int^{\frac{d}{2}+i\infty}_{\frac{d}{2}}\frac{d\Delta^{\prime}}{2\pi i}\rho_{J}(\Delta^{\prime})\mathcal{E}_{\Delta^{\prime},J}(z,\bar{z})\\
&=\frac{1}{|x_{12}|^{2\Delta_{\mathcal{O}}}|x_{34}|^{2\Delta_{\mathcal{O}}}}\sum_J\int^{\frac{d}{2}+i\infty}_{\frac{d}{2}-i\infty}\frac{d\Delta^{\prime}}{4\pi i}\rho_{J}(\Delta^{\prime})\mathcal{E}_{\Delta^{\prime},J}(z,\bar{z}),
\end{aligned}
\ee
where $z$ and $\bar{z}$ describes conformal cross ratios which are parametrized by the following expression:
\bea\label{eq:CRdef1}
&&z\bar{z}=\frac{|x_{12}|^2|x_{34}|^2}{|x_{13}|^2|x_{24}|^2},\\ \label{eq:CRdef2} &&(1-z)(1-\bar{z})=\frac{|x_{14}|^2|x_{23}|^2}{|x_{13}|^2|x_{24}|^2},
\eea
The spectral density is $\rho_J(\Delta^{\prime})$ in this case. The spectral density and the shadow symmetry of the conformal partial wave, which are provided by: 
\bea && \mathcal{E}_{\Delta^{\prime},J}=\mathcal{E}_{d-\Delta^{\prime},J}\\ &&\rho_{J}(\Delta)=\rho_{J}(d-\Delta),\eea
were utilized in the second line of \eqref{eq:EAdSCP}.
From the harmonic analysis of the Euclidean conformal group, the representations \eqref{eq:EAdSCP}, which are frequently referred to as the conformal partial wave decomposition, can be obtained directly. Refer to references. \cite{Caron-Huot:2017vep,Simmons-Duffin:2017nub,Karateev:2018oml,Hogervorst:2017sfd,Rutter:2020vpw} for further information.

The expansion of the four-point function into a discrete sum, known as the conformal block expansion, is now standard practice in conformal field theory on the boundary. To obtain such an expansion from \eqref{eq:EAdSCP}, $\mathcal{E}_{\Delta^{\prime},J}$ must first be broken down into a linear combination of conformal blocks $\mathcal{Q}$ as follows:
\be \mathcal{E}_{\Delta^{\prime},J}=\Bigg(\mathcal{C}_{d-\Delta^{\prime},J}\mathcal{Q}_{\Delta^{\prime},J}+\mathcal{C}_{\Delta^{\prime},J}\mathcal{Q}_{\Delta^{\prime},J}\Bigg).\ee
For $\Delta^{\prime}>\frac{d}{2}$, the first term decays exponentially on the right half plane, whereas the second term decays exponentially on the left half plane.
This expression is inserted into \eqref{eq:EAdSCP}, and the variables $\Delta^{\prime}\to d-\Delta^{\prime}$ in the second term are changed. The result is:
\be\nonumber
\langle \mathcal{O}(x_1)\mathcal{O}(x_2)\mathcal{O}(x_3)\mathcal{O}(x_4)\rangle=\frac{1}{|x_{12}|^{2\Delta_{\mathcal{O}}}|x_{34}|^{2\Delta_{\mathcal{O}}}}\sum_J\int^{\frac{d}{2}+i\infty}_{\frac{d}{2}-i\infty}\frac{d\Delta^{\prime}}{2\pi i}\rho_{J}(\Delta^{\prime})\mathcal{C}_{d-\Delta^{\prime},J}\mathcal{Q}_{\Delta^{\prime},J}.
\ee
Subsequently, we can shift the contour to the right and substitute the right side with the total of the poles' contributions. This gives:
\be\label{eq:confblock}
\langle \mathcal{O}(x_1)\mathcal{O}(x_2)\mathcal{O}(x_3)\mathcal{O}(x_4)\rangle=\frac{1}{|x_{12}|^{2\Delta_{\mathcal{O}}}|x_{34}|^{2\Delta_{\mathcal{O}}}}\sum_{\Delta^{\prime}}C(\Delta^{\prime})\mathcal{Q}_{\Delta^{\prime},J}(z,\bar{z}),
\ee
In this case, $\rho_{J}(\Delta^{\prime})$'s sum over poles is $\sum_{\Delta^{\prime}}$, and $C(\Delta^{\prime})$ is determined by:
\be
C(\Delta^{\prime})=-{\rm Res}_{\Delta=\Delta^{\prime}}\bigg(\rho_{J}(\Delta)\mathcal{C}_{d-\Delta,J}\bigg).
\ee
The assumption that $\rho_J(\Delta)$ is a meromorphic function of $\Delta$ was made in the derivation. In EAdS perturbation theory, this is certain. Beyond perturbation theory, however, we must use a different reasoning approach.

\section{State-operator correspondence}\label{ka8}
Without using perturbation theory, it is possible to prove the existence of the discrete expansion \eqref{eq:confblock} and its finite radius of convergence. In order to see this, we must apply the state-operator correspondence for quantum field theory in EAdS, which was initially stated in \cite{Paulos:2016fap} and is a logical extension of the corresponding statement for unitary CFT.
     
Take into consideration local operator insertions on the EAdS boundary. A hemisphere that is anchored on the boundary and encircles some of those operators is then drawn inside EAdS.  By performing a route integral inside this hemisphere, we can substitute a wave function defined on the hemisphere for the operators inside. As the AdS isometry maps this hemisphere to a constant-time slice of global AdS, this wave function corresponds exactly to a state in global AdS. Both single and multiple operator insertions can be supported by this argument: The state-operator correspondence, which offers a map between an operator at the boundary and a state in global AdS, is all that is required for a single operator. The following is a simple albeit not particularly rigorous method of understanding this: we can alter the hemisphere's radius by performing a dilatation transformation, which is one of the isometries of EAdS. Only a very small area surrounding the point where the hemisphere decreases does the wave function on the hemisphere have support. By doing this, it becomes intuitively evident that the wave function in the limit contains the same amount of information as an operator constructed at that one point.

\section{Operator Product Expansion}\label{ka9}
The operator product expansion (OPE) is obtained by running this argument to several insertions, substituting a wave function on the hemisphere for them, and then using the state-operator correspondence \cite{Rychkov:2016iqz,Simmons-Duffin:2016gjk} to convert it back to a sum of operators with definite scaling dimensions. The OPE is schematically provided by:
 \be
\mathcal{O}_1(x)\mathcal{O}_2(0)=\sum_{\mathcal{O}_3}c_{123}|x|^{\Delta_3-\Delta_1-\Delta_2}\mathcal{O}_3(0)+\cdots
\ee
Through the use of this OPE inside the four-point function and the analysis of conformal symmetry constraints, we are able to prove the presence of the expansion \eqref{eq:confblock} at a level that is completely nonequivocal. In global AdS, this also enables us to understand the sum as a sum over states. Additionally, by closely examining the series, we can demonstrate that the expansion has a finite radius of convergence. For more information, see refs.\cite{Pappadopulo:2012jk,Hogervorst:2013sma}.

\section{Analyticity of late-time correlators in dS}\label{ka10}
\subsection{Spectral amplitude}

The spectral decomposition or conformal partial wave expansion of the four-point function, which has the same shape as in EAdS, serves as the foundation for our analysis:
\be\label{eq:dSCP}
\begin{aligned}
\langle \mathcal{O}(x_1)\mathcal{O}(x_2)\mathcal{O}(x_3)\mathcal{O}(x_4)\rangle&=\frac{1}{|x_{12}|^{2\Delta_{\mathcal{O}}}|x_{34}|^{2\Delta_{\mathcal{O}}}}\sum_J\int^{\frac{d}{2}+i\infty}_{\frac{d}{2}}\frac{d\Delta^{\prime}}{2\pi i}\rho_{J}(\Delta^{\prime})\mathcal{E}_{\Delta^{\prime},J}(z,\bar{z})\\
&=\frac{1}{|x_{12}|^{2\Delta_{\mathcal{O}}}|x_{34}|^{2\Delta_{\mathcal{O}}}}\sum_J\int^{\frac{d}{2}+i\infty}_{\frac{d}{2}-i\infty}\frac{d\Delta^{\prime}}{4\pi i}\rho_{J}(\Delta^{\prime})\mathcal{E}_{\Delta^{\prime},J}(z,\bar{z}).
\end{aligned}
\ee
It is possible to demonstrate the existence of these representations both at the non-perturbative level and in perturbation theory: All of the diagrams can be mapped to EAdS in perturbation theory using the EAdS Lagrangian that was developed in the preceding section. As a result, the resulting four-point function has the same structure. We can use a representation theory of the dS isometry at a non-perturbative level. In perturbative computations, we frequently arrive at a representation that resembles the second line of \eqref{eq:dSCP}. However, $\rho_J$ is substituted with a function $f_{J}(\Delta^{\prime})$ that is not symmetric under $\Delta^{\prime} \to d-\Delta^{\prime}$. This is an intriguing difference in dS. In perturbation theory, $f_J(\Delta^{\prime})$ has better analytic properties than $\rho_J$. It is analytic on the left half plane ${\rm Re}\, \Delta<\frac{d}{2}$, with the exception of potential poles that correspond to the unitary irreducible representations of dS isometries, such as the complementary and discrete series. With the probable exception of poles that might represent bound states, this quality is similar to the partial wave amplitudes for flat space, namely the analyticity in the upper half plane. We refer to $f_{J}$ as the spectral amplitude as a result. Through simple symmetrization, it is associated with $\rho_J (\Delta)$.
\be
\rho_J(\Delta^{\prime})=\frac{1}{2}\bigg(f_{J}(\Delta^{\prime})+f_{J}(d-\Delta^{\prime})\bigg).
\ee
In contrast to the spectral density $\rho_J$, we currently lack a physical interpretation and a non-perturbative formulation for $f_J$. Connecting it to the physics of $S^{d+1}$ is one promising approach. This optimism stems from the fact that the bulk two-point function also has an analog of the spectral amplitude, which we can define non-perturbatively there by the analytic continuation of the harmonic expansion coefficients on $S^{d+1}$. We consider the physical meaning of $f_J$ and its non-perturbative characterization to be a significant outstanding problem.

\subsection{Expansion in terms of conformal block}
Using EAdS, we may drag the contour to the right half-plane and obtain a conformal block expansion if the spectral density $\rho(\Delta)$ is a meromorphic function on the right half plane. The mapping to EAdS that we developed in perturbation theory makes this unquestionably true. By definition, we have a discrete sum that resembles a conformal block expansion or, more accurately, an operator product expansion:
\be\label{eq:confblockdS}
\langle \mathcal{O}(x_1)\mathcal{O}(x_2)\mathcal{O}(x_3)\mathcal{O}(x_4)\rangle_{\bf perturbation}=\frac{1}{|x_{12}|^{2\Delta_{\mathcal{O}}}|x_{34}|^{2\Delta_{\mathcal{O}}}}\sum_{\Delta^{\prime}}C(\Delta^{\prime})\mathcal{Q}_{\Delta^{\prime},J}(z,\bar{z}),
\ee
The operators $\Delta^{\prime}$ no longer have real dimensions, which is the only difference from EAdS. The analytic structure is the same as that for EAdS except for this distinction. In this construction there are few questions exists, which needs to be addressed. First of all, one can ask about the implications and physical interpretation of the sum $\displaystyle \sum_{\Delta^{\prime}}$ appeared in \eqref{eq:confblockdS}. Further, one can ask regarding the validity of the expansion at the non-perturbative level of the computation. The existence of the state-operator correspondence in dS lies at the heart of these two queries. We will go into great depth about this since there is a significant distinction between EAdS and dS.

\subsection{State-operator correspondence in dS}
The presence of the expansion at a non-perturbative level was ensured by the fact that the conformal block expansion in EAdS could be understood as a sum over states in the global AdS. Therefore, one may wonder: Is a sum over states in dS connected to the conformal block expansion in dS? The response is no. The Poincare patch or the global dS both define states in dS on a constant-time slice. The constant-time slice in dS does not contract to a point on the late-time surface under the influence of dS isometries, in contrast to the hemisphere.
The failure of the state-operator correspondence, or at least its most basic generalization, in dS can be explained intuitively by this. Additionally, representation theory demonstrates this: A sum over operators with arbitrary complex dimensions is involved in the conformal block expansion in dS \eqref{eq:confblockdS}, as was previously mentioned. Since these operators don't match up with unitary irreducible representations of dS isometries, they can't be understood in terms of dS Hilbert space states. However, the question remains if the sum $\sum_{\Delta^{\prime}}$ can be explained by some peculiar quantization of quantum field theory in dS, which suggests a variation of a state-operator correspondence. For example, one could attempt to think about EAdS as a hemisphere analog. However, the discrepancy in the signature also makes this ineffective. The definition of the hemisphere in EAdS with its center at $(z,x^{\mu})=(0,0)$ in the Poincare coordinates is:
\be
\bigg(|x|^{2}+z^2\bigg)=R^2.
\ee
In this case, $R$ is the hemisphere's radius. The radius $R$ varies multiplicatively under the dilatation action, but it remains invariant under rotation around $x^{\mu}$. A surface described in the Poincare coordinates with comparable transformation properties in dS is as follows:
\be\label{eq:hyperboloid}
\bigg(|x|^2-\eta^2\bigg)=R^2,
\ee
The late-time surface serves as the anchor for this hyperboloid. We cannot use this to define states since the surface is time-like, even if it transforms simply under the action of dS isometries. It should be noted that the aforementioned reasons do not preclude the possibility of defining an analog of state-operator correspondence and selecting a space-like surface in dS at the expense of basic transformation features under isometries. However, as of right now, we are unaware of any tangible implementation of this concept. Rather, we examine the nature of the conformal block expansion in dS in further detail in the following paragraph, and we propose that the sum $\sum_{\Delta^{\prime}}$ can be read as a sum over quasi-normal modes in a static patch of dS.

\section{Connecting Operator Product Expansion with Quasi Normal Modes}\label{ka11}

We provide some intuitive, but not necessarily accurate, justifications for why quasi-normal modes in a static patch of dS might be important for the state-operator correspondence and OPE before delving into technical specifics. Let us first highlight a basic geometric fact: a static patch of dS is part of the causal past of a point on the late-time surface. This implies a close relationship between the physics in the static patch and the physical information present in an operator at that location.  Second, the structure of the OPE is determined by the time evolution in the static patch, which also suggests the presence of a convergence OPE expansion. Let's examine the four-point functions at late time, where two operators, $\mathcal{O}_1$ and $\mathcal{O}_2$, are far apart from the other two, $\mathcal{O}_{3}$ and $\mathcal{O}_4$. We will do this by analyzing them from the perspective of the temporal evolution of particles dual to the operators. When all of these particles are in the same past lightcone of four operators, their interactions are strongest. The dS horizon separates the particles as they move away from the shared lightcone when we time evolve such a state. Following that, the static patch Hamiltonians cause the two sets of particles to time-evolve separately, which causes the correlation between the two sets to decrease exponentially. When expressed more clearly, this results in something like:
\be\label{eq:discretesumstatic}
\langle \mathcal{O}_1\mathcal{O}_2\mathcal{O}_3\mathcal{O}_4\rangle \sim \sum_{\Delta}\exp\bigg(-2 \Delta t_{\ast}\bigg)\,,
\ee
The factor of two is derived from the independent evolutions within the two static patches, and $t_{\ast}$ is the amount of time in the static patch coordinates that has elapsed since they were last in causal touch. The discrete sum \eqref{eq:discretesumstatic} can be naturally interpreted as the summing over the quasi-normal modes since this expression originates from the time-evolution in the static patch de Sitter.  The time $t_{\ast}$ is now associated with the spatial distance between the two sets of operators $r$ as follows due to the geometric structure of de Sitter, $\exp(-t_{\ast})\sim r^{-1}$. This implies the following long distance behaviour of the four-point function in QNM expansion:
\be
\langle \mathcal{O}_1\mathcal{O}_2\mathcal{O}_3\mathcal{O}_4\rangle \sim \sum_{\Delta}r^{-2\Delta}\,.
\ee
Furthermore, the quick decay of the correlation after the particles are separated by the dS horizon implies that the expansion has a good convergence property. However, this quantitative argument is undoubtedly insufficient to prove this claim or determine the radius of convergence.

By examining the structure of correlation functions, we now try to quantitativeize these straightforward intuitive assertions. Two facts form the basis of our argument: First, the spectral function $\rho_J$ for the late-time four-point function and the spectral function for the bulk two-point function have the same poles in perturbation theory cases. Second, the frequencies of quasi-normal modes in a static patch of dS can be identified with the poles in the spectral function for the bulk two-point function, which govern the behavior at large time-like separation. In perturbation theory, the first point is self-evident: consider an exchange diagram at the tree level. Pairs of boundary points can be connected to two bulk points using bulk-to-boundary propagators in order to calculate the late-time four-point function. After then, the calculation essentially consists of calculating an integral of a bulk two-point function and a product of bulk-to-boundary propagators. These calculations will be done later. The spectral function for the four-point function shares the same pole structure as the bulk two-point function because it is expressed in terms of it. Furthermore, this is valid in several situations outside of perturbation theory, such the $O(N)$ model in the large $N$ limit. There is a technical aspect to the second point. The key points are as follows: as with the boundary four-point functions, the bulk two-point functions in dS accept a spectral decomposition in terms of an integral along: 
\be \Delta=\bigg(\frac{d}{2}+i\nu+ {\rm discrete~ sums~ from~ various~ representations}\bigg).\ee 
However, because the integral at large imaginary $\nu$ is only marginally convergent when the two-points are time-like separated, such a representation is not suitable for examining the asymptotic behavior of the time-like separated two-point function. The solution to this issue is to rewrite the integral, draw the contour to the right half plane, and substitute a sum over contributions from poles for the integral, much as we did when we derived a conformal block expansion from the spectral decomposition. Even with time-like separation, the resulting expression exhibits better convergence, and the exponential decays in a static patch, which may be recognized with quasi-normal modes, are controlled by the pole locations.

The combination of these two discoveries leads us to the conclusion that the OPE series in the late-time four-point functions may be interpreted in terms of quasi-normal modes in dS. Even if this is well-motivated by what we have learnt in a number of situations, it is still a guess at this time. Finding a non-perturbative explanation for this claim is crucial. Moreover, even if this supposition is true, a state-operator correspondence for dS is not immediately provided. The fundamental reason for this is that quasi-normal modes have complex frequencies by definition, making it impossible to interpret them in Hilbert space. As a result, they cannot be considered eigenstates of certain physical observables. However, an intriguing idea regarding the interpretation of quasi-normal modes for free scalars in Hilbert space was made \cite{Jafferis:2013qia,Anninos:2010gh}. Further development and an attempt to expand the concept to interacting quantum field theory will be crucial. 
In essence, the same thing was stated in \cite{Arkani-Hamed:2015bza}, which noted that the QNMs of the static patch are likewise connected to the power-laws that show up in the Taylor expansion of correlators at the squeezed limit in momentum space.

\section{Some possible physical implications of Analyticity}\label{ka12}
In this section we are going to briefly discuss about the possible physical implications of analyticity in the present context of discussion. The major points are appended below point-wise: 
\begin{enumerate}
    \item There have been previous arguments that the late-time CFT that holographically depicts dS should contain operators of dimension $\left(\frac{d}{2}+i\nu\right)$, that is, operators that correspond to the Euclidean conformal group's primary series representation. Operators in the boundary CFT must be members of unitary representations of the bulk isometry, according to the implicit assumption behind this. In AdS, this is unquestionably true, but not in dS. Operators in the late-time CFT do not have to be in such unitary representations since in dS, states, not operators, are characterized by unitary representations. This is because there is no state-operator relationship. We actually expect something more powerful to be true.  We assume that the late-time CFT corresponding to the principal series representation for generic interacting quantum field theory in dS does not contain any operators. To understand this, let's say that the primary series corresponds to a large scalar particle in dS. A pole on the principal series is present in the spectral decomposition of the four-point function since such a scalar appears in an exchange diagram at the tree-level. However, the particle's mass renormalizes and the pole deviates from the primary series after the interaction is activated. The process in flat space where a stable particle becomes a resonance due to the interaction is essentially a dS version of this. This is merely a dS version of a flat-space process where a stable particle undergoes an interaction to become a resonance. This nearly always occurs in flat space when a particle's pole is situated above a continuous spectral density, and we anticipate that this will also be the case in dS. However, for light particles that are part of the complementary series, the situation is different. Until it reaches $\Delta=\frac{d}{2}$, when it merges the principal series, a pole corresponding to such a particle cannot deviate off the line of the complementary series due to dS unitarity. They are comparable to bound states in flat space below the two-particle threshold in this regard.

    \item Additionally, we should note that there is some connection to two-dimensional Liouville field theory.  In Liouville field theory, normalizable modes with Liouville momentum \be \gamma=\left(\frac{Q}{2}+i\nu\right),\ee span the Hilbert space, where $Q$ is the Liouville background charge. On the other hand, vertex operators $\exp(\gamma\varphi)$ are typically used for studying correlation functions, where $\alpha$ is a real value. Similar to what we observed in dS, such vertex operators correspond to non-normalizable modes in the Hilbert space and therefore do not exhibit state-operator correspondence.

    \item Along with a single-particle state, the CFT on the boundary for a free scalar in EAdS will also include operators dual to multi-particle states. All these operators are part of unitary representations of the Lorentzian conformal group, provided that the particle mass is non-tachyonic. Even when we activate interactions, these operators remain in perturbation theory. The conformal block expansion of basic EAdS Feynman four-point diagrams, such an exchange diagram or contact diagram, can be used, for example, to verify the presence of the double-trace operators. In dS, the circumstances are different in a number of respects. Examine a free heavy scalar in dS whose dual operator is within the principal series $\left(\frac{d}{2}+i\nu\right)$. The first distinction is that three series of two-particle states will exist as a result of the fields being doubled. These states' dimensions are:
    \bea \Delta_1 &=& \bigg[2\left(\frac{d}{2}+i\nu\right)+n\bigg]=\left(d+2i\nu+n\right),\\
    \Delta_2 &=& \bigg[2\left(\frac{d}{2}-i\nu\right)+n\bigg]=\left(d-2i\nu+n\right),\\
    \Delta_3 &=& \bigg[\left(\frac{d}{2}+i\nu+\frac{d}{2}-i\nu\right)+n\bigg]=\left(d+n\right).\eea
    Here $n$ represents a positive integer in the present context of discussion. In interacting theories, these poles are also visible in Feynman diagrams. Second, there is no unitary representation for any of these two-particle states, in contrast to EAdS. Rather, they are comparable to resonances in flat space. This might seem a little counterintuitive because in flat space, resonances only appear after activating the interaction, whereas in free theory, they are already present. This can be intuitively interpreted as a result of dS expansion: In EAdS, these states are equivalent to two particles in orbit around one another. Since particles have a tendency to drift apart and the dS space is expanding, such states cannot exist as stable states in dS. This is the fundamental physical explanation for why such states are only comprehensible as resonances.

    \item In general, there is an intriguing qualitative difference between dS, flat space, and AdS. Because of the gravitational potential close to the border, which bounces back everything and prevents particle decay, all states in AdS are stable. However, particles can decay in flat space, and only a small number of states are stable, with the rest becoming resonances. In contrast, the dS expansion, as we have shown above, makes nearly all of the states in dS unstable resonances.
\end{enumerate}

\section{Positivity constraints from Unitarity}\label{ka12}

In AdS and flat Minkowski space quantum field theory, unitarity can be seen as certain positivity constraints on the fundamental observables, such as the boundary correlation functions and the S-matrix. Both finding bounds on the low-energy Wilson coefficients and developing the numerical bootstrap for these obervables depend on those positive restrictions. See refs.\cite{Bellazzini:2020cot,Tolley:2020gtv,Arkani-Hamed:2020blm,Caron-Huot:2020cmc,Sinha:2020win,Chiang:2021ziz,Caron-Huot:2021rmr,Caron-Huot:2021enk,Guerrieri:2021ivu} for further information. The unitarity requirements on the cosmological correlators are examined in this section. In contrast to the previous sections' talks, the derivation of these constraints is valid at a nonperturbative level since it does not rely on perturbation theory. Since the argument is multi-step, let's start with a sneak peek at the result. Unitarity implies positive in both dS and EAdS, but for different quantities: the conformal block expansion is a discrete series that can be created from the boundary correlation functions in EAdS:
\be\label{eq:AdSexpansion}
 \langle \mathcal{O}(x_1)\mathcal{O}(x_2)\mathcal{O}(x_3)\mathcal{O}(x_4)\rangle=\frac{1}{|x_{12}|^{2\Delta_{\mathcal{O}}}|x_{34}|^{2\Delta_{\mathcal{O}}}}\sum_{\Delta^{\prime}}C_{\Delta^{\prime}}\mathcal{Q}_{\Delta^{\prime},J}(z,\bar{z})\,.
 \ee
 This expansion requires the coefficients of the expansion $C_{\Delta^{\prime}}$ to be positive, which is equivalent to enforcing unitarity. To calculate the discrete sum in \eqref{eq:AdSexpansion}, one must add up all possible unitary representations of the Lorentzian conformal group, or AdS isometry group. A total and an integral over all conceivable unitary representations of the Euclidean conformal group, which is the isometry of dS, can be obtained using late-time correlation functions in dS, on the other hand. The final expression would be as follows:
  \bea\label{eq:EuInv}
\langle \mathcal{O}(x_1)\mathcal{O}(x_2)\mathcal{O}(x_3)\mathcal{O}(x_4)\rangle&=&
 \frac{1}{|x_{12}|^{2\Delta_{\mathcal{O}}}|x_{34}|^{2\Delta_{\mathcal{O}}}}\bigg[\sum_{J}\int_{\frac{d}{2}}^{\frac{d}{2}+i\infty} \frac{d\Delta^{\prime}}{2\pi i}\rho_J(\Delta^{\prime})\mathcal{E}_{\Delta^{\prime}}(z,\bar{z})\nonumber\\
 &&~~~~~~~~~~~~~~~~~~~~~~~~~~~~~~~~~~~~~~~~~~~~~~~~+\text{others}\bigg]\,,
\eea
The terms indicated by $\text{others}$- belong to different representations, such as the complementary series or the discrete series, whereas the terms set down below correspond to the principal series representation and may be recognized with the spectral decomposition. The coefficient's positivity then becomes the unitarity constraint, $\rho_J(\Delta^{\prime})\geq 0$. Keep in mind that an expansion such as \eqref{eq:EuInv} is also admitted by the boundary four-point functions in AdS, or more generally, the four-point functions in unitary CFT. The coefficients $\rho_J(\Delta)$ do not, however, have to be positive in that scenario.

\section{Positivity of dS two-point function}\label{ka13}
Consider the two-point function, where $X_1$ and $X_2$ are space-like separated points in dS:
 \be
 \mathcal{G}(X_1,X_2)\equiv \langle \Omega | \phi (X_1)\phi (X_2)|\Omega\rangle.
 \ee
To keep things simple, we assumed that the two operators were identical scalar operators. The Bunch-Davies vacuum, denoted by $|\Omega\rangle$, is invariant under the dS isometry. Writing a resolution of the identification is the initial stage, just like in the preceding part. In this instance, a sum and an integral of all potential unitary irreducible representations of the dS isometry provide the identity's resolution. In this case, we just require the principal series and the complementary series because $\phi$ is a scalar operator:
\be
\bigg(\text{principal}+~\text{complementary}\bigg)={\bf 1}~~~~~~~{\rm where}~~~~\text{principal}\equiv \int_{\frac{d}{2}}^{\frac{d}{2}+i\infty} \frac{d\Delta^{\prime}}{2\pi i} \mathcal{P}_{\Delta^{\prime}}\,\,.
\ee
Substituting this result to the expression for the two-point function, we get the following result:
\be\label{eq:2ptdSdecomp}
\mathcal{G}(X_1,X_2)=\int_{\frac{d}{2}}^{\frac{d}{2}+i\infty} \frac{d\Delta^{\prime}}{2\pi i} \langle \Omega | \phi (X_1)\mathcal{P}_{\Delta^{\prime}}\phi (X_2)|\Omega\rangle +\text{complementary}.
\ee
Now let's comment on this phrase. Along the Schwinger-Keldysh contour, a path integral defines the correlation functions in dS. From the perspective of the path-integral, the projector's insertion in \eqref{eq:2ptdSdecomp} is equivalent to separating the left and right contours of the in-in path integral and adding a whole set of states at the late-time point where the two contours are adhered. The projector can be inserted in the center of two dS spacetimes, each of which has a single operator, to provide a visual representation of this. Next, we apply the symmetry constraints. The dS isometry's quadratic Casimir is acted to $\phi(X_1)$ in order to do this. In other words, taking into account a double commutator:
\be\label{eq:doublecomm}
 \langle \Omega |\mathcal{C}_2\cdot \phi (X_1)\mathcal{P}_{\Delta^{\prime}}\phi (X_2)|\Omega\rangle~~~~{\rm where}~~~~\mathcal{C}_2\cdot \phi (X)\equiv- \frac{1}{2}\sum_{A,B}[\widehat{{\cal J}}^{AB},[\widehat{{\cal J}}_{AB},\phi (X)]].
\ee
Here, $\widehat{{\cal J}}_{AB}$ are the isometry's generators. With the help of the Bunch-Davies vacuum's invariance, which is given by: \be \langle \Omega|\widehat{{\cal J}}_{AB}=0,\ee and the knowledge that $\mathcal{P}_{\Delta^{\prime}}$ projects to an eigenspace of the Casimir, the equation \eqref{eq:doublecomm} may be rewritten as follows:
\be
\langle \Omega|\phi (X_1)\left(\frac{1}{2}\sum_{A,B}\widehat{{\cal J}}^{AB}\widehat{{\cal J}}_{AB}\right)\mathcal{P}_{\Delta^{\prime}}\phi (X_2)|\Omega\rangle=-\Delta^{\prime}(\Delta^{\prime}-d){\cal I}_{\Delta^{\prime}}(X_1,X_2).
\ee
where the newly defined function ${\cal I}_{\Delta^{\prime}}(X_1,X_2)$ is defined by:
\be {\cal I}_{\Delta^{\prime}}(X_1,X_2)=\langle \Omega|\phi (X_1)\mathcal{P}_{\Delta^{\prime}}\phi (X_2)|\Omega\rangle.\ee
Conversely, differential operators acting on $X$ can substitute the action of the dS isometry $\widehat{{\cal J}}_{AB}$ on $\phi (X)$. Consequently, the action of the dS Laplacian corresponds with the double-commutator in \eqref{eq:doublecomm}:
\be
\langle \Omega |\mathcal{C}_2\cdot \phi (X_1)\mathcal{P}_{\Delta^{\prime}}\phi (X_2)|\Omega\rangle=-\Box_{X_1}\langle \Omega |\phi (X_1)\mathcal{P}_{\Delta^{\prime}}\phi (X_2)|\Omega\rangle.
\ee
Hence we have:
\be
\bigg(\Box_{X_1}+\Delta^{\prime}(\Delta^{\prime}-d)\bigg){\cal I}_{\Delta^{\prime}}(X_1,X_2)=0.
\ee
This exactly merges with the equation of motion of the Green's function for dS in the present context of discussion. This further implies the following expression:
\be\label{eq:fandG}
{\cal I}_{\Delta^{\prime}}(X_1,X_2)=\rho(\Delta^{\prime})G_{\Delta^{\prime}}(X_1,X_2).
\ee
Here $G_{\Delta^{\prime}}(X_1,X_2)$ represents the Green's function, which is given by:
\be
G_{\Delta}(X_1,X_2)= \frac{\Gamma (\Delta)\Gamma(d-\Delta)}{(4\pi)^{\frac{d+1}{2}}\Gamma\left(\frac{d+1}{2}\right)}{}_2F_{1}\left(\Delta,d-\Delta,\frac{d+1}{2},\frac{1+{\cal F} (X_1,X_2)}{2}\right),
\ee
where 
\be {\cal F} (X_1,X_2)=1-\frac{\zeta(X_1,X_2)}{2}=\left(\frac{\eta_1^{2}+\eta_2^{2}-x_{12}^2}{2\eta_1\eta_2}\right).\ee
Here $\rho(\Delta^{\prime})$ playing the role of density function, which is a proportionality context in this computation. The decomposition of the two-point function is obtained by combining \eqref{eq:2ptdSdecomp} and \eqref{eq:fandG}:
\be\label{eq:deSitterdecomp2pt}
\mathcal{G}(X_1,X_2)=\int_{\frac{d}{2}}^{\frac{d}{2}+i\infty}\frac{d\Delta^{\prime}}{2\pi i}\rho(\Delta^{\prime})G_{\Delta^{\prime}}(X_1,X_2)+\text{complementary}.
\ee 
In terms of harmonic functions, the expression \eqref{eq:deSitterdecomp2pt} resembles the expansion of the two-point function in AdS. However, because the dS space is a Lorentzian manifold, performing harmonic analysis directly in dS is a difficult operation. Rather, \eqref{eq:deSitterdecomp2pt} can be related to harmonic analysis most easily by performing an analytic continuation after applying harmonic analysis to the sphere. The literature discusses this usual procedure \cite{Marolf:2010zp,Marolf:2010nz}. Deriving \eqref{eq:deSitterdecomp2pt} straight from harmonic analysis in dS is a fascinating and significant problem. The final step is to link $\rho(\Delta)$ to an overlap sum of squares. We do this by inverting the relation \eqref{eq:fandG} and writing:
\be\label{eq:Casratio}
\mathcal{G}(X_1,X_2)=\int_{\frac{d}{2}}^{\frac{d}{2}+i\infty}\frac{d\Delta^{\prime}}{2\pi i}{\cal I}_{\Delta^{\prime}}(X_1,X_2)+\text{complementary}~~~~~{\rm where}~~~~~\rho(\Delta^{\prime})=\frac{{\cal I}_{\Delta^{\prime}}(X_1,X_2)}{G_{\Delta^{\prime}}(X_1,X_2)}.
\ee 
The dependency on $X_{1,2}$ disappears when the ratio on the right side is calculated, and the equality remains true at any $X_{1,2}$. Performing the analytic continuation to a sphere and positioning the two points symmetrically across the equator is particularly convenient. This is a positive sum of squares of overlaps, as it can be understood:
\be\label{eq:denompos}
\left({\cal I}_{\Delta^{\prime}}(X_1,X_2)\right)_{\text{sphere}}=\left(\rho(\Delta^{\prime})G_{\Delta^{\prime}}(X_1,X_2)\right)_{\text{sphere}}=\sum_{\psi\in R_{\Delta^{\prime}}}\left|\langle \Omega|\phi (X)|\psi\rangle\right|^2\geq 0.
\ee
Here we use:
\be
\mathcal{P}_{\Delta^{\prime}}=\sum_{\psi\in R_{\Delta^{\prime}}}|\psi\rangle\langle \psi|.
\ee
The sum over states that belong to the principal series representation that corresponds to $\Delta^{\prime}$ is denoted by $\psi\in R_{\Delta^{\prime}}$. The sum on the right side of \eqref{eq:denompos} should diverge as $X$ gets closer to the sphere's equator since it represents a short distance limit of the two points $X_{1,2}$.
The numerator of \eqref{eq:Casratio}, however, can be increased in the small distance limit as follows:
\be
\lim_{X_1\to X_2}G_{\Delta^{\prime}}(X_1,X_2)\sim \bigg(\frac{\zeta^{\frac{1-d}{2}}}{2\pi^{\frac{d+1}{2}}(d-1)}+\cdots\bigg).
\ee
The chordal distance, $\zeta$, is playing important role here.
It is singular but positive in the limit $\zeta\to 0$, as the expansion shows, because we are approaching the singularity from a space-like direction $\zeta\to 0^{+}$. Next, \eqref{eq:Casratio} can be rewritten as:
\bea
\rho(\Delta^{\prime})&=&\lim_{X_1\to X_2=X}\lim_{X\to\,  \text{equator}}\left[\frac{{\cal I}_{\Delta^{\prime}}(X_1,X_2)}{G_{\Delta^{\prime}}(X_1,X_2)}\right]\nonumber\\
&=&\lim_{X\to\,  \text{equator}}\left[\frac{\displaystyle\sum_{\psi\in R_{\Delta^{\prime}}}\left|\langle \Omega|\phi (X)|\psi\rangle\right|^2}{\displaystyle\left(\frac{\zeta^{\frac{1-d}{2}}}{2\pi^{\frac{d+1}{2}}(d-1)}\right)}\right]\geq 0.
\eea 
The limit is defined here on a sphere, where the distance $\zeta$ vanishes and the two points approach symmetrically to the equator. As was previously stated, this expression is a limit of a positive quantity, therefore even after taking the limit, it stays positive. This proves the point we wanted to make, $\rho(\Delta^{\prime})\geq 0$.

\section{Positivity of dS four-point functions}\label{ka14}
Let us start with the late time limiting expression for the four-point function in the context of dS, which is given by:
\be
\mathcal{G}(x_1,x_2,x_3,x_4)=\langle \Omega | \mathcal{O}_1 (x_1)\mathcal{O}_2 (x_2)\mathcal{O}_3 (x_3)\mathcal{O}_4 (x_4)|\Omega\rangle.
\ee
Every operator is added at a later time $\eta_c\to 0^{-}$.
For simplicity, we assume in the following that the operators are pairwise complex conjugate, \be \mathcal{O}_1=\mathcal{O}_3^{\dagger},~~~~\mathcal{O}_2=\mathcal{O}_4^{\dagger}.\ee Additionally, we suppose that under the dS isometry, they convert into scalar conformal primaries in the complementary series and that $\Delta_j$ represents the dimensions of $\mathcal{O}_j$. This concept may be extended to heavy fields with complex dimensions in the primary series at the leading order in perturbation theory, provided that, \be \Delta_1=(\Delta_3)^{\ast},~~~~\Delta_2=(\Delta_4)^{\ast},\ee are assumed. Nonetheless, there are a number of nuances related to these hefty field insertions. To begin with, a local term proportional to $\delta^d(x-y)$ is present in a two-point function of heavy operators $\mathcal{O}^{\dagger}(x)$ and $\mathcal{O}^{\dagger}(y)$ i.e.
\be \langle \mathcal{O}^{\dagger}(x)\mathcal{O}^{\dagger}(y)\rangle \propto \delta^d(x-y).\ee Furthermore, this term depends on the operator ordering, and it is crucial to include it in order to maintain the positivity property that we will address later. This term is compatible with conformal invariance of correlators for free-theory dimensions since \be \Delta+\Delta^{\ast}=d.\ee All principal series operators, however, acquire anomalous dimensions when interactions are taken into account, meaning that the actual part is always larger than $d/2$. A unitary representation of the dS isometry group is thus even reached by them. However, it is anticipated that local terms will endure in the correlation functions. For these reasons, it is currently unclear how to maintain conformal invariance and unitarity while considering operators that correspond to heavy fields in the external states in an interaction theory. They should presumably be handled similarly to how unstable particles, or resonances, are handled in quantum physics and flat space QFTs. As with the conformal bootstrap, the argument in this section can also be extended to the case of four different operators. Similar to the last conversation, we begin by inserting an identification resolution:
\be
\sum_{J=0}^{\infty}\int^{\frac{d}{2}+i\infty}_{\frac{d}{2}}\frac{d\Delta^{\prime}}{2\pi i }\mathcal{P}_{\Delta^{\prime},J}+\text{rest~contribution}={\bf 1}.
\ee
The decomposition now includes a sum across angular momenta $J$, in contrast to the bulk two-point function covered in the preceding subsection. The terms corresponding to the primary series representation are listed here, and $\text{rest~contributions}$ represents all other unitary representations of the Euclidean conformal group, including the discrete series and complementary series. Consequently, we get a decomposition:
\be\label{eq:4ptdecompdS}
\mathcal{G}(x_1,x_2,x_3,x_4)=\sum_{J=0}^{\infty}\int_{\frac{d}{2}}^{\frac{d}{2}+i\infty}\frac{d\Delta^{\prime}}{2\pi i}\langle \Omega | \mathcal{O}_1 (x_1)\mathcal{O}_2 (x_2)\mathcal{P}_{\Delta^{\prime},J}\mathcal{O}_3 (x_3)\mathcal{O}_4 (x_4)|\Omega\rangle+\text{rest~contribution}.
\ee
The limitations resulting from symmetry are then imposed. We once more examine how the operators $\mathcal{O}_1$ and $\mathcal{O}_2$ are affected by the quadratic Casimir of the dS isometry:
\be
\langle \Omega | \left\{\mathcal{C}_2\cdot \left(\mathcal{O}_1 (x_1)\mathcal{O}_2 (x_2)\right)\right\}\mathcal{P}_{\Delta^{\prime},J}\mathcal{O}_3 (x_3)\mathcal{O}_4 (x_4)|\Omega\rangle.
\ee
In the present context of the discussion the action of the Casimir is given by:
\be
\left\{\mathcal{C}_2\cdot \left(\mathcal{O}_1 (x_1)\mathcal{O}_2 (x_2)\right)\right\}\equiv -\frac{1}{2}\sum_{A,B}[\widehat{J}^{AB},[\widehat{J}_{AB},\mathcal{O}_1(x_1)\mathcal{O}_2(x_2)]].
\ee
The action of the Casimir on $\mathcal{P}_{\Delta^{\prime},J}$ can be reformulated using the invariance of $\langle \Omega |$ under $\widehat{J}_{AB}$. Hence we get:
\be\label{eq:casimireigen4pt}
\begin{aligned}
&\langle \Omega | \mathcal{C}_2\cdot \left(\mathcal{O}_1 (x_1)\mathcal{O}_2 (x_2)\right)\mathcal{P}_{\Delta^{\prime},J}\mathcal{O}_3 (x_3)\mathcal{O}_4 (x_4)|\Omega\rangle= c^{(\Delta^{\prime},J)}_2 g_{\Delta^{\prime},J}(x_1,x_2,x_3,x_4),
\end{aligned}
\ee
where we define:
\be g_{\Delta^{\prime},J}(x_1,x_2,x_3,x_4)=\langle \Omega | \mathcal{O}_1 (x_1)\mathcal{O}_2 (x_2)\mathcal{P}_{\Delta^{\prime},J}\mathcal{O}_3 (x_3)\mathcal{O}_4 (x_4)|\Omega\rangle.\ee
In this case, the Casimir eigenvalue of \eqref{eq:eigenc} is $c_2^{\Delta^{\prime},J}$, which is provided by:
\be c_2^{(\Delta^{\prime},J)}=\bigg(\Delta^{\prime} (\Delta^{\prime}-d)+J(J+d-2)\bigg).\ee
An alternative way to express the Casimir's action is as a differential operator acting on $x_{1,2}$. To illustrate this, we write $\mathcal{O}(x)$ as follows:
\be
\mathcal{O} (x)=\exp(i\hat{P}\cdot x)\mathcal{O}(0)\exp(-i\hat{P}\cdot x).
\ee
In this case, the translation operator is $\hat{P}_{\mu}$. By computing $\widehat{J}_{AB}$ using $\exp(i\hat{P}\cdot x)$ and utilizing the fact that $\mathcal{O}$ is a conformal primary—that is, it is annihilated by the special conformal transformation—we can then recast the action of $\widehat{J}_{AB}$. We turn to standard studies \cite{Simmons-Duffin:2016gjk} for this standard manipulation in order to get the Casimir differential equation for the conformal field theory in higher dimensions. Consequently, we get:
\be
\begin{aligned}
&\langle \Omega | \left\{\mathcal{C}_2\cdot \left(\mathcal{O}_1 (x_1)\mathcal{O}_2 (x_2)\right)\right\}\mathcal{P}_{\Delta^{\prime},J}\mathcal{O}_3 (x_3)\mathcal{O}_4 (x_4)|\Omega\rangle= \mathcal{D}_{12}\cdot g_{\Delta^{\prime},J}(x_1,x_2,x_3,x_4).
\end{aligned}
\ee
The conformal Casimir is represented differentially in this case by $\mathcal{D}_{12}$. Using \eqref{eq:casimireigen4pt} to calculate this, we see that $g_{\Delta^{\prime}, J}$ meets the equation that follows:
\be\label{eq:4ptdifferential}
\mathcal{D}_{12}\cdot g_{\Delta^{\prime},J}(x_1,x_2,x_3,x_4)=c_2^{(\Delta^{\prime},J)}g_{\Delta^{\prime},J}(x_1,x_2,x_3,x_4).
\ee
The conformal partial wave \eqref{eq:difeqcpw} satisfies the differential equation in this way. The equation \eqref{eq:4ptdifferential} admits several solutions because it is a second-order differential equation involving two independent variables. We must select the solution with the correct attributes from those available options. It needs to be single-valued in $x_j$ on the late-time surface and exhibit the proper asymptotic behavior in the coincident point limit, which is $x_1\to x_2$ and $x_3\to x_4$. The conformal partial wave $\widehat{\mathcal{E}}^{\{\Delta_j\}}_{\Delta^{\prime},J}$ is exactly that. Thus, we deduce that $\widehat{\mathcal{E}}$ must be proportional to the function $g_{\Delta^{\prime},J}(x_1,x_2,x_3,x_4)$:
\be\label{eq:spectraldef4pt}
g_{\Delta^{\prime},J}(x_1,x_2,x_3,x_4)=\rho_{J}(\Delta^{\prime})\widehat{\mathcal{E}}^{\{\Delta_j\}}_{\Delta^{\prime},J}(x_1,x_2,x_3,x_4).
\ee
This calculation concludes by relating $\rho_{J}(\Delta^{\prime})$ to a positive sum of norms.
To do this, we first break down the left side of \eqref{eq:spectraldef4pt} into the translation eigenstates:
\be
g_{\Delta^{\prime},J}(x_1,x_2,x_3,x_4)=\int \frac{d^{d}p}{(2\pi)^{d}}\langle \Omega | \mathcal{O}_1 (x_1)\mathcal{O}_2 (x_2)\mathcal{P}_{\Delta^{\prime},J}^{(p)}\mathcal{O}_3 (x_3)\mathcal{O}_4 (x_4)|\Omega\rangle.
\ee
A projection to the translation eigenstates with momentum $p^{\mu}$ inside the space invariant under $\mathcal{P}_{\Delta^{\prime},J}$ is shown here by $\mathcal{P}^{(p)}_{\Delta^{\prime},J}$. Next, we examine a symmetric configuration where: \be x_3=x_1+y,~~~~x_4=x_2+y.\ee Using the following formula: \be \mathcal{O}_3(x_3)\mathcal{O}_4(x_4)=\exp(i\hat{P}\cdot y)\mathcal{O}_3(x_1)\mathcal{O}_4(x_2)\exp(-i\hat{P}\cdot y),\ee
we can find the following:
\be
\left.g_{\Delta^{\prime},J}\right|_{\rm symmetric}=\int \frac{d^{d}p}{(2\pi)^{d}}~\exp(ip\cdot y)~\langle \Omega | \mathcal{O}_1 (x_1)\mathcal{O}_2 (x_2)\mathcal{P}_{\Delta^{\prime},J}^{(p)}\mathcal{O}_3 (x_1)\mathcal{O}_4 (x_2)|\Omega\rangle.
\ee
The inverse Fourier transformation yields the following results:
\be
\int d^{d}y \, e^{-i p y}\left.g_{\Delta^{\prime},J}\right|_{\rm symmetric}=\langle \Omega | \mathcal{O}_1 (x_1)\mathcal{O}_2 (x_2)\mathcal{P}_{\Delta^{\prime},J}^{(p)}\mathcal{O}_3 (x_1)\mathcal{O}_4 (x_2)|\Omega\rangle.
\ee
The right-hand side can be written as a sum of squares of overlaps and is hence positive when the operators are pair-wise conjugate:
\be\label{eq:gfourier}
\int d^{d}y \, \exp(-i p y)\left.g_{\Delta^{\prime},J}\right|_{\rm symmetric}=\sum_{\substack{\psi\in R_{\Delta^{\prime},J},p_{\psi}=p}}\left|\langle \Omega | \mathcal{O}_1 (x_1)\mathcal{O}_2 (x_2)|\psi\rangle\right|^2\geq 0.
\ee
The same inverse Fourier transformation is then applied to $\widehat{\mathcal{E}}_{\Delta^{\prime},J}^{\{ \Delta_j\}}$. The integral representation of the conformal partial wave, which is provided by:
\be\label{eq:doubleintpartial}
\begin{aligned}
\widehat{\mathcal{E}}_{\Delta^{\prime},J}^{\{ \Delta_j\}}=&\int \frac{d^{d}p}{(2\pi)^{d}}\int \frac{d^{d}x_5\, }{|x_{12}|^{\Delta_1+\Delta_2-\Delta^{\prime}}|x_{15}|^{\Delta_1+\Delta^{\prime}-\Delta_2}|x_{25}|^{\Delta_2+\Delta^{\prime}-\Delta_1}}~\exp(-i p \cdot x_5)\\
&\times \frac{d^{d}x_6\, }{|x_{34}|^{\Delta_3+\Delta_4-(d-\Delta^{\prime})}|x_{36}|^{\Delta_3+(d-\Delta^{\prime})-\Delta_4}|x_{46}|^{\Delta_4+(d-\Delta^{\prime})-\Delta_3}}\exp(ip \cdot x_6)\times \hat{C}_{J}(\tilde{T}),
\end{aligned}
\ee
where $\tilde{T}$ is represented by:  
\be
\tilde{T}\equiv \frac{|x_{15}||x_{25}|}{|x_{12}|}\frac{|x_{36}||x_{46}|}{|x_{34}|}\left(\frac{{\bf x}_{15}}{x_{15}^2}-\frac{{\bf x}_{25}}{x_{25}^2}\right)\cdot \left(\frac{{\bf x}_{36}}{x_{36}^2}-\frac{{\bf x}_{46}}{x_{46}^2}\right).
\ee
Here  $\hat{C}_J(\eta)$ represents a Gegenbauer polynomial in the present context of discussion. Further positivity  at $\eta=1$ demmands $\hat{C}_{J}(1)\geq 0$. This may now be rebuilt in the symmetric configuration by changing the integration variable $x_6$ to $x_6+y$. Consequently, we get:
\be
\begin{aligned}
\left.\widehat{\mathcal{E}}_{\Delta^{\prime},J}^{\{ \Delta_j\}}\right|_{\rm symmetric}=&\int \frac{d^{d}p\, }{(2\pi)^{d}}\exp(i p \cdot y)~\int \frac{d^{d}x_5\, }{|x_{12}|^{\Delta_1+\Delta_2-\Delta^{\prime}}|x_{15}|^{\Delta_1+\Delta^{\prime}-\Delta_2}|x_{25}|^{\Delta_2+\Delta^{\prime}-\Delta_1}}\exp(-i p \cdot x_5)\\
&\times \frac{d^{d}x_6\,}{|x_{14}|^{\Delta_1^{\ast}+\Delta_2^{\ast}-(d-\Delta^{\prime})}|x_{16}|^{\Delta_1^{\ast}+(d-\Delta^{\prime})-\Delta_2^{\ast}}|x_{26}|^{\Delta_2^{\ast}+(d-\Delta^{\prime})-\Delta_1^{\ast}}} \exp(ip \cdot x_6)\times \hat{C}_{J}(\tilde{T}_{\rm symmetric}),
\end{aligned}
\ee
where we define:
\be
\tilde{T}_{\rm symmetric}\equiv \frac{|x_{15}||x_{25}|}{|x_{12}|}\frac{|x_{16}||x_{26}|}{|x_{12}|}\left(\frac{{\bf x}_{15}}{x_{15}^2}-\frac{{\bf x}_{25}}{x_{25}^2}\right)\cdot \left(\frac{{\bf x}_{16}}{x_{16}^2}-\frac{{\bf x}_{26}}{x_{26}^2}\right).
\ee  
After using the inverse Fourier transform, we obtain:
\be
\begin{aligned}
\int d^{d}y\, \exp(-i p y)\left.\widehat{\mathcal{E}}_{\Delta^{\prime},J}^{\{ \Delta_j\}}\right|_{\rm symmetric}=&\int \frac{d^{d}x_5\, }{|x_{12}|^{\Delta_1+\Delta_2-\Delta^{\prime}}|x_{15}|^{\Delta_1+\Delta^{\prime}-\Delta_2}|x_{25}|^{\Delta_2+\Delta^{\prime}-\Delta_1}}\exp(-i p \cdot x_5)\\
&\times \frac{d^{d}x_6\, \exp(ip \cdot x_6)}{|x_{14}|^{\Delta_1^{\ast}+\Delta_2^{\ast}-(d-\Delta^{\prime})}|x_{16}|^{\Delta_1^{\ast}+(d-\Delta^{\prime})-\Delta_2^{\ast}}|x_{26}|^{\Delta_2^{\ast}+(d-\Delta^{\prime})-\Delta_1^{\ast}}}\\
&\times \hat{C}_{J}(\tilde{T}_{\rm symmetric}).
\end{aligned}
\ee
Nevertheless, the quickly oscillating factor $\exp(i p\cdot (x_5-x_6))$ enforces $x_5\sim x_6$ and streamlines the calculation in the limit $|p|\gg 1$. Specifically, the leading behavior may be calculated by setting $x_5=x_6$ in $\hat{C}_J(\tilde{T}_{\rm symmetric})$, which is equivalent to setting $\tilde{T}_{\rm symmetric}=1$. Consequently, we discover in the limit:
\be\label{eq:Ffourier}
\begin{aligned}
&\lim_{|p|\gg 1}\int d^{d}y\, \exp(-i p y)\left.\widehat{\mathcal{E}}_{\Delta^{\prime},J}^{\{ \Delta_j\}}\right|_{\rm symmetric}\geq 0~~~~{\rm where}~~\Delta^{\prime}=(d-\Delta^{\prime})^{\ast}=\left(\frac{d}{2}+i\nu\right).
\end{aligned}
\ee
In light of this and the fact that $g_{\Delta^{\prime},J}$ is positive, we deduce that the proportionality constant $\rho_J(\Delta^{\prime})$ must likewise be positive i.e. $\rho_J(\Delta^{\prime})\geq 0$.

\section{Exchange diagram computation in EAdS}\label{ka16}
A theory with a cubic interaction and two scalar fields, $\phi$ and $\sigma$, will be our main focus:
\be
{\cal{L}(\phi,\sigma)}=-\frac{1}{2}(\partial\phi)^2-\frac{m_\phi^2}{2}\phi^2-\frac{1}{2}(\partial\sigma)^2-\frac{m_\sigma^2}{2}\sigma^2+\frac{\lambda}{2}\phi^2\sigma\,.
\ee
Here $\phi$ signifies the light sector, on the other hand, $\sigma$ characterizes the light/visible sector respectively. Let's quickly go over how the exchange diagram was calculated: 
\bea\label{eq:exchAdS}
& &\left\langle\phi(P_1)\phi(P_2)\phi(P_3)\phi(P_4)\right\rangle\supset\int dX_1 dX_2 \nonumber\\  &&~~~~~~~~~~~~~~~~~~~~~~~~~~~~~~~~~~~~~~{\cal Y}_{\nu_\phi}^{\text{AdS}}(X_1,P_1) {\cal Y}_{\nu_\phi}^{\text{AdS}}(X_1,P_2) \nonumber\\  &&~~~~~~~~~~~~~~~~~~~~~~~~~~~~~~~~~~~~~~~{\cal Y}_{\nu_\phi}^{\text{AdS}}(X_2,P_3){\cal Y}_{\nu_\phi}^{\text{AdS}}(X_2,P_4) G^{\rm AdS}_{\nu_\sigma} (X_1,X_2)\,,
\eea
where the mass parameters $\nu_\phi$ and $\nu_\sigma$ for the $\phi$ and $\sigma$ field in EAdS are represented by the following expressions:
\bea \nu_\phi &=& \sqrt{-m^2_\phi-\frac{d^2}{4}},\\
\nu_\sigma &=& \sqrt{-m^2_\sigma-\frac{d^2}{4}}.\eea
Here, additionally, it is important to note that in the present context of the discussion only the normal boundary conditions are used. If we were to utilize momentum representation for the boundary directions, it would be difficult to evaluate the integrals over the bulk points directly or only the integrals over the radial coordinates $z$. The split form of the harmonic function \cite{Moschella:2007zza, Cornalba:2008qf, Penedones:2010ue} is a simplification technique:
\be
\Omega^{\rm AdS}_\nu({\cal F}^{\rm AdS})=\frac{\nu^2 }{\pi}\sqrt{\left(\frac{\Gamma(\frac{d}{2} + i\nu)}{2 \pi^{d/2} \Gamma(1+i\nu)}\right)}\sqrt{\left(\frac{\Gamma(\frac{d}{2} - i\nu)}{2 \pi^{d/2} \Gamma(1-i\nu)}\right)}\int dP~ {\cal Y}^{\rm AdS}_\nu(X_1,P){\cal Y}^{\rm AdS}_{-\nu}(X_2,P)\,.
\ee
By doing $d X_1$ and $d X_2$ first, followed by the $dP$ integral, the diagram may now be explicitly evaluated, resulting in \cite{Penedones:2010ue, Costa:2014kfa, Carmi:2018qzm}:
\begin{align}
\begin{split}
\label{AdStree}
\left\langle\phi(P_1)\phi(P_2)\phi(P_3)\phi(P_4)\right\rangle & \approx \frac{1}{32 \pi^d x_{12}^{d+2i\nu_\phi} x_{34}^{d+2i\nu_\phi}}
\int_{-\infty}^{+\infty} \frac{d\nu'}{2\pi} \\
&~~~~~~~~~~\left(\frac{1}{\nu'^2-\nu_\sigma^2}\right)\frac{\G^2\left(\frac{d+4i\nu_\phi\pm2i\nu'}{4}\right)\G^2\left(\frac{d\pm2i\nu'}{4}\right)}{\G^2\left(\frac{d}{2}+i\nu_\phi\right)\G^2\left(1+i\nu_\phi\right)\G\left(\pm i\nu'\right)}{\cal E^{\{\nu_\phi\}}_{\nu'}}(z,\bar z).
\end{split}
\end{align}

\section{Computation of four point function at the tree level}\label{ka17}

In order to contribute to the four-point function $\left\langle\phi(x_1,\eta_c)\phi(x_2,\eta_c)\phi(x_3,\eta_c)\phi(x_4,\eta_c)\right\rangle$, we first compute the tree-level s-channel exchange diagram. Assume for the time being that $\sigma$ is also a light field, $m_\sigma<\frac{d}{2}$, or $i\nu_\sigma>0$. Since each of the two interaction vertices might originate from either the left or the right portion of the contour, there are a total of four different diagrams. Clearly, summing them all at the tree level we have:
\bea
\label{dSDiagrams}
  {\cal A}^{s,{\rm tree}}_{\nu_\sigma}
 &=&\lambda^2\left(-\eta_c\right)^{4\left(\frac{d}{2}-i\nu_\phi\right)}\times\left(\frac{\displaystyle \frac{1}{4 \pi^{\frac{d}{2}+1}} \G\left(\frac{d}{2}-i\nu_\phi\right)\G(i \nu_\phi)}{\displaystyle\sqrt{\left(\frac{\Gamma(\frac{d}{2} - i\nu_\phi)}{2 \pi^{d/2} \Gamma(1-i\nu_\phi)}\right)}}\right)^4\nonumber\\
 &&~~~~~~~~~~~~\times\Bigg[\G\left(\pm i\nu_\sigma\right) 4 \sin^2\frac{\pi}{2}\left(\frac{d}{2}-i\nu_\sigma-2 i \nu_\phi\right){\cal M}_\Omega \nonumber\\ &&~~~~~~~~~~~~~~~~~~~~~~~~~~~~~~~ -2 \sin \pi \left(\frac{d}{2}-2i \nu_\phi\right){\cal M}_G\Bigg]\,,
\eea
where ${\cal M}_\Omega$ and ${\cal M}_G$ are given by the following expressions:
\bea
 {\cal M}_\Omega &=&\int dX_1 dX_2 {\cal Y}_{-\nu_\phi}^{\text{AdS}}(X_1,P_1) {\cal Y}_{-\nu_\phi}^{\text{AdS}}(X_1,P_2) {\cal Y}_{-\nu_\phi}^{\text{AdS}}(X_2,P_3){\cal Y}_{-\nu_\phi}^{\text{AdS}}(X_2,P_4) \Omega^{\rm AdS}_{\nu_\sigma} (X_1,X_2)\,\\
 &=&\frac{1}{x_{12}^{d-2i\nu_\phi} x_{34}^{d-2i\nu_\phi}}
\frac{\G^2\left(\frac{d-4i\nu_\phi\pm2i\nu_\sigma}{4}\right)\G^2\left(\frac{d\pm2i\nu_\sigma}{4}\right)}{64 \pi^{d+1}\G^2\left(\frac{d}{2}-i\nu_\phi\right)\G^2\left(1-i\nu_\phi\right)\G\left(\pm i\nu_\sigma\right)}{\cal E^{\{-\nu_\phi\}}_{\nu_\sigma}}(z,\bar z)~~~~~~~~~~~~~~~~~~~~~~,\nonumber\eea
\bea {\cal M}_G &=&\int dX_1 dX_2 {\cal Y}_{-\nu_\phi}^{\text{AdS}}(X_1,P_1) {\cal Y}_{-\nu_\phi}^{\text{AdS}}(X_1,P_2) {\cal Y}_{-\nu_\phi}^{\text{AdS}}(X_2,P_3){\cal Y}_{-\nu_\phi}^{\text{AdS}}(X_2,P_4) G^{\rm AdS}_{\nu_\sigma} (X_1,X_2)\,\\
&=&\frac{1}{x_{12}^{d-2i\nu_\phi} x_{34}^{d-2i\nu_\phi}}
\int \frac{d\nu'}{2\pi}\left(\frac{1}{\nu'^2-\nu_\sigma^2}\right)\frac{\G^2\left(\frac{d-4i\nu_\phi\pm2i\nu'}{4}\right)\G^2\left(\frac{d\pm2i\nu'}{4}\right)}{32 \pi^d\G^2\left(\frac{d}{2}-i\nu_\phi\right)\G^2\left(1-i\nu_\phi\right)\G\left(\pm i\nu'\right)}{\cal E^{\{-\nu_\phi\}}_{\nu'}}(z,\bar z)~.\nonumber
\eea
After further simplification at the tree-level $s$-channel we get the following result:
\bea 
  {\cal A}^{s,{\rm tree}}_{\nu_\sigma}
 &=&\lambda^2\left( \frac{\eta_c}{x_{12}}\right)^{d-2i\nu_\phi} \left(\frac{\eta_c}{x_{34}}\right)^{d-2i\nu_\phi}\frac{\G^4(i\nu_\phi)}{2^{11}\pi^{2d+5}}\nonumber \\
&&\nonumber \times\Bigg[2 \sin^2\frac{\pi}{2}\left(\frac{d}{2}-i\nu_\sigma-2 i \nu_\phi\right)\G^2\left(\frac{d-4i\nu_\phi\pm2i\nu_\sigma}{4}\right)\G^2\left(\frac{d\pm2i\nu_\sigma}{4}\right)   {\cal E^{\{-\nu_\phi\}}_{\nu_\sigma}}(z,\bar z)  \\
&&- \sin \pi \left(\frac{d}{2}-2i \nu_\phi\right) \int^{+\infty}_{-\infty} \frac{d\nu'}{\nu'^2-\nu_\sigma^2}\frac{\G^2\left(\frac{d-4i\nu_\phi\pm2i\nu'}{4}\right)\G^2\left(\tfrac{d\pm2i\nu'}{4}\right)}{\G\left(\pm i\nu'\right)}{\cal E^{\{-\nu_\phi\}}_{\nu'}}(z,\bar z)\Bigg]~.~~~~~~~~~~~ \eea

\section{Resummation technique}\label{ka18}

Resuming the bubble diagrams is necessary to acquire the full propagator of the $\sigma$ particle, which includes real values corresponding to a heavy $\sigma$ particle, as well as a physically reasonable solution for all values of $\nu_\sigma$. We take advantage of the fact that the spectral representation transfers convolutions to products to resum the propagator of $\sigma$. The link to the EAdS spectral representation or the analytic continuation from the sphere can be used to obtain this characteristic. Regardless, we are left with the simple task of adding up a geometric series that involves the products of the bubble diagram and the propagator's spectral representation:
\bea
\label{sigma2pt}
\left\langle\sigma(X^{\alpha_1})\sigma(Y^{\alpha_2})\right\rangle =\int_{-\infty-i\left(\frac{d}{2} -\eps\right)}^{+\infty-i\left(\frac{d}{2} -\eps\right)} d \nu \,\frac{\nu}{\pi i} \underbrace{\left(\frac{1}{\displaystyle \nu^2-\nu_\sigma^2 - \frac{\lambda^2}{2} \hat{C}^{(d)}_{\nu_\phi}(\nu)}\right)}_{\bf Resummed~propagator}~G_\nu^{\alpha_1\alpha_2}({\cal F})\,
\eea
where the bubble function $\hat{C}^{(d)}_{\nu_\phi}(\nu)$ for $d=2$ is described by the following expression:
\be
\hat{C}^{(2)}_{\nu_\phi}(\nu)={\frac{i}{8 \pi \nu}}\left[\pi+\cot{\pi i \nu_\phi}\left\{\psi\left(\frac{1}{2}+\frac{i\nu}{2}-i\nu_\phi\right)-\psi\left(\frac{1}{2}+\frac{i\nu}{2}+i\nu_\phi\right)\right\}\right]\,.
\ee
In a theory of a free scalar $\phi$, the dS spectral density for the two-point function of the operator $\phi^2$ is interpreted by the function $\hat{C}^{(d)}_{\nu_\phi}(\nu)$. For real positive $\nu$, observe that the odd, or equivalently, the imaginary part of $\hat{C}^{(d)}_{\nu_\phi}(\nu)$ is positive because of unitarity, which requires, $-i(\hat{C}^{(d)}_{\nu_\phi}(\nu)-\hat{C}^{(d)}_{\nu_\phi}(-\nu))>0$. With the possible exception of the interval $i\nu\in(0,\frac{d}{2})$, the bubble function is holomorphic in the lower half of the $\nu$ plane in this case. In the upper half-plane, it has poles at the locations of double-trace operators in the mean field theory of the operator associated with the free field $\phi$. These three pole families are as follows:
\bea
&&\nu_{n++}=2\nu_\phi + i \left(\frac{d}{2}+2n\right)~,\\
&& \nu_{n+-}=i \left(\frac{d}{2}+2n\right)~,\\
&& \nu_{n--}=-2\nu_\phi + i \left(\frac{d}{2}+2n\right)~~~{\rm where}~~~~n=\mathbb{Z}_{\geq 0}
\eea
The bubble function gives the resummed propagator an intriguing analytical structure. With the possible exception of poles for $i\nu\in(0,\frac{d}{2})$, it remains holomorphic on the lower-half $\nu$ plane, since it is itself the spectral density of a two-point function. We decided to convey the contour underneath each of the resummed propagator's potential poles. Alternatively, we may select the genuine $\nu$ contour and explicitly incorporate the contribution from every residue of the resummed propagator for $i\nu\in(0,\frac{d}{2})$. If $\sigma$ and $\phi$ are light enough, they can correspond to either single-$\sigma$-particle poles or two-$\phi$-particle poles. We do not take into account particles that are exactly massless, thus it is sufficient to assume that $\eps$ is a tiny positive integer. The resummation shifts the poles at $\nu =\pm \nu_\sigma$ from the real $\nu$ axis to the upper-half plane, which is noteworthy when $\sigma$ is a heavy particle. Currently, we swap a composite object defined by the spectral density for a single particle with weight $\nu_\sigma$ instead:
\bea
{\cal Z}^{s,{\rm loop}}&=&\int_{-\infty-i\left(\frac{d}{2}-\epsilon\right)}^{+\infty-i\left(\frac{d}{2} -\epsilon\right)}d\nu~\frac{\nu}{\pi i}~\underbrace{\left(\frac{1}{\displaystyle \nu^2-\nu_\sigma^2 - \frac{g^2}{2} \hat{C}^{(d)}_{\nu_\phi}(\nu)}\right)}_{\bf Resummed~propagator}~ {\cal Z}^{s,{\rm tree}}_{\nu}\,\nonumber\\
&=& \lambda^2\left( \frac{\eta_c}{x_{12}}\right)^{d-2i\nu_\phi} \left(\frac{\eta_c}{x_{34}}\right)^{d-2i\nu_\phi} \frac{\G^4(i\nu_\phi)}{2^{10}\pi^{2d+5}}\nonumber\\
&&~~~~~~~~\times\int_{-\infty-i\left(\frac{d}{2}-\epsilon\right)}^{+\infty-i\left(\frac{d}{2} -\epsilon\right)} d \nu \, \sin^2\frac{\pi}{2}\left(\frac{d}{2}-i\nu-2 i \nu_\phi\right)\nonumber\\
&&~~~~~~~~~~~~~~~~~~~~~~~~\times\G^2\left(\frac{d-4i\nu_\phi\pm2i\nu}{4}\right)\G^2\left(\frac{d\pm2i\nu}{4}\right)\nonumber\\
&&~~~~~~~~~~~~~~~~~~~~~~~~\times \frac{\nu}{\pi i}~\underbrace{\left(\frac{1}{\displaystyle \nu^2-\nu_\sigma^2 - \frac{g^2}{2} \hat{C}^{(d)}_{\nu_\phi}(\nu)}\right)}_{\bf Resummed~propagator}~ {\cal E^{\{-\nu_\phi\}}_\nu}(z,\bar z).
\eea
\section{Resonanace in four-point function}\label{ka19}

The four-point function of four identical operators that correspond to light external fields must have a positive spectral density, meaning that $\nu_\phi$ must be imaginary. Let's see if our one-loop outcome actually demonstrates this. Remember that the symmetrization of the four-point resummed contribution under $\nu\to-\nu$ is the spectral density with the positivity property. We therefore want to demonstrate the positivity of the following quantity:
\begin{align}\label{eq:resrho}
\begin{split}
\rho(\nu)  
& = \rho_{\rm f}(\nu) +\frac{\lambda^2}{2} \left[\frac{{\cal I}(\nu)}{\displaystyle\nu^2 -\nu_\sigma^2 -\frac{\lambda^2}{2} \hat{C}^{(d)}_{\nu_\phi}(\nu)}+ \frac{{\cal I}(-\nu)}{\displaystyle\nu^2 -\nu_\sigma^2 -\frac{\lambda^2}{2} \hat{C}^{(d)}_{\nu_\phi}(-\nu)}\right]~.
\end{split}
\end{align}
Here $\rho_{\rm f}(\nu)$ and ${\cal I}(\nu)$ are given by the following expressions:
\bea \rho_{\rm f}(\nu)&=&\frac{\G\left(\frac{d}{2}\right) \G^4\left(i\nu_\phi\right)}{64 \pi^{\frac{3 d}{2}+3}}
\frac{\G\left(\frac{d}{2}\pm i\nu\right)\G\left(\frac{d-4i\nu_\phi\pm2i\nu}{4}\right)}{\G(\pm i\nu)\G\left(\frac{d+4i\nu_\phi\pm2i\nu}{4}\right)}\,,\\
{\cal I}(\nu)&=&\frac{\G^4(i\nu_\phi)}{2^{10}\pi^{2d+5}}\sin^2\frac{\pi}{2}\left(\frac{d}{2}-i\nu-2 i \nu_\phi\right)\G^2\left(\frac{d-4i\nu_\phi\pm2i\nu}{4}\right)\G^2\left(\frac{d\pm2i\nu}{4}\right) \frac{\nu}{\pi i}\,.~~~~~~\eea
Since the correction is subleading with regard to the positive free field contribution, this condition is clearly satisfied away from the resonance at, $\nu =\pm \nu_\sigma$. On the other hand, the resonant contributions for real $\nu_\sigma$ and $\nu\approx \pm \sigma$ are order one in terms of $\lambda^2$ counting and are not sign definite by themselves. As a result, the positive check is no longer simple. The real and imaginary parts of $\hat{C}^{(d)}_{\nu_\phi}(\nu)$ are first separated, and for real $\nu$, they exhibit the parity properties, 
${\rm Re}~ \left[\hat{C}^{(d)}_{\nu_\phi}(\nu)\right]  = {\rm Re}~ \left[\hat{C}^{(d)}_{\nu_\phi}(-\nu)\right]~,$ and $
{\rm Im}~ \left[\hat{C}^{(d)}_{\nu_\phi}(\nu)\right]  = -{\rm Im}~ \left[\hat{C}^{(d)}_{\nu_\phi}(-\nu)\right]~$~\footnote{We observe that the real part of $C^{(d)}_{\nu_\phi}(\nu)$ close to the resonance equates to order $\lambda^2$ renormalization of the $\sigma$ particle's mass; that is, the poles at $\pm \nu_\sigma$ migrate to,$\pm\left(\nu_\sigma + \frac{\lambda^2}{2}{\rm Re}~ \left[\hat{C}^{(d)}_{\nu_\phi}(\nu_\sigma)\right]\right)$.}. Next, we zoom to the vicinity of the resonance by plugging $\nu = \nu_\sigma + \frac{\lambda^2}{2}x$ and keeping only the leading term in \eqref{eq:resrho} for tiny $\lambda^2$. We acquire the following result:
\begin{align}
\begin{split}
\label{unit1}
\rho(\nu)=\rho_{\rm f}(\nu_\sigma)& \left[1+\frac{ 1}{32 \pi^{\frac{d}{2}+3}\G\left(\frac{d}{2}\right)}
\frac{\G(\pm i\nu_\sigma)\G\left(\frac{d\pm4i\nu_\phi\pm2i\nu_\sigma}{4}\right)\G^2\left(\frac{d\pm2i\nu_\sigma}{4}\right)}{\G\left(\frac{d}{2}\pm i\nu_\sigma\right)} \right.\\
& \times \left.\left(\left(1-\cos\pi\left(\frac{d}{2}-2 i \nu_\phi\right)\cosh\pi\nu_\sigma  \right)\frac{\Xi}{x^2 + \Xi^2} \right.\right.\\ & \left.\left. \phantom{\frac{\G(\pm i\nu_\sigma)}{\G\left(\frac{d}{2}\pm i\nu_\sigma\right)}}-  \sin\pi\left(\frac{d}{2}-2 i \nu_\phi\right)\sinh \pi \nu_\sigma \frac{x}{x^2 + \Xi^2}\right )\right] +O(\lambda^2)~,
\end{split}
\end{align}
where the expression for $\Xi$ is given by the following equation~\footnote{Here it is important to note that the width of the resonance is given by $\lambda^2\Xi$ in the present context of the discussion.}:
\bea
\Xi&=&\frac{1}{2 \nu_\sigma}\text{Im}~\left[\hat{C}^{(d)}_{\nu_\phi}(\nu_\sigma)\right]\nonumber\\
&=& \frac{  \Gamma\left(\frac{d}{4} \pm i \frac{\nu_\sigma}{2}\right)^2  \Gamma\left(\frac{d}{4} \pm i \frac{\nu_\sigma -2 \nu_\phi}{2}\right)  \Gamma\left(\frac{d}{4} \pm i \frac{\nu_\sigma +2 \nu_\phi}{2}\right) }{64 \pi ^{\frac{d}{2}+1}\Gamma\left(\frac{d}{2}\right)\,\nu_\sigma^2\, \G\l(\pm i \nu_\sigma\r) \Gamma\left(\frac{d}{2}\pm i \nu_\sigma\right)}\,.
\eea
Let's factor out $\Xi$ in the numerator and denominator and declare: \bea && \tilde x= \frac{x}{\Xi},\\
&& \alpha_\phi=\pi\left(\frac{d}{2}-2 i \nu_\phi\right)\eea to further simplify the expression. Following these adjustments, the majority of the long overall components cancel out, and we ultimately obtain:
\be
\label{ressimple}
\rho(\nu)=\rho_{\rm f}(\nu_\sigma)\left[1+2\left(\frac{1-\cosh\pi\nu_\sigma\cos\alpha_\phi}{\sinh^2\pi\nu_\sigma}\frac{1}{\tilde x^2 + 1}-\frac{\sin\alpha_\phi}{\sinh\pi\nu_\sigma}\frac{\tilde x}{\tilde x^2 + 1}\right)\right]\,.
\ee
Depending on the sign of $\alpha_\phi$, this expression approaches the positive value $\rho_{\rm f}(\nu_\sigma)$ for $\tilde{x}\to-\infty$ from above or below, and it approaches the same value for $\tilde{x}\to+\infty$ from the opposite direction. Additionally, it is extremized at positions $\tilde{x}_1$ and $\tilde{x}_2$, which are represented by the following expressions:
\bea\label{eq:extrema1}
&& \tilde x_1:=-\cot\left(\frac{\alpha_\phi}{2}\right)\tanh\left(\frac{\pi \nu_\sigma}{2}\right),\\
&&  \tilde x_2:=\tan\left(\frac{\alpha_\phi}{2}\right)\coth\left(\frac{\pi \nu_\sigma}{2}\right)\,.
\eea
At these values $\tilde{x}_1$ and $\tilde{x}_2$ the spectral density function takes the following simplified form:
\bea
&&\rho_1(\nu)=\left[\rho(\nu)\right]_{\tilde{x}=\tilde{x}_1}=\rho_{\rm f}(\nu_\sigma)\left(\cosh\pi\nu_\sigma-\cos\alpha_\phi\right)\times{\rm cosech}^{2}\left(\frac{\pi\nu_\sigma}{2}\right)>0\,,\\
&&\rho_2(\nu)=\left[\rho(\nu)\right]_{\tilde{x}=\tilde{x}_2}=\rho_{\rm f}(\nu_\sigma)\left(\cosh\pi\nu_\sigma-\cos\alpha_\phi\right)\times{\rm sech}^{2}\left(\frac{\pi\nu_\sigma}{2}\right)>0\,,
\eea
for the eral valued $\nu_\sigma$ and $\alpha_\phi$. A straightforward expression for the behavior of the spectral density at the resonance can be found in equation \eqref{ressimple}. The interacting component of the signal is of order one in the coupling constant, or, to put it another way, comparable to the gaussian free theory part, when the spectral parameter is on the resonance, that is, \be |\tilde x | \approx 1~~~~~{\rm or}~~~~ |\nu-\nu_\sigma|\approx \lambda^2\Xi.\ee This characteristic of our representation offers a potentially helpful method of searching experimental data for the non-gaussian features linked to such resonances. Naturally, a lot more work needs to be done before our method can be applied to practical cosmological settings. For the time being, we might assume that the resonance's qualitative shape endures the required changes. The resonance's shape is then only dependent on the parameter $\alpha$, which is equal to $-\frac{3 \pi}{2}$ for the massless external particle in four-dimensional dS. When the particle is heavy, the signal is muted by $\exp(-\pi \nu_\sigma)$. Naturally, if we could determine the mass of the swapped particle in Hubble units, the resonant search would be most effective.

\section{Conclusion}
\label{ka20}

We wrap up by summarizing our results and outlining potential future directions and generalizations. We examined several facets of QFT on a dS backdrop in this study, with a particular emphasis on correlation functions of huge scalar fields situated on the future boundary. With the aid of an auxiliary EAdS theory with the field content doubled in comparison to the original one, we created a formalism that enables the calculation of such correlation functions. This method shows significant analytic features of the correlators and significantly simplifies the computations. Specifically, we anticipate that dS correlators in perturbation theory will have the same meromorphic structure as AdS or, similarly, CFT correlation functions to all orders. This is rather unexpected as the states are designated by the continuum rather than discrete parameters and the volume of the spatial slice of dS is not constrained. As a result, one could have anticipated a more complex analytic structure similar to that found in flat space scattering amplitudes. Indeed, as we discussed, states in the dS Hilbert space do not correspond to the OPE that correlators satisfy. We also determined the requirements on the correlators imposed by the unitarity of the QFT in dS. The positivity of the spectral density in the CPW decomposition of the four-point function is our most potent finding on this. The unitarity condition does not depend in any way on perturbation theory and is not dependent on the EAdS method. 
In the categorization of \cite{Gorbenko:2018ncu}, the resulting correlation function structure is that of a real non-unitary CFT. Operators appear in complex conjugate pairs and have complex dimensions, which is entirely typical for such theories \cite{Hogervorst:2015akt}. The aforementioned positivity requirement is another characteristic of such cosmic CFT. Additionally, by resuming the bubble diagrams for the propagator of the exchanged particle, we examined the exchange diagrams at the tree level and subsequently at the one-loop level using our novel methods. Using the complete propagator is especially important for preserving the proper structure near a narrow resonance when dealing with heavy exchanged particles.

Finally, we cover several intriguing open problems. We feel it handy to include them here, arranged in many approximate categories, even if a number of them are addressed throughout the main text. These problems are appended below point-wise:
\begin{enumerate}
    \item Within the scope of large QFT on a dS background, there are various avenues to investigate. Establishing the analytic structure discovered in this paper at the non-perturbative level is one of them. Although it is tempting to speculate that the validity of the OPE and the meromorphic features of spectral densities remain beyond perturbation theory, we currently lack the means to demonstrate this. Because our EAdS Lagrangian and accompanying Euclidean action are not constrained from below, they cannot be used in the standard route integral operations. The convergence of OPE, which is connected to the asymptotic behaviors of the spectral densities used in our computations, is another related problem. As of right now, we are only certain that OPE for the four-point function will converge in our explicit cases. For this aim, it could be quite beneficial to comprehend the physical meaning of operators in the bulk language. Their relationship to the QNMs of the static patch theory is the most promising path. This motivates us to look into the relationships between our findings and those of \cite{Jafferis:2013qia,Anninos:2020hfj}.

  \item   The following is another query about analyticity and representations that appear as intermediate states. Unlike the comparable decomposition in EAdS, the decomposition of the scalar dS Greens function into irreps of the isometry group requires some extra representations. However, our final responses do not include these extra representations. These claims do not contradict each other, but they might be a clue to some of the aspects of QFT in dS that we have yet to discover.

  \item IR divergences are another topic that is frequently discussed in relation to scalar fields in dS. In fact, we made the assumption that perturbation theory holds true throughout the computation in this study, even though it is known that this is not the case once the fields get light enough. The method of Starobinski \cite{Starobinsky:1984,Starobinsky:1994bd} that was developed into a systematic framework in \cite{Gorbenko:2019rza} is the only known formalism for handling this issue. In our method, the poles of the spectral density for $\nu\approx-i\frac{d}{2}$ will exhibit the IR divergences of perturbation theory, whose residues are expected to be high and increase with the perturbation theory order. One can question whether our methods can offer a different framework for handling these divergences.

  \item A large-$N$ expansion is an effective tool for studying non-perturbative characteristics of QFTs. As an example, we can think about a large-$N$ $O(N)$ model on dS space at a finite coupling. Both methods should be effective when the non-perturbativity is caused by IR divergences and in the light physical mass regime. For $i \nu\in (0,\frac{d}{2})$, the eigenvalues of the Starobinski operator should then appear as poles in the spectral density, computed to a subleading order in \cite{Gorbenko:2019rza,Mirbabayi:2020vyt,Cohen:2021fzf}.

  \item Beyond perturbation theory, there are additional well-known methods for handling QFTs, such as analytic and numerical bootstrap. It has been effectively used to confine flat space, AdS, and CFT observables in a variety of scenarios during the past few years. The bootstrap of cosmic observables was started in a relatively recent study \cite{Hogervorst:2021uvp}. In this work, the positivity of the spectral densities and the continuum set of states in dS are used as inputs for the bootstrap process. Seeing how this software develops in the future is exciting. The discrete set of operators might alternatively be used as a basis, in which case the CFT bootstrap version suitable for non-unitary theories would be required \cite{Gliozzi:2013ysa,El-Showk:2016mxr}.

  \item The other is to investigate gravity duals of huge $N$ field theories in dS that are strongly linked using holography. Since the dual spacetime involves a FRW-like geometry with a singularity \cite{Maldacena:2010un,Hertog:2004rz,Hertog:2005hu,Turok:2007ry,Craps:2007ch,Barbon:2011ta,Barbon:2013nta,Kumar:2015jxa}, this arrangement is intriguing on its own. There are two obvious questions to answer in this regard. The first is to use the dual description to calculate the late-time correlators and QNMs in dS and examine their analytic structures. The second is to comprehend a signature of the bulk singularity in the boundary's late-time correlators.

  \item Despite the fact that the choice of state for the QFT calculations is a topic that is frequently debated, we attempted to be concise in this study. The choice of state is merely a component of the theory's specification for our QFT calculations. This definition can be applied in a variety of ways, such as by using $i\eps$ prescriptions in other coordinate systems, the no-boundary contour of \cite{Hartle:1983ai}, continuing the calculations on the sphere, etc., all of which provide the same outcomes. For modes shorter than the Horizon scale, the correlation functions reduce to the Minkowski ones while respecting dS invariance. As a result, we have a clear problem and a methodical framework for solving it. When we attempt to construct a realistic model of cosmology or when we choose to experiment with different toy models for cosmology to test our theoretical understanding, the question of what state is best for our universe is a different matter. Our intuition, common sense, and the experimental data are now all we have for the choice of state. Ideally, we would wish for a comprehensive theory for the real-world universe that would also predict the beginning state, but we are far from having such a theory. At least at the level of free theory, it would be fascinating to observe how our debate changes when we take into account other states, such as so-called alpha-states \cite{Mottola:1984ar,Allen:1985ux,Adhikari:2021ked,Choudhury:2021tuu,Choudhury:2020yaa,Choudhury:2018ppd,Choudhury:2017qyl,Choudhury:2017glj} that are likewise dS invariant. Naturally, when gravity is dynamical, the discussion of the initial state becomes extremely important. In this instance, the Hartle-Hawking state is the closed analog of the state under consideration. Other semiclassical additions to the gravitational path integral, known as bra-ket wormholes, have recently been recognized as potentially significant, at least in certain lower-dimensional toy models of cosmology \cite{Chen:2020tes,Penington:2019kki,Dong:2020uxp}. Therefore, it is crucial to use a generalization of our methods to examine QFT correlators on the bra-ket wormhole backdrops.

  \item Let's discuss a change of state that is unquestionably required to use our methods for the inflationary phenomenology: the clock field's breaking of dS isometries, which is present in all inflation models \cite{Cheung:2007st,Baumann:2019ghk,Choudhury:2025hnu,Choudhury:2025vso,Choudhury:2011sq,Choudhury:2012yh,Choudhury:2013zna,Choudhury:2013jya,Choudhury:2013iaa,Choudhury:2014sxa,Choudhury:2014uxa,Choudhury:2014kma,Choudhury:2014sua,Choudhury:2015pqa,Choudhury:2015hvr,Choudhury:2017cos,Choudhury:2025kxg,Choudhury:2024kjj,Choudhury:2024aji,Choudhury:2024dzw,Choudhury:2024dei,Choudhury:2024jlz,Choudhury:2024ybk,Choudhury:2024one,Choudhury:2023fjs,Choudhury:2023fwk,Choudhury:2023hfm,Choudhury:2023kdb,Choudhury:2023hvf,Choudhury:2023rks,Choudhury:2023jlt,Choudhury:2023vuj,Choudhury:2013woa,Choudhury:2011jt}, particularly in the context of the EFT of single field inflation \cite{Cheung:2007st,Choudhury:2017glj,Choudhury:2025pzp}. This breaking can be applied at several levels. When scale invariance is maintained while only specific conformal transformations are disrupted, this is a crucial limit to take into account. In this instance, the propagator may have a varied sound speed but essentially has the same form. Additionally, there is no issue with the interaction vertices continuing. The scale is additionally broken by eventually realistic inflationary backdrops, or, to put it another way, time-translation invariance by different slow-roll adjustments. Similar to \cite{Konstantinidis:2016nio,McFadden:2011kk,Bzowski:2012ih}, we think our methods have a reasonable probability of working in many simple scenarios where the time-dependence of the inflationary background is sufficiently smooth. However, models where time-dependence incorporates so-called features \cite{Chluba_2015} can need extra consideration and a case-by-case analysis.

  \item We will receive a wealth of new cosmological data in the next few years from both Large Scale Structure \cite{Schlegel:2019eqc,2018AAS...23135423D} and Cosmic Microwave Background \cite{Ade_2019,CMB-S4:2020lpa} experiments. These investigations aim to find non-gaussian characteristics in the initial inflationary fluctuations. This translates to differentiating between an interacting and free theory that generates late time correlators in the context of our paper. Experimental evidence currently supports a free theory \cite{Planck:2019kim}, although interactions must exist at some level. We primarily addressed more formal features of QFT in the context of cosmology in this study, but it's intriguing to consider whether the advancements we achieved here could be useful for analyzing experimental data. Finding narrow resonances is a logical first step. By substituting the inflaton backdrop for one of the external legs of a four-point function in QFT in dS, an inflationary three-point function can be extracted. This work examines the resonance's effects on inflationary observables in momentum space, where it appears as specific oscillations in the squeezed limit. No matter how weak the coupling is, our spectral representation transforms these oscillations into a resonant peak so that the departure from free theory is order one when the spectral parameter is near the spectral parameter of the resonance. With the spectral parameter $\nu$ being comparable to the center of mass energy, our parametrization of the correlator by the spectral function is comparable to the momentum representation of the flat space scattering amplitude. Therefore, it is conceivable that searching for resonances in the spectrum representation could be more effective or trustworthy. In reality, this would mean that instead of using the standard expansion in spherical harmonics, the cosmic data would be subjected to an analog of the CPW decomposition. The feasibility of this process is still to be determined, and as previously said, further work is required to take into account the symmetries of realistic inflationary models. The spectral density must meet certain requirements, which we covered in detail above, even when it is not near the resonance. This can be helpful in separating the primordial signal from different background sources whose spectral densities are not similarly limited. The features of large resonances with spin are also studied including such particles is a logical extension of our formalism.

\end{enumerate}



	\subsection*{Acknowledgements}
SC would like to thank the work
friendly environment of The Thanu Padmanabhan Centre For Cosmology and Science Popularization (CCSP), 
Shree Guru Gobind Singh Tricentenary (SGT) University,  Gurugram,  Delhi-NCR for providing tremendous support in research and offer the Assistant Professor (Senior Grade) position.  SC would like to thank The North American Nanohertz Observatory for Gravitational
Waves (NANOGrav) collaboration and the National Academy of Sciences (NASI), Prayagraj, India, for being elected as an associate member and the member of the academy
respectively. SC also thanks
all the members of virtual international
non-profit consortium Quantum Aspects of the Space Time \& Matter (QASTM) for elaborative discussions.  Last but not
least, we would like to acknowledge our debt to the people belonging to the various parts of the world for their
generous and steady support for research in natural sciences.
	
\subsection*{Data availability Statement}
Data sharing not applicable to this article as no datasets were generated or analysed during the current study.

\clearpage

\addcontentsline{toc}{section}{References}
\bibliographystyle{utphys}
\bibliography{references2}

\providecommand{\href}[2]{#2}\begingroup\raggedright\begin{thebibliography}{100}

\bibitem{Arkani-Hamed:2015bza}
N.~Arkani-Hamed and J.~Maldacena, ``{Cosmological Collider Physics},''
  \href{http://arxiv.org/abs/1503.08043}{{\ttfamily arXiv:1503.08043
  [hep-th]}}.

\bibitem{Sleight:2020obc}
C.~Sleight and M.~Taronna, ``{From AdS to dS Exchanges: Spectral
  Representation, Mellin Amplitudes and Crossing},''
  \href{http://arxiv.org/abs/2007.09993}{{\ttfamily arXiv:2007.09993
  [hep-th]}}.

\bibitem{Gorbenko:2019rza}
V.~Gorbenko and L.~Senatore, ``{$\lambda \phi^4$ in dS},''
  \href{http://arxiv.org/abs/1911.00022}{{\ttfamily arXiv:1911.00022
  [hep-th]}}.

\bibitem{Maldacena:2010un}
J.~Maldacena, ``{Vacuum decay into Anti de Sitter space},''
  \href{http://arxiv.org/abs/1012.0274}{{\ttfamily arXiv:1012.0274 [hep-th]}}.

\bibitem{Hertog:2004rz}
T.~Hertog and G.~T. Horowitz, ``{Towards a big crunch dual},''
  \href{http://dx.doi.org/10.1088/1126-6708/2004/07/073}{{\em JHEP} {\bfseries
  07} (2004) 073}, \href{http://arxiv.org/abs/hep-th/0406134}{{\ttfamily
  arXiv:hep-th/0406134}}.

\bibitem{Hertog:2005hu}
T.~Hertog and G.~T. Horowitz, ``{Holographic description of AdS cosmologies},''
  \href{http://dx.doi.org/10.1088/1126-6708/2005/04/005}{{\em JHEP} {\bfseries
  04} (2005) 005}, \href{http://arxiv.org/abs/hep-th/0503071}{{\ttfamily
  arXiv:hep-th/0503071}}.

\bibitem{Turok:2007ry}
N.~Turok, B.~Craps, and T.~Hertog, ``{From big crunch to big bang with
  AdS/CFT},'' \href{http://arxiv.org/abs/0711.1824}{{\ttfamily arXiv:0711.1824
  [hep-th]}}.

\bibitem{Craps:2007ch}
B.~Craps, T.~Hertog, and N.~Turok, ``{On the Quantum Resolution of Cosmological
  Singularities using AdS/CFT},''
  \href{http://dx.doi.org/10.1103/PhysRevD.86.043513}{{\em Phys. Rev. D}
  {\bfseries 86} (2012) 043513},
  \href{http://arxiv.org/abs/0712.4180}{{\ttfamily arXiv:0712.4180 [hep-th]}}.

\bibitem{Barbon:2011ta}
J.~L.~F. Barbon and E.~Rabinovici, ``{AdS Crunches, CFT Falls And Cosmological
  Complementarity},'' \href{http://dx.doi.org/10.1007/JHEP04(2011)044}{{\em
  JHEP} {\bfseries 04} (2011) 044},
  \href{http://arxiv.org/abs/1102.3015}{{\ttfamily arXiv:1102.3015 [hep-th]}}.

\bibitem{Barbon:2013nta}
J.~L.~F. Barb\'on and E.~Rabinovici, ``{Conformal Complementarity Maps},''
  \href{http://dx.doi.org/10.1007/JHEP12(2013)023}{{\em JHEP} {\bfseries 12}
  (2013) 023}, \href{http://arxiv.org/abs/1308.1921}{{\ttfamily arXiv:1308.1921
  [hep-th]}}.

\bibitem{Kumar:2015jxa}
S.~P. Kumar and V.~Vaganov, ``{Probing crunching AdS cosmologies},''
  \href{http://dx.doi.org/10.1007/JHEP02(2016)026}{{\em JHEP} {\bfseries 02}
  (2016) 026}, \href{http://arxiv.org/abs/1510.03281}{{\ttfamily
  arXiv:1510.03281 [hep-th]}}.

\bibitem{Spradlin:2001pw}
M.~Spradlin, A.~Strominger, and A.~Volovich, ``{Les Houches lectures on de
  Sitter space},'' in {\em {Les Houches Summer School: Session 76: Euro Summer
  School on Unity of Fundamental Physics: Gravity, Gauge Theory and Strings}}.
\newblock 10, 2001.
\newblock \href{http://arxiv.org/abs/hep-th/0110007}{{\ttfamily
  arXiv:hep-th/0110007}}.

\bibitem{Anninos:2012qw}
D.~Anninos, ``{De Sitter Musings},''
  \href{http://dx.doi.org/10.1142/S0217751X1230013X}{{\em Int. J. Mod. Phys. A}
  {\bfseries 27} (2012) 1230013},
  \href{http://arxiv.org/abs/1205.3855}{{\ttfamily arXiv:1205.3855 [hep-th]}}.

\bibitem{newton1950note}
T.~Newton, ``A note on the representations of the de sitter group,'' {\em
  Annals of Mathematics} (1950) 730--733.

\bibitem{thomas1941unitary}
L.~Thomas, ``On unitary representations of the group of de sitter space,'' {\em
  Annals of mathematics} (1941) 113--126.

\bibitem{Dobrev:1976vr}
V.~K. Dobrev, G.~Mack, I.~T. Todorov, V.~B. Petkova, and S.~G. Petrova, ``{On
  the Clebsch-Gordan Expansion for the Lorentz Group in n Dimensions},''
  \href{http://dx.doi.org/10.1016/0034-4877(76)90057-4}{{\em Rept. Math. Phys.}
  {\bfseries 9} (1976) 219--246}.

\bibitem{dobrev1977harmonic}
V.~Dobrev {\em et~al.}, ``Harmonic analysis on the n-dimensional lorentz group
  and its application to conformal quantum field theory.''.

\bibitem{Baumann:2017jvh}
D.~Baumann, G.~Goon, H.~Lee, and G.~L. Pimentel, ``{Partially Massless Fields
  During Inflation},'' \href{http://dx.doi.org/10.1007/JHEP04(2018)140}{{\em
  JHEP} {\bfseries 04} (2018) 140},
  \href{http://arxiv.org/abs/1712.06624}{{\ttfamily arXiv:1712.06624
  [hep-th]}}.

\bibitem{Witten:2001kn}
E.~Witten, ``{Quantum gravity in de Sitter space},'' in {\em {Strings 2001:
  International Conference}}.
\newblock 6, 2001.
\newblock \href{http://arxiv.org/abs/hep-th/0106109}{{\ttfamily
  arXiv:hep-th/0106109}}.

\bibitem{Strominger:2001pn}
A.~Strominger, ``{The dS / CFT correspondence},''
  \href{http://dx.doi.org/10.1088/1126-6708/2001/10/034}{{\em JHEP} {\bfseries
  10} (2001) 034}, \href{http://arxiv.org/abs/hep-th/0106113}{{\ttfamily
  arXiv:hep-th/0106113}}.

\bibitem{Hartle:1983ai}
J.~B. Hartle and S.~W. Hawking, ``{Wave Function of the Universe},''
  \href{http://dx.doi.org/10.1103/PhysRevD.28.2960}{{\em Phys. Rev. D}
  {\bfseries 28} (1983) 2960--2975}.

\bibitem{Maldacena:2002vr}
J.~M. Maldacena, ``{Non-Gaussian features of primordial fluctuations in single
  field inflationary models},''
  \href{http://dx.doi.org/10.1088/1126-6708/2003/05/013}{{\em JHEP} {\bfseries
  05} (2003) 013}, \href{http://arxiv.org/abs/astro-ph/0210603}{{\ttfamily
  arXiv:astro-ph/0210603}}.

\bibitem{Choudhury:2026pus}
S.~Choudhury, ``{EFT Perspective On de-Sitter S-Matrix},''
  \href{http://arxiv.org/abs/2601.18101}{{\ttfamily arXiv:2601.18101
  [hep-th]}}.

\bibitem{Choudhury:2025pzp}
S.~Choudhury, ``{Notes On de-Sitter Mellin Barnes Amplitudes},''
  \href{http://arxiv.org/abs/2512.09175}{{\ttfamily arXiv:2512.09175
  [hep-th]}}.

\bibitem{Marolf:2012kh}
D.~Marolf, I.~A. Morrison, and M.~Srednicki, ``{Perturbative S-matrix for
  massive scalar fields in global de Sitter space},''
  \href{http://dx.doi.org/10.1088/0264-9381/30/15/155023}{{\em Class. Quant.
  Grav.} {\bfseries 30} (2013) 155023},
  \href{http://arxiv.org/abs/1209.6039}{{\ttfamily arXiv:1209.6039 [hep-th]}}.

\bibitem{Cotler:2019dcj}
J.~Cotler and K.~Jensen, ``{Emergent unitarity in de Sitter from matrix
  integrals},'' \href{http://arxiv.org/abs/1911.12358}{{\ttfamily
  arXiv:1911.12358 [hep-th]}}.

\bibitem{Albrychiewicz:2020ruh}
E.~Albrychiewicz and Y.~Neiman, ``{Scattering in the static patch of de Sitter
  space},'' \href{http://dx.doi.org/10.1103/PhysRevD.103.065014}{{\em Phys.
  Rev. D} {\bfseries 103} no.~6, (2021) 065014},
  \href{http://arxiv.org/abs/2012.13584}{{\ttfamily arXiv:2012.13584
  [hep-th]}}.

\bibitem{Arkani-Hamed:2017fdk}
N.~Arkani-Hamed, P.~Benincasa, and A.~Postnikov, ``{Cosmological Polytopes and
  the Wavefunction of the Universe},''
  \href{http://arxiv.org/abs/1709.02813}{{\ttfamily arXiv:1709.02813
  [hep-th]}}.

\bibitem{Hillman:2019wgh}
A.~Hillman, ``{Symbol Recursion for the dS Wave Function},''
  \href{http://arxiv.org/abs/1912.09450}{{\ttfamily arXiv:1912.09450
  [hep-th]}}.

\bibitem{Benincasa:2019vqr}
P.~Benincasa, ``{Cosmological Polytopes and the Wavefuncton of the Universe for
  Light States},'' \href{http://arxiv.org/abs/1909.02517}{{\ttfamily
  arXiv:1909.02517 [hep-th]}}.

\bibitem{Meltzer:2021zin}
D.~Meltzer, ``{The Inflationary Wavefunction from Analyticity and
  Factorization},'' \href{http://arxiv.org/abs/2107.10266}{{\ttfamily
  arXiv:2107.10266 [hep-th]}}.

\bibitem{Pajer:2020wnj}
E.~Pajer, D.~Stefanyszyn, and J.~Supe, ``{The Boostless Bootstrap: Amplitudes
  without Lorentz boosts},''
  \href{http://dx.doi.org/10.1007/JHEP12(2020)198}{{\em JHEP} {\bfseries 12}
  (2020) 198}, \href{http://arxiv.org/abs/2007.00027}{{\ttfamily
  arXiv:2007.00027 [hep-th]}}.

\bibitem{Cespedes:2020xqq}
S.~C\'espedes, A.-C. Davis, and S.~Melville, ``{On the time evolution of
  cosmological correlators},''
  \href{http://dx.doi.org/10.1007/JHEP02(2021)012}{{\em JHEP} {\bfseries 02}
  (2021) 012}, \href{http://arxiv.org/abs/2009.07874}{{\ttfamily
  arXiv:2009.07874 [hep-th]}}.

\bibitem{Goodhew:2021oqg}
H.~Goodhew, S.~Jazayeri, M.~H. Gordon~Lee, and E.~Pajer, ``{Cutting
  Cosmological Correlators},''
  \href{http://arxiv.org/abs/2104.06587}{{\ttfamily arXiv:2104.06587
  [hep-th]}}.

\bibitem{Goodhew:2020hob}
H.~Goodhew, S.~Jazayeri, and E.~Pajer, ``{The Cosmological Optical Theorem},''
  \href{http://dx.doi.org/10.1088/1475-7516/2021/04/021}{{\em JCAP} {\bfseries
  04} (2021) 021}, \href{http://arxiv.org/abs/2009.02898}{{\ttfamily
  arXiv:2009.02898 [hep-th]}}.

\bibitem{Jazayeri:2021fvk}
S.~Jazayeri, E.~Pajer, and D.~Stefanyszyn, ``{From Locality and Unitarity to
  Cosmological Correlators},''
  \href{http://arxiv.org/abs/2103.08649}{{\ttfamily arXiv:2103.08649
  [hep-th]}}.

\bibitem{Baumann:2021fxj}
D.~Baumann, W.-M. Chen, C.~Duaso~Pueyo, A.~Joyce, H.~Lee, and G.~L. Pimentel,
  ``{Linking the Singularities of Cosmological Correlators},''
  \href{http://arxiv.org/abs/2106.05294}{{\ttfamily arXiv:2106.05294
  [hep-th]}}.

\bibitem{Anninos:2017eib}
D.~Anninos, F.~Denef, R.~Monten, and Z.~Sun, ``{Higher Spin de Sitter Hilbert
  Space},'' \href{http://dx.doi.org/10.1007/JHEP10(2019)071}{{\em JHEP}
  {\bfseries 10} (2019) 071}, \href{http://arxiv.org/abs/1711.10037}{{\ttfamily
  arXiv:1711.10037 [hep-th]}}.

\bibitem{Anninos:2011af}
D.~Anninos, S.~A. Hartnoll, and D.~M. Hofman, ``{Static Patch Solipsism:
  Conformal Symmetry of the de Sitter Worldline},''
  \href{http://dx.doi.org/10.1088/0264-9381/29/7/075002}{{\em Class. Quant.
  Grav.} {\bfseries 29} (2012) 075002},
  \href{http://arxiv.org/abs/1109.4942}{{\ttfamily arXiv:1109.4942 [hep-th]}}.

\bibitem{Isono:2020qew}
H.~Isono, H.~M. Liu, and T.~Noumi, ``{Wavefunctions in dS/CFT revisited:
  principal series and double-trace deformations},''
  \href{http://dx.doi.org/10.1007/JHEP04(2021)166}{{\em JHEP} {\bfseries 04}
  (2021) 166}, \href{http://arxiv.org/abs/2011.09479}{{\ttfamily
  arXiv:2011.09479 [hep-th]}}.

\bibitem{Witten:2001ua}
E.~Witten, ``{Multitrace operators, boundary conditions, and AdS / CFT
  correspondence},'' \href{http://arxiv.org/abs/hep-th/0112258}{{\ttfamily
  arXiv:hep-th/0112258}}.

\bibitem{Klebanov:1999tb}
I.~R. Klebanov and E.~Witten, ``{AdS / CFT correspondence and symmetry
  breaking},'' \href{http://dx.doi.org/10.1016/S0550-3213(99)00387-9}{{\em
  Nucl. Phys. B} {\bfseries 556} (1999) 89--114},
  \href{http://arxiv.org/abs/hep-th/9905104}{{\ttfamily arXiv:hep-th/9905104}}.

\bibitem{DiPietro:2021sjt}
L.~Di~Pietro, V.~Gorbenko, and S.~Komatsu, ``{Analyticity and unitarity for
  cosmological correlators},''
  \href{http://dx.doi.org/10.1007/JHEP03(2022)023}{{\em JHEP} {\bfseries 03}
  (2022) 023}, \href{http://arxiv.org/abs/2108.01695}{{\ttfamily
  arXiv:2108.01695 [hep-th]}}.

\bibitem{Gorbenko:2018ncu}
V.~Gorbenko, S.~Rychkov, and B.~Zan, ``{Walking, Weak first-order transitions,
  and Complex CFTs},'' \href{http://dx.doi.org/10.1007/JHEP10(2018)108}{{\em
  JHEP} {\bfseries 10} (2018) 108},
  \href{http://arxiv.org/abs/1807.11512}{{\ttfamily arXiv:1807.11512
  [hep-th]}}.

\bibitem{Leigh:2003ez}
R.~G. Leigh and A.~C. Petkou, ``{SL(2,Z) action on three-dimensional CFTs and
  holography},'' \href{http://dx.doi.org/10.1088/1126-6708/2003/12/020}{{\em
  JHEP} {\bfseries 12} (2003) 020},
  \href{http://arxiv.org/abs/hep-th/0309177}{{\ttfamily arXiv:hep-th/0309177}}.

\bibitem{Compere:2008us}
G.~Compere and D.~Marolf, ``{Setting the boundary free in AdS/CFT},''
  \href{http://dx.doi.org/10.1088/0264-9381/25/19/195014}{{\em Class. Quant.
  Grav.} {\bfseries 25} (2008) 195014},
  \href{http://arxiv.org/abs/0805.1902}{{\ttfamily arXiv:0805.1902 [hep-th]}}.

\bibitem{Giombi:2013yva}
S.~Giombi, I.~R. Klebanov, S.~S. Pufu, B.~R. Safdi, and G.~Tarnopolsky, ``{AdS
  Description of Induced Higher-Spin Gauge Theory},''
  \href{http://dx.doi.org/10.1007/JHEP10(2013)016}{{\em JHEP} {\bfseries 10}
  (2013) 016}, \href{http://arxiv.org/abs/1306.5242}{{\ttfamily arXiv:1306.5242
  [hep-th]}}.

\bibitem{Goel:2020yxl}
A.~Goel, L.~V. Iliesiu, J.~Kruthoff, and Z.~Yang, ``{Classifying boundary
  conditions in JT gravity: from energy-branes to $\alpha$-branes},''
  \href{http://dx.doi.org/10.1007/JHEP04(2021)069}{{\em JHEP} {\bfseries 04}
  (2021) 069}, \href{http://arxiv.org/abs/2010.12592}{{\ttfamily
  arXiv:2010.12592 [hep-th]}}.

\bibitem{Cotler:2019nbi}
J.~Cotler, K.~Jensen, and A.~Maloney, ``{Low-dimensional de Sitter quantum
  gravity},'' \href{http://dx.doi.org/10.1007/JHEP06(2020)048}{{\em JHEP}
  {\bfseries 06} (2020) 048}, \href{http://arxiv.org/abs/1905.03780}{{\ttfamily
  arXiv:1905.03780 [hep-th]}}.

\bibitem{Weinberg:2005vy}
S.~Weinberg, ``{Quantum contributions to cosmological correlations},''
  \href{http://dx.doi.org/10.1103/PhysRevD.72.043514}{{\em Phys. Rev. D}
  {\bfseries 72} (2005) 043514},
  \href{http://arxiv.org/abs/hep-th/0506236}{{\ttfamily arXiv:hep-th/0506236}}.

\bibitem{Seery:2007we}
D.~Seery, ``{One-loop corrections to a scalar field during inflation},''
  \href{http://dx.doi.org/10.1088/1475-7516/2007/11/025}{{\em JCAP} {\bfseries
  11} (2007) 025}, \href{http://arxiv.org/abs/0707.3377}{{\ttfamily
  arXiv:0707.3377 [astro-ph]}}.

\bibitem{Adshead:2009cb}
P.~Adshead, R.~Easther, and E.~A. Lim, ``{The 'in-in' Formalism and
  Cosmological Perturbations},''
  \href{http://dx.doi.org/10.1103/PhysRevD.80.083521}{{\em Phys. Rev. D}
  {\bfseries 80} (2009) 083521},
  \href{http://arxiv.org/abs/0904.4207}{{\ttfamily arXiv:0904.4207 [hep-th]}}.

\bibitem{Senatore:2009cf}
L.~Senatore and M.~Zaldarriaga, ``{On Loops in Inflation},''
  \href{http://dx.doi.org/10.1007/JHEP12(2010)008}{{\em JHEP} {\bfseries 12}
  (2010) 008}, \href{http://arxiv.org/abs/0912.2734}{{\ttfamily arXiv:0912.2734
  [hep-th]}}.

\bibitem{Sleight:2019mgd}
C.~Sleight, ``{A Mellin Space Approach to Cosmological Correlators},''
  \href{http://dx.doi.org/10.1007/JHEP01(2020)090}{{\em JHEP} {\bfseries 01}
  (2020) 090}, \href{http://arxiv.org/abs/1906.12302}{{\ttfamily
  arXiv:1906.12302 [hep-th]}}.

\bibitem{Sleight:2019hfp}
C.~Sleight and M.~Taronna, ``{Bootstrapping Inflationary Correlators in Mellin
  Space},'' \href{http://dx.doi.org/10.1007/JHEP02(2020)098}{{\em JHEP}
  {\bfseries 02} (2020) 098}, \href{http://arxiv.org/abs/1907.01143}{{\ttfamily
  arXiv:1907.01143 [hep-th]}}.

\bibitem{Sleight:2021iix}
C.~Sleight and M.~Taronna, ``{On the consistency of (partially-)massless matter
  couplings in de Sitter space},''
  \href{http://arxiv.org/abs/2106.00366}{{\ttfamily arXiv:2106.00366
  [hep-th]}}.

\bibitem{Arkani-Hamed:2018kmz}
N.~Arkani-Hamed, D.~Baumann, H.~Lee, and G.~L. Pimentel, ``{The Cosmological
  Bootstrap: Inflationary Correlators from Symmetries and Singularities},''
  \href{http://dx.doi.org/10.1007/JHEP04(2020)105}{{\em JHEP} {\bfseries 04}
  (2020) 105}, \href{http://arxiv.org/abs/1811.00024}{{\ttfamily
  arXiv:1811.00024 [hep-th]}}.

\bibitem{Baumann:2019oyu}
D.~Baumann, C.~Duaso~Pueyo, A.~Joyce, H.~Lee, and G.~L. Pimentel, ``{The
  cosmological bootstrap: weight-shifting operators and scalar seeds},''
  \href{http://dx.doi.org/10.1007/JHEP12(2020)204}{{\em JHEP} {\bfseries 12}
  (2020) 204}, \href{http://arxiv.org/abs/1910.14051}{{\ttfamily
  arXiv:1910.14051 [hep-th]}}.

\bibitem{Baumann:2020dch}
D.~Baumann, C.~Duaso~Pueyo, A.~Joyce, H.~Lee, and G.~L. Pimentel, ``{The
  Cosmological Bootstrap: Spinning Correlators from Symmetries and
  Factorization},'' \href{http://arxiv.org/abs/2005.04234}{{\ttfamily
  arXiv:2005.04234 [hep-th]}}.

\bibitem{Marolf:2010zp}
D.~Marolf and I.~A. Morrison, ``{The IR stability of de Sitter: Loop
  corrections to scalar propagators},''
  \href{http://dx.doi.org/10.1103/PhysRevD.82.105032}{{\em Phys. Rev. D}
  {\bfseries 82} (2010) 105032},
  \href{http://arxiv.org/abs/1006.0035}{{\ttfamily arXiv:1006.0035 [gr-qc]}}.

\bibitem{Marolf:2010nz}
D.~Marolf and I.~A. Morrison, ``{The IR stability of de Sitter QFT: results at
  all orders},'' \href{http://dx.doi.org/10.1103/PhysRevD.84.044040}{{\em Phys.
  Rev. D} {\bfseries 84} (2011) 044040},
  \href{http://arxiv.org/abs/1010.5327}{{\ttfamily arXiv:1010.5327 [gr-qc]}}.

\bibitem{Marolf:2011sh}
D.~Marolf and I.~A. Morrison, ``{The IR stability of de Sitter QFT: Physical
  initial conditions},''
  \href{http://dx.doi.org/10.1007/s10714-011-1233-3}{{\em Gen. Rel. Grav.}
  {\bfseries 43} (2011) 3497--3530},
  \href{http://arxiv.org/abs/1104.4343}{{\ttfamily arXiv:1104.4343 [gr-qc]}}.

\bibitem{Harlow:2011ke}
D.~Harlow and D.~Stanford, ``{Operator Dictionaries and Wave Functions in
  AdS/CFT and dS/CFT},'' \href{http://arxiv.org/abs/1104.2621}{{\ttfamily
  arXiv:1104.2621 [hep-th]}}.

\bibitem{Mata:2012bx}
I.~Mata, S.~Raju, and S.~Trivedi, ``{CMB from CFT},''
  \href{http://dx.doi.org/10.1007/JHEP07(2013)015}{{\em JHEP} {\bfseries 07}
  (2013) 015}, \href{http://arxiv.org/abs/1211.5482}{{\ttfamily arXiv:1211.5482
  [hep-th]}}.

\bibitem{Anninos:2014lwa}
D.~Anninos, T.~Anous, D.~Z. Freedman, and G.~Konstantinidis, ``{Late-time
  Structure of the Bunch-Davies De Sitter Wavefunction},''
  \href{http://dx.doi.org/10.1088/1475-7516/2015/11/048}{{\em JCAP} {\bfseries
  11} (2015) 048}, \href{http://arxiv.org/abs/1406.5490}{{\ttfamily
  arXiv:1406.5490 [hep-th]}}.

\bibitem{Carmi:2018qzm}
D.~Carmi, L.~Di~Pietro, and S.~Komatsu, ``{A Study of Quantum Field Theories in
  AdS at Finite Coupling},''
  \href{http://dx.doi.org/10.1007/JHEP01(2019)200}{{\em JHEP} {\bfseries 01}
  (2019) 200}, \href{http://arxiv.org/abs/1810.04185}{{\ttfamily
  arXiv:1810.04185 [hep-th]}}.

\bibitem{Caron-Huot:2017vep}
S.~Caron-Huot, ``{Analyticity in Spin in Conformal Theories},''
  \href{http://dx.doi.org/10.1007/JHEP09(2017)078}{{\em JHEP} {\bfseries 09}
  (2017) 078}, \href{http://arxiv.org/abs/1703.00278}{{\ttfamily
  arXiv:1703.00278 [hep-th]}}.

\bibitem{Simmons-Duffin:2017nub}
D.~Simmons-Duffin, D.~Stanford, and E.~Witten, ``{A spacetime derivation of the
  Lorentzian OPE inversion formula},''
  \href{http://dx.doi.org/10.1007/JHEP07(2018)085}{{\em JHEP} {\bfseries 07}
  (2018) 085}, \href{http://arxiv.org/abs/1711.03816}{{\ttfamily
  arXiv:1711.03816 [hep-th]}}.

\bibitem{Karateev:2018oml}
D.~Karateev, P.~Kravchuk, and D.~Simmons-Duffin, ``{Harmonic Analysis and Mean
  Field Theory},'' \href{http://dx.doi.org/10.1007/JHEP10(2019)217}{{\em JHEP}
  {\bfseries 10} (2019) 217}, \href{http://arxiv.org/abs/1809.05111}{{\ttfamily
  arXiv:1809.05111 [hep-th]}}.

\bibitem{Hogervorst:2017sfd}
M.~Hogervorst and B.~C. van Rees, ``{Crossing symmetry in alpha space},''
  \href{http://dx.doi.org/10.1007/JHEP11(2017)193}{{\em JHEP} {\bfseries 11}
  (2017) 193}, \href{http://arxiv.org/abs/1702.08471}{{\ttfamily
  arXiv:1702.08471 [hep-th]}}.

\bibitem{Rutter:2020vpw}
D.~Rutter and B.~C. Van~Rees, ``{Applications of Alpha Space},''
  \href{http://dx.doi.org/10.1007/JHEP12(2020)048}{{\em JHEP} {\bfseries 12}
  (2020) 048}, \href{http://arxiv.org/abs/2003.07964}{{\ttfamily
  arXiv:2003.07964 [hep-th]}}.

\bibitem{Paulos:2016fap}
M.~F. Paulos, J.~Penedones, J.~Toledo, B.~C. van Rees, and P.~Vieira, ``{The
  S-matrix bootstrap. Part I: QFT in AdS},''
  \href{http://dx.doi.org/10.1007/JHEP11(2017)133}{{\em JHEP} {\bfseries 11}
  (2017) 133}, \href{http://arxiv.org/abs/1607.06109}{{\ttfamily
  arXiv:1607.06109 [hep-th]}}.

\bibitem{Rychkov:2016iqz}
S.~Rychkov, \href{http://dx.doi.org/10.1007/978-3-319-43626-5}{{\em {EPFL
  Lectures on Conformal Field Theory in D\ensuremath{>}= 3 Dimensions}}}.
\newblock SpringerBriefs in Physics. 1, 2016.
\newblock \href{http://arxiv.org/abs/1601.05000}{{\ttfamily arXiv:1601.05000
  [hep-th]}}.

\bibitem{Simmons-Duffin:2016gjk}
D.~Simmons-Duffin, \href{http://dx.doi.org/10.1142/9789813149441_0001}{``{The
  Conformal Bootstrap},''} in {\em {Theoretical Advanced Study Institute in
  Elementary Particle Physics}: {New Frontiers in Fields and Strings}}.
\newblock 2, 2016.
\newblock \href{http://arxiv.org/abs/1602.07982}{{\ttfamily arXiv:1602.07982
  [hep-th]}}.

\bibitem{Pappadopulo:2012jk}
D.~Pappadopulo, S.~Rychkov, J.~Espin, and R.~Rattazzi, ``{OPE Convergence in
  Conformal Field Theory},''
  \href{http://dx.doi.org/10.1103/PhysRevD.86.105043}{{\em Phys. Rev. D}
  {\bfseries 86} (2012) 105043},
  \href{http://arxiv.org/abs/1208.6449}{{\ttfamily arXiv:1208.6449 [hep-th]}}.

\bibitem{Hogervorst:2013sma}
M.~Hogervorst and S.~Rychkov, ``{Radial Coordinates for Conformal Blocks},''
  \href{http://dx.doi.org/10.1103/PhysRevD.87.106004}{{\em Phys. Rev. D}
  {\bfseries 87} (2013) 106004},
  \href{http://arxiv.org/abs/1303.1111}{{\ttfamily arXiv:1303.1111 [hep-th]}}.

\bibitem{Jafferis:2013qia}
D.~L. Jafferis, A.~Lupsasca, V.~Lysov, G.~S. Ng, and A.~Strominger,
  ``{Quasinormal quantization in de Sitter spacetime},''
  \href{http://dx.doi.org/10.1007/JHEP01(2015)004}{{\em JHEP} {\bfseries 01}
  (2015) 004}, \href{http://arxiv.org/abs/1305.5523}{{\ttfamily arXiv:1305.5523
  [hep-th]}}.

\bibitem{Anninos:2010gh}
D.~Anninos and T.~Anous, ``{A de Sitter Hoedown},''
  \href{http://dx.doi.org/10.1007/JHEP08(2010)131}{{\em JHEP} {\bfseries 08}
  (2010) 131}, \href{http://arxiv.org/abs/1002.1717}{{\ttfamily arXiv:1002.1717
  [hep-th]}}.

\bibitem{Bellazzini:2020cot}
B.~Bellazzini, J.~Elias~Mir\'o, R.~Rattazzi, M.~Riembau, and F.~Riva,
  ``{Positive Moments for Scattering Amplitudes},''
  \href{http://arxiv.org/abs/2011.00037}{{\ttfamily arXiv:2011.00037
  [hep-th]}}.

\bibitem{Tolley:2020gtv}
A.~J. Tolley, Z.-Y. Wang, and S.-Y. Zhou, ``{New positivity bounds from full
  crossing symmetry},'' \href{http://dx.doi.org/10.1007/JHEP05(2021)255}{{\em
  JHEP} {\bfseries 05} (2021) 255},
  \href{http://arxiv.org/abs/2011.02400}{{\ttfamily arXiv:2011.02400
  [hep-th]}}.

\bibitem{Arkani-Hamed:2020blm}
N.~Arkani-Hamed, T.-C. Huang, and Y.-T. Huang, ``{The EFT-Hedron},''
  \href{http://dx.doi.org/10.1007/JHEP05(2021)259}{{\em JHEP} {\bfseries 05}
  (2021) 259}, \href{http://arxiv.org/abs/2012.15849}{{\ttfamily
  arXiv:2012.15849 [hep-th]}}.

\bibitem{Caron-Huot:2020cmc}
S.~Caron-Huot and V.~Van~Duong, ``{Extremal Effective Field Theories},''
  \href{http://dx.doi.org/10.1007/JHEP05(2021)280}{{\em JHEP} {\bfseries 05}
  (2021) 280}, \href{http://arxiv.org/abs/2011.02957}{{\ttfamily
  arXiv:2011.02957 [hep-th]}}.

\bibitem{Sinha:2020win}
A.~Sinha and A.~Zahed, ``{Crossing Symmetric Dispersion Relations in Quantum
  Field Theories},''
  \href{http://dx.doi.org/10.1103/PhysRevLett.126.181601}{{\em Phys. Rev.
  Lett.} {\bfseries 126} no.~18, (2021) 181601},
  \href{http://arxiv.org/abs/2012.04877}{{\ttfamily arXiv:2012.04877
  [hep-th]}}.

\bibitem{Chiang:2021ziz}
L.-Y. Chiang, Y.-t. Huang, W.~Li, L.~Rodina, and H.-C. Weng, ``{Into the
  EFThedron and UV constraints from IR consistency},''
  \href{http://arxiv.org/abs/2105.02862}{{\ttfamily arXiv:2105.02862
  [hep-th]}}.

\bibitem{Caron-Huot:2021rmr}
S.~Caron-Huot, D.~Mazac, L.~Rastelli, and D.~Simmons-Duffin, ``{Sharp
  Boundaries for the Swampland},''
  \href{http://arxiv.org/abs/2102.08951}{{\ttfamily arXiv:2102.08951
  [hep-th]}}.

\bibitem{Caron-Huot:2021enk}
S.~Caron-Huot, D.~Mazac, L.~Rastelli, and D.~Simmons-Duffin, ``{AdS Bulk
  Locality from Sharp CFT Bounds},''
  \href{http://arxiv.org/abs/2106.10274}{{\ttfamily arXiv:2106.10274
  [hep-th]}}.

\bibitem{Guerrieri:2021ivu}
A.~Guerrieri, J.~Penedones, and P.~Vieira, ``{Where is String Theory?},''
  \href{http://arxiv.org/abs/2102.02847}{{\ttfamily arXiv:2102.02847
  [hep-th]}}.

\bibitem{Moschella:2007zza}
U.~Moschella and R.~Schaeffer, ``{Quantum theory on Lobatchevski spaces},''
  \href{http://dx.doi.org/10.1088/0264-9381/24/14/003}{{\em Class. Quant.
  Grav.} {\bfseries 24} (2007) 3571--3602},
  \href{http://arxiv.org/abs/0709.2795}{{\ttfamily arXiv:0709.2795 [hep-th]}}.

\bibitem{Cornalba:2008qf}
L.~Cornalba, M.~S. Costa, and J.~Penedones, ``{Eikonal Methods in AdS/CFT: BFKL
  Pomeron at Weak Coupling},''
  \href{http://dx.doi.org/10.1088/1126-6708/2008/06/048}{{\em JHEP} {\bfseries
  06} (2008) 048}, \href{http://arxiv.org/abs/0801.3002}{{\ttfamily
  arXiv:0801.3002 [hep-th]}}.

\bibitem{Penedones:2010ue}
J.~Penedones, ``{Writing CFT correlation functions as AdS scattering
  amplitudes},'' \href{http://dx.doi.org/10.1007/JHEP03(2011)025}{{\em JHEP}
  {\bfseries 03} (2011) 025}, \href{http://arxiv.org/abs/1011.1485}{{\ttfamily
  arXiv:1011.1485 [hep-th]}}.

\bibitem{Costa:2014kfa}
M.~S. Costa, V.~Gon\c{c}alves, and J.~a. Penedones, ``{Spinning AdS
  Propagators},'' \href{http://dx.doi.org/10.1007/JHEP09(2014)064}{{\em JHEP}
  {\bfseries 09} (2014) 064}, \href{http://arxiv.org/abs/1404.5625}{{\ttfamily
  arXiv:1404.5625 [hep-th]}}.

\bibitem{Hogervorst:2015akt}
M.~Hogervorst, S.~Rychkov, and B.~C. van Rees, ``{Unitarity violation at the
  Wilson-Fisher fixed point in 4-$\epsilon$ dimensions},''
  \href{http://dx.doi.org/10.1103/PhysRevD.93.125025}{{\em Phys. Rev. D}
  {\bfseries 93} no.~12, (2016) 125025},
  \href{http://arxiv.org/abs/1512.00013}{{\ttfamily arXiv:1512.00013
  [hep-th]}}.

\bibitem{Anninos:2020hfj}
D.~Anninos, F.~Denef, Y.~T.~A. Law, and Z.~Sun, ``{Quantum de Sitter horizon
  entropy from quasicanonical bulk, edge, sphere and topological string
  partition functions},'' \href{http://arxiv.org/abs/2009.12464}{{\ttfamily
  arXiv:2009.12464 [hep-th]}}.

\bibitem{Starobinsky:1984}
A.~A. Starobinsky, ``{In Fundamental Interactions, MGPI Press, Moscow, 1984
  },''.

\bibitem{Starobinsky:1994bd}
A.~A. Starobinsky and J.~Yokoyama, ``{Equilibrium state of a selfinteracting
  scalar field in the De Sitter background},''
  \href{http://dx.doi.org/10.1103/PhysRevD.50.6357}{{\em Phys. Rev.} {\bfseries
  D50} (1994) 6357--6368},
\href{http://arxiv.org/abs/astro-ph/9407016}{{\ttfamily arXiv:astro-ph/9407016
  [astro-ph]}}.

\bibitem{Mirbabayi:2020vyt}
M.~Mirbabayi, ``{Markovian Dynamics in de Sitter},''
  \href{http://arxiv.org/abs/2010.06604}{{\ttfamily arXiv:2010.06604
  [hep-th]}}.

\bibitem{Cohen:2021fzf}
T.~Cohen, D.~Green, A.~Premkumar, and A.~Ridgway, ``{Stochastic Inflation at
  NNLO},'' \href{http://arxiv.org/abs/2106.09728}{{\ttfamily arXiv:2106.09728
  [hep-th]}}.

\bibitem{Hogervorst:2021uvp}
M.~Hogervorst, J.~a. Penedones, and K.~S. Vaziri, ``{Towards the
  non-perturbative cosmological bootstrap},''
  \href{http://arxiv.org/abs/2107.13871}{{\ttfamily arXiv:2107.13871
  [hep-th]}}.

\bibitem{Gliozzi:2013ysa}
F.~Gliozzi, ``{More constraining conformal bootstrap},''
  \href{http://dx.doi.org/10.1103/PhysRevLett.111.161602}{{\em Phys. Rev.
  Lett.} {\bfseries 111} (2013) 161602},
  \href{http://arxiv.org/abs/1307.3111}{{\ttfamily arXiv:1307.3111 [hep-th]}}.

\bibitem{El-Showk:2016mxr}
S.~El-Showk and M.~F. Paulos, ``{Extremal bootstrapping: go with the flow},''
  \href{http://dx.doi.org/10.1007/JHEP03(2018)148}{{\em JHEP} {\bfseries 03}
  (2018) 148}, \href{http://arxiv.org/abs/1605.08087}{{\ttfamily
  arXiv:1605.08087 [hep-th]}}.

\bibitem{Mottola:1984ar}
E.~Mottola, ``{Particle Creation in de Sitter Space},''
  \href{http://dx.doi.org/10.1103/PhysRevD.31.754}{{\em Phys. Rev. D}
  {\bfseries 31} (1985) 754}.

\bibitem{Allen:1985ux}
B.~Allen, ``{Vacuum States in de Sitter Space},''
  \href{http://dx.doi.org/10.1103/PhysRevD.32.3136}{{\em Phys. Rev. D}
  {\bfseries 32} (1985) 3136}.

\bibitem{Adhikari:2021ked}
K.~Adhikari, S.~Choudhury, H.~N. Pandya, and R.~Srivastava, ``{Primordial
  Gravitational Wave Circuit Complexity},''
  \href{http://dx.doi.org/10.3390/sym15030664}{{\em Symmetry} {\bfseries 15}
  no.~3, (2023) 664}, \href{http://arxiv.org/abs/2108.10334}{{\ttfamily
  arXiv:2108.10334 [gr-qc]}}.

\bibitem{Choudhury:2021tuu}
S.~Choudhury, ``{The Cosmological OTOC: A New Proposal for Quantifying
  Auto-correlated Random Non-chaotic Primordial Fluctuations},''
  \href{http://dx.doi.org/10.20944/preprints202102.0616.v1}{{\em Symmetry}
  {\bfseries 13} no.~4, (2021) 599},
  \href{http://arxiv.org/abs/2106.01305}{{\ttfamily arXiv:2106.01305
  [physics.gen-ph]}}.

\bibitem{Choudhury:2020yaa}
S.~Choudhury, ``{The Cosmological OTOC: Formulating new cosmological
  micro-canonical correlation functions for random chaotic fluctuations in
  Out-of-Equilibrium Quantum Statistical Field Theory},''
  \href{http://dx.doi.org/10.3390/sym12091527}{{\em Symmetry} {\bfseries 12}
  no.~9, (2020) 1527}, \href{http://arxiv.org/abs/2005.11750}{{\ttfamily
  arXiv:2005.11750 [hep-th]}}.

\bibitem{Choudhury:2018ppd}
S.~Choudhury and S.~Panda, ``{Cosmological Spectrum of Two-Point Correlation
  Function from Vacuum Fluctuation of Stringy Axion Field in De Sitter Space: A
  Study of the Role of Quantum Entanglement},''
  \href{http://dx.doi.org/10.3390/universe6060079}{{\em Universe} {\bfseries 6}
  no.~6, (2020) 79}, \href{http://arxiv.org/abs/1809.02905}{{\ttfamily
  arXiv:1809.02905 [hep-th]}}.

\bibitem{Choudhury:2017qyl}
S.~Choudhury and S.~Panda, ``{Quantum entanglement in de Sitter space from
  stringy axion: An analysis using $\alpha$ vacua},''
  \href{http://dx.doi.org/10.1016/j.nuclphysb.2019.03.018}{{\em Nucl. Phys. B}
  {\bfseries 943} (2019) 114606},
  \href{http://arxiv.org/abs/1712.08299}{{\ttfamily arXiv:1712.08299
  [hep-th]}}.

\bibitem{Choudhury:2017glj}
S.~Choudhury, ``{CMB from EFT},''
  \href{http://dx.doi.org/10.3390/universe5060155}{{\em Universe} {\bfseries 5}
  no.~6, (2019) 155}, \href{http://arxiv.org/abs/1712.04766}{{\ttfamily
  arXiv:1712.04766 [hep-th]}}.

\bibitem{Chen:2020tes}
Y.~Chen, V.~Gorbenko, and J.~Maldacena, ``{Bra-ket wormholes in gravitationally
  prepared states},'' \href{http://dx.doi.org/10.1007/JHEP02(2021)009}{{\em
  JHEP} {\bfseries 02} (2021) 009},
  \href{http://arxiv.org/abs/2007.16091}{{\ttfamily arXiv:2007.16091
  [hep-th]}}.

\bibitem{Penington:2019kki}
G.~Penington, S.~H. Shenker, D.~Stanford, and Z.~Yang, ``{Replica wormholes and
  the black hole interior},'' \href{http://arxiv.org/abs/1911.11977}{{\ttfamily
  arXiv:1911.11977 [hep-th]}}.

\bibitem{Dong:2020uxp}
X.~Dong, X.-L. Qi, Z.~Shangnan, and Z.~Yang, ``{Effective entropy of quantum
  fields coupled with gravity},''
  \href{http://dx.doi.org/10.1007/JHEP10(2020)052}{{\em JHEP} {\bfseries 10}
  (2020) 052}, \href{http://arxiv.org/abs/2007.02987}{{\ttfamily
  arXiv:2007.02987 [hep-th]}}.

\bibitem{Cheung:2007st}
C.~Cheung, P.~Creminelli, A.~L. Fitzpatrick, J.~Kaplan, and L.~Senatore, ``{The
  Effective Field Theory of Inflation},''
  \href{http://dx.doi.org/10.1088/1126-6708/2008/03/014}{{\em JHEP} {\bfseries
  03} (2008) 014}, \href{http://arxiv.org/abs/0709.0293}{{\ttfamily
  arXiv:0709.0293 [hep-th]}}.

\bibitem{Baumann:2019ghk}
D.~Baumann, D.~Green, and T.~Hartman, ``{Dynamical Constraints on RG Flows and
  Cosmology},'' \href{http://dx.doi.org/10.1007/JHEP12(2019)134}{{\em JHEP}
  {\bfseries 12} (2019) 134}, \href{http://arxiv.org/abs/1906.10226}{{\ttfamily
  arXiv:1906.10226 [hep-th]}}.

\bibitem{Choudhury:2025hnu}
S.~Choudhury, S.~K. Singh, and S.~K. Sahoo, ``{Quintessential Inflation in
  Light of ACT DR6},'' \href{http://arxiv.org/abs/2511.19898}{{\ttfamily
  arXiv:2511.19898 [gr-qc]}}.

\bibitem{Choudhury:2025vso}
S.~Choudhury, G.~Bauyrzhan, S.~K. Singh, and K.~Yerzhanov, ``{What new physics
  can we extract from inflation using the ACT DR6 and DESI DR2
  Observations?},'' \href{http://arxiv.org/abs/2506.15407}{{\ttfamily
  arXiv:2506.15407 [astro-ph.CO]}}.

\bibitem{Choudhury:2011sq}
S.~Choudhury and S.~Pal, ``{Brane inflation in background supergravity},''
  \href{http://dx.doi.org/10.1103/PhysRevD.85.043529}{{\em Phys. Rev. D}
  {\bfseries 85} (2012) 043529},
  \href{http://arxiv.org/abs/1102.4206}{{\ttfamily arXiv:1102.4206 [hep-th]}}.

\bibitem{Choudhury:2012yh}
S.~Choudhury and S.~Pal, ``{DBI Galileon inflation in background SUGRA},''
  \href{http://dx.doi.org/10.1016/j.nuclphysb.2013.05.010}{{\em Nucl. Phys. B}
  {\bfseries 874} (2013) 85--114},
  \href{http://arxiv.org/abs/1208.4433}{{\ttfamily arXiv:1208.4433 [hep-th]}}.

\bibitem{Choudhury:2013zna}
S.~Choudhury, T.~Chakraborty, and S.~Pal, ``{Higgs inflation from new
  K{\"a}hler potential},''
  \href{http://dx.doi.org/10.1016/j.nuclphysb.2014.01.002}{{\em Nucl. Phys. B}
  {\bfseries 880} (2014) 155--174},
  \href{http://arxiv.org/abs/1305.0981}{{\ttfamily arXiv:1305.0981 [hep-th]}}.

\bibitem{Choudhury:2013jya}
S.~Choudhury, A.~Mazumdar, and S.~Pal, ``{Low {\&} High scale MSSM inflation,
  gravitational waves and constraints from Planck},''
  \href{http://dx.doi.org/10.1088/1475-7516/2013/07/041}{{\em JCAP} {\bfseries
  07} (2013) 041}, \href{http://arxiv.org/abs/1305.6398}{{\ttfamily
  arXiv:1305.6398 [hep-ph]}}.

\bibitem{Choudhury:2013iaa}
S.~Choudhury and A.~Mazumdar, ``{An accurate bound on tensor-to-scalar ratio
  and the scale of inflation},''
  \href{http://dx.doi.org/10.1016/j.nuclphysb.2014.03.005}{{\em Nucl. Phys. B}
  {\bfseries 882} (2014) 386--396},
  \href{http://arxiv.org/abs/1306.4496}{{\ttfamily arXiv:1306.4496 [hep-ph]}}.

\bibitem{Choudhury:2014sxa}
S.~Choudhury, A.~Mazumdar, and E.~Pukartas, ``{Constraining ${\cal N}=1$
  supergravity inflationary framework with non-minimal K{\"a}hler operators},''
  \href{http://dx.doi.org/10.1007/JHEP04(2014)077}{{\em JHEP} {\bfseries 04}
  (2014) 077}, \href{http://arxiv.org/abs/1402.1227}{{\ttfamily arXiv:1402.1227
  [hep-th]}}.

\bibitem{Choudhury:2014uxa}
S.~Choudhury, ``{Constraining N = 1 supergravity inflation with non-minimal
  Kaehler operators using $\delta$N formalism},''
  \href{http://dx.doi.org/10.1007/JHEP04(2014)105}{{\em JHEP} {\bfseries 04}
  (2014) 105}, \href{http://arxiv.org/abs/1402.1251}{{\ttfamily arXiv:1402.1251
  [hep-th]}}.

\bibitem{Choudhury:2014kma}
S.~Choudhury and A.~Mazumdar, ``{Reconstructing inflationary potential from
  BICEP2 and running of tensor modes},''
  \href{http://arxiv.org/abs/1403.5549}{{\ttfamily arXiv:1403.5549 [hep-th]}}.

\bibitem{Choudhury:2014sua}
S.~Choudhury, ``{Can Effective Field Theory of inflation generate large
  tensor-to-scalar ratio within Randall{\textendash}Sundrum single
  braneworld?},'' \href{http://dx.doi.org/10.1016/j.nuclphysb.2015.02.024}{{\em
  Nucl. Phys. B} {\bfseries 894} (2015) 29--55},
  \href{http://arxiv.org/abs/1406.7618}{{\ttfamily arXiv:1406.7618 [hep-th]}}.

\bibitem{Choudhury:2015pqa}
S.~Choudhury, ``{Reconstructing inflationary paradigm within Effective Field
  Theory framework},'' \href{http://dx.doi.org/10.1016/j.dark.2015.11.003}{{\em
  Phys. Dark Univ.} {\bfseries 11} (2016) 16--48},
  \href{http://arxiv.org/abs/1508.00269}{{\ttfamily arXiv:1508.00269
  [astro-ph.CO]}}.

\bibitem{Choudhury:2015hvr}
S.~Choudhury and S.~Panda, ``{COSMOS-e{\textquoteright}-GTachyon from string
  theory},'' \href{http://dx.doi.org/10.1140/epjc/s10052-016-4072-2}{{\em Eur.
  Phys. J. C} {\bfseries 76} no.~5, (2016) 278},
  \href{http://arxiv.org/abs/1511.05734}{{\ttfamily arXiv:1511.05734
  [hep-th]}}.

\bibitem{Choudhury:2017cos}
S.~Choudhury, ``{COSMOS-$e'$- soft Higgsotic attractors},''
  \href{http://dx.doi.org/10.1140/epjc/s10052-017-5001-8}{{\em Eur. Phys. J. C}
  {\bfseries 77} no.~7, (2017) 469},
  \href{http://arxiv.org/abs/1703.01750}{{\ttfamily arXiv:1703.01750
  [hep-th]}}.

\bibitem{Choudhury:2025kxg}
S.~Choudhury, ``{Stochastic origin of primordial fluctuations in the sky},''
  \href{http://dx.doi.org/10.1142/S0218271825440237}{{\em Int. J. Mod. Phys. D}
  {\bfseries 34} no.~16, (2025) 2544023},
  \href{http://arxiv.org/abs/2503.17635}{{\ttfamily arXiv:2503.17635 [gr-qc]}}.

\bibitem{Choudhury:2024kjj}
S.~Choudhury, K.~Dey, S.~Ganguly, A.~Karde, S.~K. Singh, and P.~Tiwari,
  ``{Negative non-Gaussianity as a salvager for PBHs with PTAs in bounce},''
  \href{http://dx.doi.org/10.1140/epjc/s10052-025-14176-z}{{\em Eur. Phys. J.
  C} {\bfseries 85} no.~4, (2025) 472},
  \href{http://arxiv.org/abs/2409.18983}{{\ttfamily arXiv:2409.18983
  [astro-ph.CO]}}.

\bibitem{Choudhury:2024aji}
S.~Choudhury and M.~Sami, ``{Large fluctuations and primordial black holes},''
  \href{http://dx.doi.org/10.1016/j.physrep.2024.10.007}{{\em Phys. Rept.}
  {\bfseries 1103} (2025) 1--276},
  \href{http://arxiv.org/abs/2407.17006}{{\ttfamily arXiv:2407.17006 [gr-qc]}}.

\bibitem{Choudhury:2024dzw}
S.~Choudhury, S.~Ganguly, S.~Panda, S.~SenGupta, and P.~Tiwari, ``{Obviating
  PBH overproduction for SIGWs generated by pulsar timing arrays in loop
  corrected EFT of bounce},''
  \href{http://dx.doi.org/10.1088/1475-7516/2024/09/013}{{\em JCAP} {\bfseries
  09} (2024) 013}, \href{http://arxiv.org/abs/2407.18976}{{\ttfamily
  arXiv:2407.18976 [astro-ph.CO]}}.

\bibitem{Choudhury:2024dei}
S.~Choudhury, A.~Karde, S.~Panda, and S.~SenGupta,
  ``{Regularized-renormalized-resummed loop corrected power spectrum of
  non-singular bounce with Primordial Black Hole formation},''
  \href{http://dx.doi.org/10.1140/epjc/s10052-024-13460-8}{{\em Eur. Phys. J.
  C} {\bfseries 84} no.~11, (2024) 1149},
  \href{http://arxiv.org/abs/2405.06882}{{\ttfamily arXiv:2405.06882
  [astro-ph.CO]}}.

\bibitem{Choudhury:2024jlz}
S.~Choudhury, A.~Karde, P.~Padiyar, and M.~Sami, ``{Primordial black holes from
  effective field theory of stochastic single field inflation at NNNLO},''
  \href{http://dx.doi.org/10.1140/epjc/s10052-024-13644-2}{{\em Eur. Phys. J.
  C} {\bfseries 85} no.~1, (2025) 21},
  \href{http://arxiv.org/abs/2403.13484}{{\ttfamily arXiv:2403.13484
  [astro-ph.CO]}}.

\bibitem{Choudhury:2024ybk}
S.~Choudhury, ``{Large fluctuations in the sky},''
  \href{http://dx.doi.org/10.1142/S0218271824410074}{{\em Int. J. Mod. Phys. D}
  {\bfseries 33} no.~15, (2024) 2441007},
  \href{http://arxiv.org/abs/2403.07343}{{\ttfamily arXiv:2403.07343
  [astro-ph.CO]}}.

\bibitem{Choudhury:2024one}
S.~Choudhury, A.~Karde, S.~Panda, and M.~Sami, ``{Realisation of the ultra-slow
  roll phase in Galileon inflation and PBH overproduction},''
  \href{http://dx.doi.org/10.1088/1475-7516/2024/07/034}{{\em JCAP} {\bfseries
  07} (2024) 034}, \href{http://arxiv.org/abs/2401.10925}{{\ttfamily
  arXiv:2401.10925 [astro-ph.CO]}}.

\bibitem{Choudhury:2023fjs}
S.~Choudhury, K.~Dey, and A.~Karde, ``{Untangling PBH overproduction in
  $w$-SIGWs generated by Pulsar Timing Arrays for MST-EFT of single field
  inflation},'' \href{http://arxiv.org/abs/2311.15065}{{\ttfamily
  arXiv:2311.15065 [astro-ph.CO]}}.

\bibitem{Choudhury:2023fwk}
S.~Choudhury, K.~Dey, A.~Karde, S.~Panda, and M.~Sami, ``{Primordial
  non-Gaussianity as a saviour for PBH overproduction in SIGWs generated by
  pulsar timing arrays for Galileon inflation},''
  \href{http://dx.doi.org/10.1016/j.physletb.2024.138925}{{\em Phys. Lett. B}
  {\bfseries 856} (2024) 138925},
  \href{http://arxiv.org/abs/2310.11034}{{\ttfamily arXiv:2310.11034
  [astro-ph.CO]}}.

\bibitem{Choudhury:2023hfm}
S.~Choudhury, A.~Karde, S.~Panda, and M.~Sami, ``{Scalar induced gravity waves
  from ultra slow-roll galileon inflation},''
  \href{http://dx.doi.org/10.1016/j.nuclphysb.2024.116678}{{\em Nucl. Phys. B}
  {\bfseries 1007} (2024) 116678},
  \href{http://arxiv.org/abs/2308.09273}{{\ttfamily arXiv:2308.09273
  [astro-ph.CO]}}.

\bibitem{Choudhury:2023kdb}
S.~Choudhury, A.~Karde, S.~Panda, and M.~Sami, ``{Primordial non-Gaussianity
  from ultra slow-roll Galileon inflation},''
  \href{http://dx.doi.org/10.1088/1475-7516/2024/01/012}{{\em JCAP} {\bfseries
  01} (2024) 012}, \href{http://arxiv.org/abs/2306.12334}{{\ttfamily
  arXiv:2306.12334 [astro-ph.CO]}}.

\bibitem{Choudhury:2023hvf}
S.~Choudhury, S.~Panda, and M.~Sami, ``{Galileon inflation evades the no-go for
  PBH formation in the single-field framework},''
  \href{http://dx.doi.org/10.1088/1475-7516/2023/08/078}{{\em JCAP} {\bfseries
  08} (2023) 078}, \href{http://arxiv.org/abs/2304.04065}{{\ttfamily
  arXiv:2304.04065 [astro-ph.CO]}}.

\bibitem{Choudhury:2023rks}
S.~Choudhury, S.~Panda, and M.~Sami, ``{Quantum loop effects on the power
  spectrum and constraints on primordial black holes},''
  \href{http://dx.doi.org/10.1088/1475-7516/2023/11/066}{{\em JCAP} {\bfseries
  11} (2023) 066}, \href{http://arxiv.org/abs/2303.06066}{{\ttfamily
  arXiv:2303.06066 [astro-ph.CO]}}.

\bibitem{Choudhury:2023jlt}
S.~Choudhury, S.~Panda, and M.~Sami, ``{PBH formation in EFT of single field
  inflation with sharp transition},''
  \href{http://dx.doi.org/10.1016/j.physletb.2023.138123}{{\em Phys. Lett. B}
  {\bfseries 845} (2023) 138123},
  \href{http://arxiv.org/abs/2302.05655}{{\ttfamily arXiv:2302.05655
  [astro-ph.CO]}}.

\bibitem{Choudhury:2023vuj}
S.~Choudhury, M.~R. Gangopadhyay, and M.~Sami, ``{No-go for the formation of
  heavy mass Primordial Black Holes in Single Field Inflation},''
  \href{http://dx.doi.org/10.1140/epjc/s10052-024-13218-2}{{\em Eur. Phys. J.
  C} {\bfseries 84} no.~9, (2024) 884},
  \href{http://arxiv.org/abs/2301.10000}{{\ttfamily arXiv:2301.10000
  [astro-ph.CO]}}.

\bibitem{Choudhury:2013woa}
S.~Choudhury and A.~Mazumdar, ``{Primordial blackholes and gravitational waves
  for an inflection-point model of inflation},''
  \href{http://dx.doi.org/10.1016/j.physletb.2014.04.050}{{\em Phys. Lett. B}
  {\bfseries 733} (2014) 270--275},
  \href{http://arxiv.org/abs/1307.5119}{{\ttfamily arXiv:1307.5119
  [astro-ph.CO]}}.

\bibitem{Choudhury:2011jt}
S.~Choudhury and S.~Pal, ``{Fourth level MSSM inflation from new flat
  directions},'' \href{http://dx.doi.org/10.1088/1475-7516/2012/04/018}{{\em
  JCAP} {\bfseries 04} (2012) 018},
  \href{http://arxiv.org/abs/1111.3441}{{\ttfamily arXiv:1111.3441 [hep-ph]}}.

\bibitem{Konstantinidis:2016nio}
G.~Konstantinidis, R.~Mahajan, and E.~Shaghoulian, ``{Late-time Structure of
  the Bunch-Davies FRW Wavefunction},''
  \href{http://dx.doi.org/10.1007/JHEP10(2016)103}{{\em JHEP} {\bfseries 10}
  (2016) 103}, \href{http://arxiv.org/abs/1608.06163}{{\ttfamily
  arXiv:1608.06163 [hep-th]}}.

\bibitem{McFadden:2011kk}
P.~McFadden and K.~Skenderis, ``{Cosmological 3-point correlators from
  holography},'' \href{http://dx.doi.org/10.1088/1475-7516/2011/06/030}{{\em
  JCAP} {\bfseries 06} (2011) 030},
  \href{http://arxiv.org/abs/1104.3894}{{\ttfamily arXiv:1104.3894 [hep-th]}}.

\bibitem{Bzowski:2012ih}
A.~Bzowski, P.~McFadden, and K.~Skenderis, ``{Holography for inflation using
  conformal perturbation theory},''
  \href{http://dx.doi.org/10.1007/JHEP04(2013)047}{{\em JHEP} {\bfseries 04}
  (2013) 047}, \href{http://arxiv.org/abs/1211.4550}{{\ttfamily arXiv:1211.4550
  [hep-th]}}.

\bibitem{Chluba_2015}
J.~Chluba, J.~Hamann, and S.~P. Patil, ``Features and new physical scales in
  primordial observables: Theory and observation,''
  \href{http://dx.doi.org/10.1142/s0218271815300232}{{\em International Journal
  of Modern Physics D} {\bfseries 24} no.~10, (Aug, 2015) 1530023}.
  \url{http://dx.doi.org/10.1142/S0218271815300232}.

\bibitem{Schlegel:2019eqc}
D.~J. Schlegel {\em et~al.}, ``{Astro2020 APC White Paper: The MegaMapper: a z
  \ensuremath{>} 2 Spectroscopic Instrument for the Study of Inflation and Dark
  Energy},'' \href{http://arxiv.org/abs/1907.11171}{{\ttfamily arXiv:1907.11171
  [astro-ph.IM]}}.

\bibitem{2018AAS...23135423D}
O.~{Dore} and {SPHEREx Science Team}, ``{SPHEREx: Probing the Physics of
  Inflation with an All-Sky Spectroscopic Galaxy Survey},'' in {\em American
  Astronomical Society Meeting Abstracts \#231}, vol.~231 of {\em American
  Astronomical Society Meeting Abstracts}, p.~354.23.
\newblock Jan., 2018.

\bibitem{Ade_2019}
P.~Ade, J.~Aguirre, Z.~Ahmed, S.~Aiola, A.~Ali, D.~Alonso, M.~A. Alvarez,
  K.~Arnold, P.~Ashton, J.~Austermann, and et~al., ``The simons observatory:
  science goals and forecasts,''
  \href{http://dx.doi.org/10.1088/1475-7516/2019/02/056}{{\em Journal of
  Cosmology and Astroparticle Physics} {\bfseries 2019} no.~02, (Feb, 2019)
  056–056}. \url{http://dx.doi.org/10.1088/1475-7516/2019/02/056}.

\bibitem{CMB-S4:2020lpa}
{\bfseries CMB-S4} Collaboration, K.~Abazajian {\em et~al.}, ``{CMB-S4:
  Forecasting Constraints on Primordial Gravitational Waves},''
  \href{http://arxiv.org/abs/2008.12619}{{\ttfamily arXiv:2008.12619
  [astro-ph.CO]}}.

\bibitem{Planck:2019kim}
{\bfseries Planck} Collaboration, Y.~Akrami {\em et~al.}, ``{Planck 2018
  results. IX. Constraints on primordial non-Gaussianity},''
  \href{http://dx.doi.org/10.1051/0004-6361/201935891}{{\em Astron. Astrophys.}
  {\bfseries 641} (2020) A9}, \href{http://arxiv.org/abs/1905.05697}{{\ttfamily
  arXiv:1905.05697 [astro-ph.CO]}}.

\end{thebibliography}\endgroup

\end{document}